\documentclass[journal,onecolumn]{IEEEtran}

\ifCLASSINFOpdf
\else
\fi

\usepackage{mathtools}
\usepackage{amsmath,amsfonts} 
\usepackage{algorithmic}
\usepackage{graphicx}
\usepackage{textcomp}
\usepackage{dirtytalk}
\usepackage{listings}
\usepackage[dvipsnames]{xcolor}
\usepackage{siunitx}

\usepackage{mdframed}

\usepackage{threeparttable}
\usepackage{pifont} 
\newcommand{\orbit}[1]{\raisebox{.5pt}{\textcircled{\raisebox{-.9pt}{#1}}}}

\definecolor{codeblue}{rgb}{0.0, 0.0, 0.5}
\definecolor{codegreen}{rgb}{0.0, 0.5, 0.0}
\definecolor{codegray}{rgb}{0.5, 0.5, 0.5}
\definecolor{codepurple}{rgb}{0.58, 0.0, 0.82}

\lstdefinestyle{mystyle}{
backgroundcolor=\color{lightgray!15},
commentstyle=\color{codegreen},
keywordstyle=\color{codeblue},
numberstyle=\tiny\color{codegray},
stringstyle=\color{codepurple},
basicstyle=\ttfamily\tiny, 
breakatwhitespace=false,
breaklines=true,
captionpos=b,
keepspaces=true,
numbers=left,
numbersep=5pt,
showspaces=false,
showstringspaces=false,
showtabs=false,
tabsize=2
}

\definecolor{delim}{RGB}{20,105,176}
\definecolor{numb}{RGB}{106, 109, 32}
\definecolor{string}{rgb}{0.64,0.08,0.08}

\lstdefinelanguage{json}{
    numbers=left,
    numberstyle=\small,
    frame=single,
    rulecolor=\color{black},
    showspaces=false,
    showtabs=false,
    breaklines=true,
    postbreak=\raisebox{0ex}[0ex][0ex]{\ensuremath{\color{gray}\hookrightarrow\space}},
    breakatwhitespace=true,
    basicstyle=\ttfamily\small,
    upquote=true,
    morestring=[b]",
    stringstyle=\color{string},
    literate=
     *{0}{{{\color{numb}0}}}{1}
      {1}{{{\color{numb}1}}}{1}
      {2}{{{\color{numb}2}}}{1}
      {3}{{{\color{numb}3}}}{1}
      {4}{{{\color{numb}4}}}{1}
      {5}{{{\color{numb}5}}}{1}
      {6}{{{\color{numb}6}}}{1}
      {7}{{{\color{numb}7}}}{1}
      {8}{{{\color{numb}8}}}{1}
      {9}{{{\color{numb}9}}}{1}
      {\{}{{{\color{delim}{\{}}}}{1}
      {\}}{{{\color{delim}{\}}}}}{1}
      {[}{{{\color{delim}{[}}}}{1}
      {]}{{{\color{delim}{]}}}}{1},
}

\usepackage{enumitem}
\usepackage{booktabs}        
\usepackage{multirow}
\usepackage{array, colortbl}
\usepackage{ulem, comment}
\usepackage{lscape}
\usepackage{longtable}
\usepackage{adjustbox}
\usepackage{slashbox}
\usepackage{csquotes}
\renewcommand{\mkbegdispquote}[2]{\itshape}
\usepackage{soul,color}
\usepackage{tabularray}
\usepackage{tabularx}
\usepackage{ragged2e} 

\usepackage{hyperref}

\usepackage{caption}
\usepackage{subcaption}

\usepackage{ifthen}

\usepackage{url}
 \usepackage{amssymb} 
\newboolean{showcomments}
\setboolean{showcomments}{true} 
\ifthenelse{\boolean{showcomments}}
{\newcommand{\nb}[2]{
		\fcolorbox{black}{yellow}{\bfseries\sffamily\scriptsize#1}
		{\sf\small$\blacktriangleright$\textit{#2}$\blacktriangleleft$}
	}
	
}
{\newcommand{\nb}[2]{}
	
}

\usepackage{amsmath}

\usepackage[compact]{titlesec}
\newcommand{\rqone}{\textbf{What are the most prominent threat TTPs and their common entry points in ML attack scenarios?  }}

\newcommand{\rqtwo}{\textbf{What is the effect of threat TTPs on different ML phases and models?  }}

\newcommand{\rqthree}{\textbf{ 
What previously undocumented security threats can be identified in the AI Incident Database, the literature, and ML repositories that are missing from the ATLAS database? 
}}

\usepackage[many]{tcolorbox}

\DeclareMathOperator*{\argmax}{arg\,max}

\newcommand{\tosemrev}[1]{\textcolor{black}{#1}}

\newcommand{\majorev}[1]{\textcolor{black}{#1}}
\newcommand{\minor}[1]{\textcolor{black}{#1}}

\newcommand{\etal}{\textit{et al.}}
\newtcolorbox{boxblock}[2][]{
top=0.15in,left=4pt,right=4pt,bottom=4pt,
fonttitle=\bfseries,
colbacktitle=gray,
colback=gray!5,
colframe=gray!40!black,
enhanced,
attach boxed title to top left={xshift=1.5em,yshift=-\tcboxedtitleheight/2},
boxed title style={size=small},
drop shadow={black!50!white},
title=#2,#1}

\definecolor{codebg}{RGB}{248,248,250}      
\lstdefinestyle{cotstyle}{
  language=tex,
  basicstyle=\scriptsize\ttfamily,
  backgroundcolor=\color{codebg},
  frame=single,
  rulecolor=\color{black!20}
}

\AtBeginDocument{%
  \providecommand\BibTeX{{%
    \normalfont B\kern-0.5em{\scshape i\kern-0.25em b}\kern-0.8em\TeX}}}

\usepackage{stfloats}

\begin{document}
%
\title{Multi-Agent Framework for Threat Mitigation and Resilience in AI–Based Systems}
%
%

\author{Armstrong~Foundjem,~\IEEEmembership{sMIEEE,}
       Lionel~Nganyewou~Tidjon,~\IEEEmembership{sMIEEE,}
       Leuson~Da~Silva,
        and~Foutse~Khomh,~\IEEEmembership{sMIEEE}
\thanks{All authors are affiliated with the Department
of Computer and Software Engineering, Polytechnique Montreal, QC H3T 0A3, e-mail: \{a.foundjem,lionel.tidjon,leuson-mario-pedro.da-silva,foutse.khomh\}@polymtl.ca.}
}

\maketitle

\begin{abstract}
\noindent
Machine learning (ML) increasingly underpins foundation models and autonomous pipelines in high-stakes domains such as finance, healthcare, and national infrastructure, rendering these systems prime targets for sophisticated adversarial threats. Attackers now leverage advanced Tactics, Techniques, and Procedures (TTPs) spanning data poisoning, model extraction, prompt injection, automated jailbreaking, training data
exfiltration, and---more recently---\textbf{preference-guided black-box optimization} that exploits models' own comparative judgments to craft successful attacks iteratively. These emerging text-only, query-based methods demonstrate that larger and better-calibrated models can be paradoxically more vulnerable to introspection-driven jailbreaks and cross-modal manipulations. While traditional cybersecurity frameworks
offer partial mitigation, they lack ML-specific threat modeling and fail to capture evolving attack vectors across foundation, multimodal, and federated settings.  \textbf{Objective:}
This research empirically characterizes modern ML security risks by identifying dominant attacker TTPs,
exposed vulnerabilities, and lifecycle stages most frequently targeted in foundation-model, multimodal,
and retrieval-augmented (RAG) pipelines. The study also assesses the scalability of current defenses against
generative and introspection-based attacks, highlighting the need for adaptive, ML-aware security mechanisms. \textbf{Methods:}
We conduct a large-scale empirical analysis of ML security, extracting 93 distinct threats from multiple
sources: real-world incidents in MITRE ATLAS (26), the AI Incident Database (12), and peer-reviewed
literature (55), supplemented by 854 ML repositories from GitHub and the Python Advisory database.
A multi-agent reasoning system with enhanced Retrieval-Augmented Generation (RAG)---powered by
ChatGPT-4o (temperature 0.4)---automatically extracts TTPs, vulnerabilities, and lifecycle stages from
over 300 scientific articles using evidence-grounded reasoning. The resulting ontology-driven threat graph
supports cross-source validation and lifecycle mapping. \textbf{Results:}
Our analysis uncovers multiple unreported threats beyond current ATLAS coverage, including
model-stealing attacks against commercial LLM APIs, data leakage through parameter memorization,
and preference-guided query optimization enabling text-only jailbreaks and multimodal adversarial examples.
\texttt{Gradient-based obstinate attacks}, \texttt{MASTERKEY automated jailbreaking},
\texttt{federated learning poisoning}, \texttt{diffusion backdoor embedding},
and \texttt{preference-oriented optimization leakage} emerge as dominant TTPs,
disproportionately impacting pretraining and inference.
Graph-based dependency analysis shows that specific ML libraries and model hubs exhibit dense
vulnerability clusters lacking effective issue-tracking and patch-propagation mechanisms.  \textbf{Conclusion:}
This study underscores the urgent need for adaptive, ML-specific security frameworks that address
introspection-based and preference-guided attacks alongside classical adversarial vectors.
Robust dependency management, automated threat intelligence, and continuous monitoring are essential
to mitigate supply-chain and inference-time risks throughout the ML lifecycle.
By unifying empirical evidence from incidents, literature, and repositories, this research delivers
a comprehensive threat landscape for next-generation AI systems and establishes a foundation for
proactive, multi-agent security governance in the era of large-scale and generative AI.
\end{abstract}


\IEEEpeerreviewmaketitle

\section{Introduction}\label{sec:intro}

Nowadays, Machine Learning (ML) is achieving significant success in dealing with various complex problems in safety-critical domains such as healthcare~\cite{kourou2015machine}
aviation~\cite{8903554}, automotive~\cite{kuutti2020survey}, railways~\cite{10.1007/978-3-030-18744-6_6}, and space~\cite{girimonte2007artificial}. ML has also been applied in cybersecurity to detect threatening anomalous behaviors such as spam, malware, and malicious URLs~\cite{8735821}, allowing a system to respond to real-time inputs containing both normal and suspicious data and learn to reject malicious behavior. 
While ML is strengthening defense systems, it also helps threat actors improve their tactics, techniques, and procedures (TTPs) and expand their attack surface. Attackers leverage the black-box nature of ML models and manipulate input data to affect their performance~\cite{carlini2021extracting,jagielski2018manipulating, biggio2018wild, 8294186}.

Early work~\cite{8294186,carlini2021extracting,arp2022and,morris2020textattack,pierazzi2020intriguing,tramer2020adaptive,carlini2019evaluating,abdullah2019practical,eykholt2018robust,biggio2018wild} outlined ML attacks and defenses targeting different phases of the ML lifecycle, i.e., input data, training, inference, and monitoring. ML-based systems are also often deployed on-premise or on cloud service providers, which increases attack vectors and makes them vulnerable to traditional attacks at different layers, like software, system, and network levels. At the software level, ML-based systems are vulnerable to operating system (OS) attacks since attackers can exploit the OS. At the system level, ML-based systems are vulnerable to attacks, including CPU side-channel~\cite{7163050} and memory-based~\cite{197211}. 
Finally, at the network level, ML-based systems can be compromised under attacks~\cite{8735821}, including Denial of Service (DoS), botnets, and ransomware. To achieve their goals, ML threat actors can poison data and fool ML-based systems using different strategies, like 
\mbox{evasion~\cite{8294186,biggio2018wild,morris2020textattack,abdullah2019practical}}, \mbox{extraction~\cite{carlini2021extracting,jagielski2020high,orekondy2019knockoff}}, inference\mbox{~\cite{shokri2017membership,8294186}}, and poisoning\mbox{~\cite{biggio2018wild,jagielski2018manipulating,morris2020textattack, 8294186,abdullah2019practical}}. To defend against such threats, adversarial defenses have been proposed \mbox{~\cite{arp2022and,tramer2020adaptive}}. Usually, threat TTPs and mitigations are reported in a threat assessment framework to help conduct attack and defense operations. Unfortunately, there is a lack of concrete applications of threat assessment in the ML field that provide a broader overview of ML threats, ML tool vulnerabilities, and mitigation solutions.
\tosemrev{
The goal of this study is to systematically characterize ML security threats, assess their impact on ML components (phases, models, tools), and identify effective mitigation strategies. While ML enhances various domains, its black-box nature and deployment across software, system, and network layers expose it to adversarial attacks such as evasion, extraction, inference, and poisoning. Existing research lacks a unified threat assessment framework that maps ML vulnerabilities, attack tactics, and mitigation strategies across different lifecycle stages. To achieve this goal, we conduct an empirical investigation, integrating real-world threat intelligence to analyze ML-specific security risks, classify TTPs, and propose structured mitigation solutions, enhancing ML security frameworks against evolving threats.
}

\tosemrev{Thus, we asked the following research questions (RQs):
\begin{enumerate}
    \item \rqone
    \item \rqtwo
    \item \rqthree
\end{enumerate}
}
Results suggest that Convolutional neural networks (e.g., GPT2, Fisheye, Copycat, ResNet) are one of the most targeted models in attack scenarios. ML repositories such as \textit{TensorFlow}, \textit{OpenCV}, \textit{Notebook}, and \textit{Numpy} have the largest vulnerability prominence. The most severe dependencies that caused the vulnerabilities include \textit{tensorflow}, \textit{linux\_kernel}, \textit{vim}, \textit{openssl}, \textit{magemagick}, and 
\textit{pillow}. \textit{DoS}, \textit{improper input validation}, and \textit{buffer-overflow} were the most frequent in ML repositories. Our examinations of vulnerabilities and attacks reveal that \textit{testing}, \textit{inference}, \textit{training},
and \textit{data collection} are the most targeted ML phases by threat actors. The mitigation of these vulnerabilities and threats includes adversarial~\cite{arp2022and,tramer2020adaptive,aigrain2019detecting} and traditional defenses such as software updates, and cloud security policies (e.g., zero trust). Leveraging our findings, ML red/blue teams can take advantage of the ATLAS TTPs and the newly identified TTPs from the AI incident database to better conduct attacks/defenses using the most exploited TTPs and models for more impact.

Since ML-based systems are increasingly in production, ML practitioners can leverage these results to prevent vulnerabilities and threats in ML products during their lifecycle. Researchers can also use the results to propose theories and algorithms for strengthening defenses.

\minor{
\textbf{Contributions.}
This paper advances ML threat assessment by integrating attacker Tactics, Techniques, and Procedures (TTPs)
from frameworks such as MITRE ATLAS and ATT\&CK with real-world vulnerabilities across the ML lifecycle.
Our main contributions are:
\begin{enumerate}
    \item \textbf{Lifecycle-Centric Threat Framework.}
    A unified mapping of TTPs to vulnerabilities across data, software, and system layers—spanning
    collection, training, deployment, and inference—enabling holistic reasoning on cascading risks.
    \item \textbf{State-of-the-Art Model Analysis.}
    Extension of threat assessment to foundation and multimodal models
    (\textsc{GPT-4o}, \textsc{Claude-3.5}, \textsc{Gemini-1.5}, \textsc{LLaMA-3.2}, \textsc{DeepSeek-R1}, etc.),
    revealing composite backdoors, tokenization exploits, and prompt-injection jailbreaks
    that generalize across modalities.
    \item \textbf{Scalable Multi-Agent Mapping.}
    A modular three-step pipeline combining retrieval-augmented reasoning, ontology alignment,
    and heterogeneous GNNs to scale threat mapping to RAG and RLHF pipelines.
    \item \textbf{Graph-Based Severity Estimation.}
    A heterogeneous GNN fusing code, model-artifact, and threat-intelligence features to learn
    a severity score~$\hat{s}$ validated against real-world \textsc{CVE}s and incident costs.
    \item \textbf{Automated Repository Mining.}
    Extraction of \textsc{CVE–CPE–tool} relations from GitHub/PyPI and construction of dependency
    graphs linking vulnerabilities to ML toolchains, exposing supply-chain risks.
    \item \textbf{Open Practitioner Toolkit.}
    A reproducible toolkit supporting cluster drill-downs and visual lifecycle mapping
    (stage~$\rightarrow$~vulnerability~$\rightarrow$~stakeholder) for researchers and practitioners.
\end{enumerate}
\noindent Together, these contributions deliver a scalable, evidence-driven foundation for analyzing and mitigating
threats in modern ML ecosystems, bridging traditional cybersecurity taxonomies and emerging AI security practice.
}
The rest of this paper is structured as follows. In Section~\ref{background}, we define basic concepts such as vulnerabilities and threats, and review the related literature. Section~\ref{methodology} describes the study methodology for threat assessment, defines some research questions, and presents a formal definition of an ML threat. In Section~\ref{sec:results}, we present results while answering the defined research questions. Section~\ref{sec:mitigation} proposes mitigation solutions for the observed threats and vulnerabilities. Section~\ref{sec:discussions} discusses the results and the application in corporate settings. In Section~\ref{threat2valid}, we present threats that could affect the validity of the reported results. Section~\ref{conclusion} concludes the paper and outlines avenues for future work.

\section{Background and related work}\label{background}

Before diving into ML threat assessment, generic security concepts such as assets, vulnerabilities, and threats must be defined. This section provides an overview of security concepts and related work.

\subsection{Assets}
\begin{figure*}[!ht]
\centering
\includegraphics[width=1\textwidth]{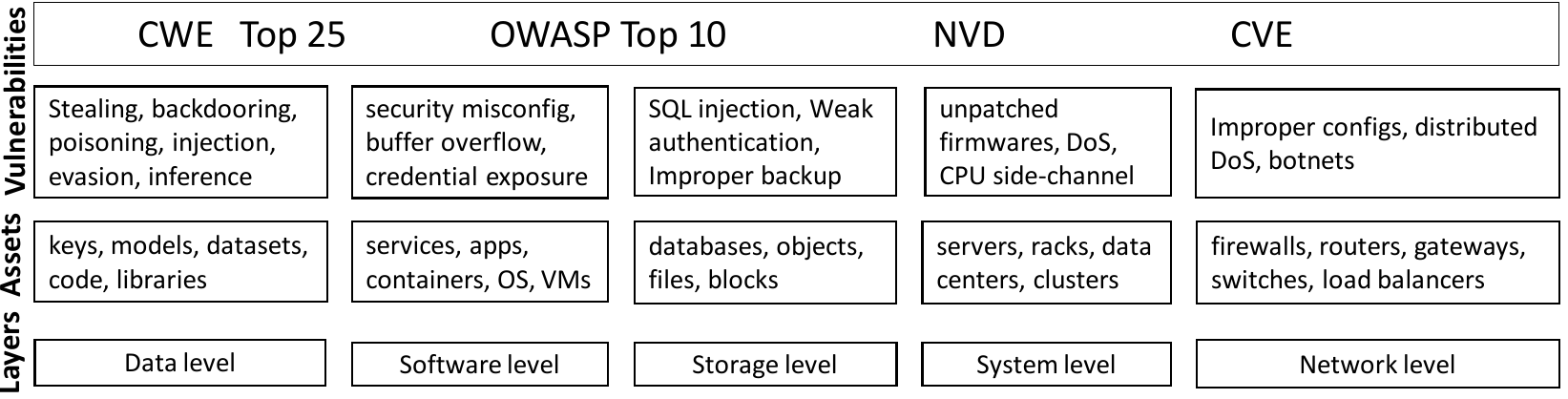}
\caption{\majorev{\textbf{Assets and vulnerabilities across ML/AI system layers.}  
Mapping five infrastructure layers—data, software, storage, system, and 
network, to their representative assets and prevalent vulnerabilities, drawing from 
CWE~Top~25, OWASP~Top~10, NVD, and CVE taxonomies.}} 
\label{fig:vuln_arch}
\end{figure*}

In computer security, an \emph{asset} is any valuable logical or physical resource 
owned by an organization, such as data, software, hardware, storage, or network 
infrastructure (see Fig.~\ref{fig:vuln_arch}).  
In ML systems, assets span five layers:  
\textbf{Data level}—access credentials (tokens, passwords, cryptographic keys, 
certificates), datasets, models and parameters, source code, and libraries.  
\textbf{Software level}—Machine-Learning-as-a-Service (MaaS) APIs, production ML 
applications, containers, and virtual machines (VMs).  
\textbf{Storage level}—databases, object stores (e.g., buckets), files, and block 
storage hosting training data, models, or code.  
\textbf{System level}—servers, racks, data centers, and compute clusters.  
\textbf{Network level}—firewalls, routers, gateways, switches, and load balancers.

Data-level assets (models, datasets, code, keys) face threats such as theft, poisoning, backdooring, and evasion. Software-level components (services, apps, OS, VMs) are susceptible to misconfiguration, buffer overflow, and credential exposure. Storage systems (databases, files, blocks) are vulnerable to SQL injection, weak authentication, and improper backups. System-level hardware (servers, clusters) can be exploited via unpatched firmware, DoS, and side-channel attacks. Network devices (firewalls, routers) are vulnerable to misconfiguration, DDoS attacks, and botnet infiltration. This layered view supports targeted risk assessment and defense prioritization.

\subsection{Vulnerabilities}

A vulnerability is a software or hardware flaw that threat actors can exploit to execute malicious command-and-control (C2) operations, such as data theft and destruction. 

\subsubsection*{\textbf{Types of vulnerabilities.}}

Vulnerabilities occur at different levels: \textit{data}, \textit{software}, \textit{storage}, \textit{system}, and \textit{network} (see \mbox{Fig~\ref{fig:vuln_arch}}). At the \textit{data} level, data assets are vulnerable to model stealing, backdooring~\cite{gao2020backdoor}, poisoning, injection, evasion, and inference. At the \textit{software} level, threat actors look for errors or bugs in ML apps, such as buffer overflows, exposed credentials, and security misconfigurations. At the \textit{storage} level, ML databases and cloud storage are vulnerable to weak authentication, improper backup, and SQL injection attacks. At the \textit{system} level, they exploit several hardware vulnerabilities, including firmware unpatching and CPU side-channel attacks~\cite {7163050}, to launch attacks such as DoS, affecting the ML cloud infrastructure where ML apps are managed. 
At \textit{network} level, threat actors can exploit improper configurations of the network, making it vulnerable to distributed DoS and botnet attacks. These vulnerabilities are reported in the Common Weaknesses Exposure (CWE) Top 25~\cite{mitrecwe}, the OWASP Top 10~\cite{owasp}, the National Vulnerability Database (NVD), and the Common Vulnerability and Exposures (CVE) standards. \tosemrev{ Lastly, CPE dependency in cybersecurity refers to a vulnerability’s reliance on specific Common Platform Enumeration (CPE) components that describe hardware, software, or firmware. The CPE name and version indicate the affected systems, which is essential for identifying and managing associated risks.}

\subsection{Threats}\label{sota_attack}

A threat exploits a given vulnerability to damage and/or destroy a target asset. Threats can be of two types: \textit{insider} threats and \textit{outsider} threats. \textit{Insider} threats originate from the internal system, and they are more often executed by a trusted entity of the system (e.g., employee). \textit{Outsider} threats are operated from the remote/external system. In the following, we distinguish between traditional threats and recent machine learning threats. 

\subsubsection*{\textbf{Traditional threats.}}\label{trad_attacks}

Adversarial Tactics, Techniques, and Common Knowledge (ATT\&CK)~\cite{attack_desc} is a public and standard knowledge database of attack TTPs. Traditional attack phases are divided into two groups: conventional pre-attack phases and attack phases.

\textit{\underline{Pre-attack}}. The pre-attack phase consists of two tactics: reconnaissance and resource development~\cite{attack_desc}. During reconnaissance, attackers use several techniques, including network scanning to identify a victim's open ports and OS version (e.g., nmap, censys), and phishing to embed malicious links in emails or SMS messages. During resource development, attackers use several techniques, including acquiring resources to support C2 operations (e.g., domains), purchasing a network of compromised systems (e.g., a botnet) for C2, developing tools (e.g., crawlers, exploit toolkits), and phishing.

\textit{\underline{Attack.}} 
Once the pre-attack phase is complete, attackers will attempt to gain initial access to the target victim host or network by delivering a malicious file or link through phishing, or by exploiting vulnerabilities in the websites/software used by victims. 
Also, attackers manipulate software dependencies and development tools before they are delivered to the final consumer. Upon successful initial access, they will execute malicious code on the victim host/network. After execution, they will attempt to persist on the target by modifying registries (e.g., Run Keys/Startup Folder), and automatically executing at boot. In addition, attackers will try to gain high-level permissions (e.g., as root/administrator). To hide their malicious activities, they will ensure they remain undetected by installing antivirus or Endpoint Detection Response (EDR) tools. An attacker can also execute lateral movement techniques, such as exploiting remote services, to spread to other hosts or networks and achieve greater impact. 

\subsubsection*{Machine learning threats}

Adversarial Threat Landscape for Artificial Intelligence Systems (ATLAS)~\cite{mitreatlas} is a public and standard knowledge database of adversarial TTPs for ML-based systems~\cite{mitreatlas}.  
ML attack phases are divided into two groups: ML pre-attack phases and attack phases.

\textit{\underline{ML Pre-attack.}}
ML pre-attack tactics are similar to those used in traditional threats, but with new techniques and procedures adapted to the ML context~\cite{mitreatlas}. During reconnaissance, Threat actors will search for the victim's publicly available research materials, such as technical blogs and pre-print repositories, and search for public ML artifacts, such as development tools (e.g., TensorFlow). For resource development, they will also acquire adversarial ML attack implementations such as adversarial robustness~\cite{pmlr-v97-engstrom19a} toolbox (ART).

\mbox{\textit{\underline{ML Attack.}}} ML systems are vulnerable to traditional attacks and other kinds of attacks that turn their normal behaviors into threatening behaviors called adversarial attacks. Like traditional threats, ML threats target the confidentiality, integrity, and availability of data. To achieve the goal, attackers may have full knowledge (white-box), partial knowledge (gray-box), or no knowledge (black-box) of the targeted ML system. In black-box settings, attackers do not have access to the training dataset, the model, or the executing code (since assets are hosted on a private corporate network). Still, they can access the public ML API as a legitimate user. This allows them to only perform queries and observe \mbox{outputs~\cite{mitreatlas101}}.

In white-box settings, attackers have knowledge of the model architecture and can access the training dataset or model to manipulate the training process. In gray-box settings, they have either a partial knowledge of the model architecture or some information about the training process. Whether white-box, gray-box, or black-box attacks, they can be targeted (focused on a particular class/sample) or untargeted (applied to any class/sample with no specific choice) to cause models to misclassify inputs. Different attack techniques are used: poisoning, evasion, extraction, and inference. During poisoning~\cite{jagielski2018manipulating, biggio2018wild, eykholt2018robust, chen2021badnl, morris2020textattack, abdullah2021hear, abdullah2019practical, abdullah2021sok}, attackers inject false training data to corrupt the learning model (even allowing it to be \mbox{backdoored~\cite{gao2020backdoor}}) to achieve an expected goal at inference time. During evasion~\cite{biggio2013evasion, biggio2018wild, pierazzi2020intriguing, kwon2019selective}, attackers iteratively and carefully modify ML API queries and observe the output at inference \mbox{time~\cite{mitreatlas101}}. The queries seem normal, but are misclassified by ML models. During extraction~\cite{carlini2021extracting, jagielski2020high, orekondy2019knockoff, schonherr2018adversarial}, attackers iteratively query the online model~\cite{mitreatlas101} allowing them to extract information about the model. Then, they use this information to gradually train a substitute model that mimics the target model's predictive behavior. During inference~\cite{shokri2017membership, choquette2021label, nasr2019comprehensive}, attackers probe the online model with different queries. Based on the results, they can infer whether features are used to train the model, which may compromise private data.

The adversarial models used~\cite{8294186} for attack include (1) \textit{fast gradient sign method (FGSM)} which consists in adding noise with the same direction as the gradient of the cost function w.r.t to data, (2) \textit{DeepFool} efficiently computes perturbations that fool deep networks, (3) \textit{Carlini and Wagner (C\&W)} is a set of three attacks against defensive distilled neural networks~\cite{papernot2016distillation}, 
(4) \textit{Jacobian-based saliency map (JSMA)} saturates a few pixels in an image to their maximum/minimum values, (5) \textit{universal adversarial perturbations} are agnostic-image perturbations that can fool a network on \textit{any} image with high probability, (6) \textit{Basic Iterative Method (BIM)} is an iterative version of the FGSM, (7) \textit{one pixel} is when a single pixel in the image is changed to fool classifiers, (8) \textit{Iterative Least-likely Class Method (ILCM)} is an extension of BIM where an image label is replaced by a target label of the least-likely class predicted by a classifier, and (9) \textit{Adversarial Transformation Networks (ATNs)} turns any input into an adversarial attack on the target network, while disrupting the original inputs and outputs of the target network as little as possible. Table~\ref{tab:traditional-vs-mlthreats} provides a comparison where conventional security controls might suffice versus where novel defenses—such as adversarial training, robust model architectures, and differential privacy—are required. 

\begin{table}[h]
    \centering
    \caption{Comparison of Traditional vs. ML-Specific Threats. Each property is contrasted to highlight the unique nature of ML-based systems and where traditional methods need adaptation.}
    \label{tab:traditional-vs-mlthreats}
    \resizebox{1.0\textwidth}{!}{%
    \begin{tabular}{p{0.25\textwidth}p{0.70\textwidth}p{0.80\textwidth}}
    \toprule
    \textbf{Property} & \textbf{Traditional Threats} & \textbf{ML/AI-Specific Threats}\\
    \midrule
    \textbf{Attack Surface} 
      & Mainly network endpoints, OS vulnerabilities, user credentials, etc. 
      & Training data pipelines, learned model parameters, inference-time inputs \\
    \addlinespace[3pt]
    \textbf{Adversary’s Knowledge} 
      & Typically partial or zero-knowledge about system internals (black-box) 
      & Varies from black-box to full white-box of ML model (depends on threat model)\\
    \addlinespace[3pt]
    \textbf{Attack Goal} 
      & Exfiltrate data, disrupt services, gain unauthorized system access 
      & Misclassify outputs, extract model IP, infer membership, degrade model performance \\
    \addlinespace[3pt]
    \textbf{Stealth Mechanism} 
      & Malware obfuscation, phishing, network-level exploits 
      & Imperceptible perturbations to inputs, or subtle data poisoning manipulations \\
    \addlinespace[3pt]
    \textbf{Impact on System} 
      & Potential data loss, financial damage, operational downtime 
      & Degraded accuracy, privacy leakage, model unavailability, or IP theft \\
    \addlinespace[3pt]
    \textbf{Required Expertise} 
      & Skilled in OS/network exploits, social engineering 
      & Data science \& ML knowledge plus exploit expertise \\
    \addlinespace[3pt]
    \textbf{Common Defenses} 
      & Firewalls, antivirus, patching, intrusion detection systems 
      & Adversarial training, differential privacy, secure model architectures \\
    \addlinespace[3pt]
    \textbf{Lifecycle Complexity} 
      & Security is mostly at the network, OS, or application layer 
      & Vulnerabilities spread across data collection, training, inference phases \\
    \bottomrule
    \end{tabular}}
\end{table}

\subsubsection{Threat, Vulnerability, and Incident Databases\\}\label{sec:threat-vul-incident-db}
\tosemrev{
The MITRE ATT\&CK framework, ATLAS, and the AI Incident Database are essential resources in the field of security for machine learning (ML) and AI systems. Each of these databases contains valuable information that helps understand, mitigate, and analyze threats to AI and ML systems.\\
\textbf{MITRE ATT\&CK. }
The MITRE ATT\&CK framework is a globally recognized, structured knowledge base that catalogs adversarial tactics, techniques, and procedures (TTPs). It focuses on real-world observations of threat actor behavior targeting various systems, including enterprise IT, cloud platforms, and industrial control systems. The framework provides detailed mappings between tactics (adversarial goals like privilege escalation) and techniques (specific methods to achieve those goals). The dataset is available via MITRE's TAXII server\footnote{\url{https://attack.mitre.org/resources/attack-data-and-tools/}} and in STIX format, allowing researchers to access structured data programmatically for vulnerability analysis and threat modeling.\\
\textbf{ATLAS (Adversarial Threat Landscape for Artificial-Intelligence Systems). }
ATLAS is a specialized knowledge base designed to document and catalog adversarial threats specific to AI and ML systems. Developed by MITRE, ATLAS builds upon the ATT\&CK framework but focuses exclusively on AI-related attack scenarios. It maps threats to particular phases of the ML lifecycle (e.g., data poisoning~\cite{huang2021data,Wang23:data-poisoning,Zhang21:dataPoisoning-attack} during training or adversarial attacks during deployment) and includes real-world examples of adversarial tactics. The ATLAS datasets can be accessed through its website using programmatic access (API) and may require scraping or specialized tools.\\
\textbf{AI Incident Database. }
The AI Incident Database is a community-driven repository that catalogs real-world incidents involving the failure or exploitation of AI systems. It documents issues such as data bias, adversarial attacks, and safety-critical errors. Unlike ATT\&CK and ATLAS, which focus on tactics and techniques, this database emphasizes the broader impacts of AI failures, including their societal and ethical consequences. The database is publicly accessible through its website, allowing users to browse or download records for research purposes. Data extraction can typically be performed through web scraping or API integration.\\
}

\subsubsection{ML Life-cycle stages documented in literature and production.\\ }\label{sec:ml_lifecycle_stages}
Following widely-adopted process models such as CRISP--DM, ISO/IEC~5338, and ISO/IEC~23053, and integrating modern post-training and operational practices, the machine learning life-cycle~\cite{Shankar2024Lifecycle,ashmore2021lifecycle,Schlegel2023Lifecycle,yu2020cloudleak} can be described as: 
\emph{(1)} Problem definition \& requirements elicitation; 
\emph{(2)} Data acquisition/collection; 
\emph{(3)} Data labeling/annotation; 
\emph{(4)} Data governance \& security (including PII handling and access control); 
\emph{(5)} Data preprocessing/augmentation; 
\emph{(6)} Feature engineering or tokenization; 
\emph{(7)} Pre-training (foundation model training); 
\emph{(8)} Fine-tuning or parameter-efficient adaptation (e.g., LoRA, adapters); 
\emph{(9)} Alignment through reinforcement learning from human feedback (RLHF) or AI feedback (RLAIF); 
\emph{(10)} Evaluation \& validation (including robustness, fairness, and reliability tests); 
\emph{(11)} Security testing/ red-teaming (e.g., jailbreaks, poisoning, extraction); 
\emph{(12)} Packaging \& registration (model artifacts, registry, and versioning); 
\emph{(13)} Deployment/serving (batch, online, or edge); 
\emph{(14)} Inference-time augmentation (retrieval-augmented generation, tool use, or agent integration); 
\emph{(15)} Monitoring \& observability (concept drift, data quality, latency, cost); 
\emph{(16)} Guardrails \& policy enforcement (filters, rate limiting, authorization); 
\emph{(17)} Incident response \& rollback; 
\emph{(18)} Continuous learning or re-training (data refresh, hotfixes); and 
\emph{(19)} Archival \& decommissioning. 
For empirical mapping in this work, these stages are collapsed into five macro-phases: \text{Data preparation}, \text{Pre-training}, \text{Fine-tuning (incl.\ PEFT/LoRA)}, \text{RLHF/Alignment}, and \text{Deployment/Inference (incl.\ agents/RAG)}.

\subsection{Threat Model}\label{sec:threat-model}
\tosemrev{A threat model for ML systems identifies the critical assets, potential adversaries, vulnerabilities, attack vectors, and mitigations. This model provides a comprehensive understanding of the threats that ML systems face during their lifecycle and across different layers, such as data, models, infrastructure, and APIs.\\ \textbf{Goal.} Attackers aim to affect the confidentiality, integrity, and availability of data (e.g., training data, features, model) depending on threat objectives. Poisoning attacks can affect data integrity. Extraction attacks can enable the theft of models or features, thereby compromising confidentiality. \textit{Key elements of the threat model include Entities and Assets. }\\
\textbf{Entities.}
 \texttt{\underline{Users:}} Legitimate actors interacting with the system. For example, data scientists, ML engineers, or application users querying the model.
\texttt{\underline{Adversaries}:} Malicious actors targeting the system. For example, Competitors attempting to steal a model, cybercriminals exploiting APIs, or insiders corrupting datasets. \\\textbf{Assets.} 
\texttt{\underline{Data:}} Training, validation, and test datasets are critical to model performance. For example, A dataset of medical records is used to train a diagnostic model.
\texttt{\underline{Models:}} The core algorithms and their parameters, including deployed and pre-trained models. For example, A fraud detection model is used in real-time transaction monitoring. \texttt{\underline{Infrastructure:}} Hardware, servers, APIs, and cloud environments hosting ML systems. For example, TensorFlow Serving for real-time inference, GPUs, or cloud-based data storage. \texttt{\underline{APIs}} Interfaces for model inference or interaction. For example, REST APIs for querying a sentiment analysis model.
\subsubsection*{ML threat scenarios}
\textbf{Data Poisoning.} Adversaries inject malicious samples into the training data to manipulate the model's behavior. \texttt{\underline{Use-case}:} A dataset for spam detection is poisoned with mislabeled spam emails as non-spam. The trained model allows spam emails to bypass the filter. \textbf{Impact:} Degrades the model's accuracy and reliability, impacting predictions.\\ \textbf{Mitigation.} Validate datasets for anomalies, use differential privacy, and apply robust data-cleaning techniques. \\
\noindent\textbf{Infrastructure Exploitation. } Exploit vulnerabilities in hardware, configurations, or hosting environments. \texttt{\underline{Use-case}:} An attacker exploits unpatched vulnerabilities in TensorFlow Serving (e.g., CVE-2020-15208) to inject a malicious model.\\ \textbf{Impact.} Backdoor insertion compromises the integrity of predictions. \\ \textbf{Mitigation.} Patch vulnerabilities regularly, use role-based access control (RBAC), and host models in secure environments like trusted execution environments (TEEs).
}

\subsection*{Mathematical Representations of Threats}

Let $a \in A$ be an asset, where $A$ is a set of assets from the system $\mathcal{S}$. An asset $a$ can be owned or accessed by an entity (e.g., a user, a user group, a program, or a set of programs) denoted as $E$. $\mathcal{E}$ denotes the set of all entities and $E \in \mathcal{E}$. Let $AC_{\mathcal{S}}: A \times \mathcal{E} \rightarrow R$ be a function that defines the level of privilege that an entity $E$ has on an asset $a$ or an asset group $A_g \subseteq A$, under the system $\mathcal{S}$. $R$ is a set of right access, and it can take values (1) $R=\{\textsf{none} \; ( \varnothing), \textsf{user}, \textsf{root}\}$ meaning that entities can have either no privilege ($\textsf{none}$), user access on $A$ ($\textsf{user}$), and full access on $A$ ($\textsf{root}$); or (2) $R=\{\textsf{none}, \textsf{read}, \textsf{write}\}$ meaning that entities can have no privilege ($\textsf{none}$), read access on $A$ ($\textsf{read}$), and write access on $A$ ($\textsf{write}$). When $AC_{AWS}('model.pkl', 'ml\_api') = \textsf{root}$, it means that the Amazon Web Service (AWS) ML API service $ml\_api$ has full access to the pickled model file $model.pkl$. When $AC_{VM}('training.csv', 'John') = \textsf{write}$, it means that user $John$ can modify or delete the training data file $training.csv$ in the virtual machine $VM$.

Let $P_1,...P_n$ be a set of premises and $C$ the goal to achieve. This relation is represented by:
\[
   \frac{P_1,...P_n}{C}
\]
It also means that $C$ can succeed when properties $P_1,...P_n$ are satisfied. Based on \cite{kindred1999theory}, we define the following notations. The notation
\[
   \xmapsto[]{\text{a}}E
\]
means that $a$ is $E$'s asset. Given $k_E \in K$ a protection property (e.g., encryption key, certificate, token), the notation 
\[
   \{a\}_{k_E}
\]
means that the protection $k_E$ is enforced on asset $a$ by an entity $E$. Let $E1, E2 \in \mathcal{E}$ be two entities that share an asset $a$. The notation
\[
   E1 \xleftrightarrow[]{\text{a}} E2
\]
means that $a$ is shared by $E1$ and $E2$. The sharing is satisfied when $E1$ send $a$ to $E2$ and $E2$ send $a$ to $E1$ as follows,
\[
   \frac{E1 \xrightarrow[]{\text{a}} E2,E2 \xrightarrow[]{\text{a}} E1}{E1 \xleftrightarrow[]{\text{a}} E2}
\]
Let $m: A \rightarrow C$ be a model function that takes data in $A$ and returns decisions in $C$ based on the inputs. $C$ can be two classes (i.e., $\{c_1, c_2\}$) or multiple classes (i.e., $\{c_1, c_2, ...,c_n\}$), where $c_1, c_2, ...,c_n \in C$.

\paragraph{\textbf{Knowledge}.} 

\tosemrev{The attacker's knowledge of the target ML system determines their strategy and attack feasibility. Attack models fall into three categories: \textit{black-box}, \textit{gray-box}, and \textit{white-box}, each reflecting different levels of access to model parameters, training data, and system components. These attacks exist within a broader \textit{adversarial setting}, which defines the overall context of adversarial manipulation, including attack objectives, specificity, and defensive considerations.\\
}

\textbf{\textit{In black-box settings}}, an attacker $AT$ does not have direct access~\cite{mitreatlas101} to ML assets $A_V$ of the target victim $V$ (i.e., model, executing code, datasets), i.e.,  $\forall a_V \in A_V, AC_V(a_V, AT) =  \varnothing.$ They have only access to the ML inference API using an access token $k_V$ obtained as a legitimate user from the victim's platform $V$, i.e.,
\[
   \xmapsto[]{\{a_{AT}\}} AT
\]
where $a_{AT} \in A$ are data crafted offline by attacker $AT$ to be sent via API. During the attack, $AT$ performs queries using the victim's ML inference API and observes outputs. To do so, $AT$ sends an online request with the crafted data $a_{AT}$ using access token $k_{V}$, i.e., 
\[
   AT \xrightarrow[]{\{a_{AT}\}_{k_{V}}} V
\]
Then, $AT$ will receive prediction responses and analyze them to further improve its data for attack, i.e.,
\[
   V \xrightarrow[]{\{m_V(a_{AT})\}} AT
\]
where $m_V$ is the executed model behind the ML inference API of the target victim $V$. \\
\tosemrev{
\textbf{\textit{In gray-box settings}}, an attacker $AT$ has partial access to some ML assets $\tilde{A}_V \subseteq A_V$ of the target victim $V$. This partial access could include knowledge of the model architecture, hyperparameters, or a subset of training data, but not full access to model parameters or gradients. Formally, 
\[
\exists a_V \in \tilde{A}_V, AC_V(a_V, AT) = \text{partial}
\]
In this scenario, the attacker $AT$ can leverage transfer learning, metadata analysis, or limited model responses to refine their adversarial strategies. The attack process can involve training a shadow model $\hat{M}$ to approximate the target model $M_V$:
\[
   \hat{M} \approx M_V, \quad \text{where} \quad \mathcal{D}_{partial} \subset \mathcal{D}_{train}.
\]
Using this approximated model, the attacker can estimate gradients and generate adversarial examples:
\[
    x_{adv} = x + \delta, \quad \text{such that} \quad M_V(x_{adv}) = y_t, \quad \|\delta\|_p \leq \epsilon.
\]
Where $y_t$ is the targeted misclassification.\\
}
\textbf{\textit{In white-box settings}}, attacker $AT$ may have internal access to some ML assets $\tilde{A}_V \subseteq A_V$ of the target victim $V$ (e.g., model, training data), i.e.,  
\[
\forall a_V \in \tilde{A}_V, AC_V(a_V, AT) \in \{\textsf{read},\textsf{write}\}
\] 
Then, $AT$ can perform several state-of-the-art attack techniques such as poisoning, evasion, extraction, and inference (see Section~\ref{sota_attack}).

\paragraph{\textbf{Specificity}.} In adversarial settings, ML threats can target a specific class/sample for misclassification (adversarially targeted) or any class/sample for misclassification (adversarially untargeted). The goal of $AT$ is to maximize the loss $L$ so that model $m_V$ misclassifies input data,

\[
   \argmax_{a} L(m_V(a), c)
\]

where $a \in A$ is an input data, $c \in C$ is a target class, and $m_V(a)$ is the predicted target data given $a$. To achieve the goal, $AT$ can execute a targeted or untargeted attack affecting the integrity and confidentiality of data~\cite{barreno2010security, jagielski2018manipulating}. 

When attack is targeted, $AT$ substitutes the predicted class $c$ by adding a small pertubation $\theta_u(a, c)$ so that 
\[
m_V(a_{AT}) = c,
\]
 where $a_{AT} = a + \theta_u(a, c)$ is an adversarial sample. 

In untargeted attack, $AT$ adds a small pertubation $\theta_t(a)$ to input $a$ so that 
\[
m_V(a_{AT}) \neq m_V(a),
\]
 where $a_{AT} = a + \theta_t(a)$ is an adversarial sample. ML threats can also leverage traditional TTPs to achieve their goals. 

\tosemrev{
In traditional settings, threat actors can either actively pursue and compromise specific targets while maintaining anonymity (traditionally targeted) or spread indiscriminately across the network without a predefined objective (traditionally untargeted). The terms \textit{Adversarially} and \textit{Traditionally} are used to distinguish between attack specificity in adversarial settings and broader, less targeted approaches in traditional settings. In traditional attacks, the attacker $AT$ typically targets critical assets such as user accounts, servers, virtual machines, databases, and networks. By exploiting vulnerabilities and bypassing authentication mechanisms or firewalls, $AT$ gains unauthorized access to these assets. Once inside, $AT$ can escalate privileges to gain full control of the ML assets belonging to the victim $V$, denoted as $\forall a_V \in A_V, AC_V(a_V, AT) = \textsf{root}$. This unrestricted access enables the attacker to cause extensive damage, such as exfiltrating sensitive data, corrupting models, or disrupting system operations.
}

\tosemrev{
\textbf{Capability.} Threat actors employ a variety of tactics to execute machine learning (ML) attacks effectively\mbox{~\cite{mitreatlas}}. These tactics include \textit{Reconnaissance, Resource Development, Initial Access, ML Model Access, Execution, Persistence, Defense Evasion, Discovery, Collection, ML Attack Staging, Exfiltration,} and \textit{Impact}. During \texttt{\underline{Reconnaissance and Resource Development}}, attackers gather intelligence about the target system by analyzing publicly available resources, such as papers, repositories, or technical documentation. Simultaneously, they establish command-and-control (C2) infrastructure to facilitate the attack. In the \texttt{\underline{Initial Access}} phase, attackers attempt to infiltrate the victim’s infrastructure, focusing on entry points containing ML artifacts such as datasets, models, or APIs. Once inside, they escalate their activities to gain deeper access to model internals and physical environments (\textit{ML Model Access}). During \texttt{\underline{Execution}}, attackers deploy remote-controlled malicious code to extract sensitive data or disrupt normal operations. To maintain access, they rely on \textit{Persistence} techniques such as implanting backdoor ML models or preserving compromised access channels. Attackers use \texttt{\underline{Evasion}} strategies to bypass detection mechanisms such as classifiers and intrusion detection systems\mbox{~\cite{biggio2013evasion,biggio2018wild,pierazzi2020intriguing,kwon2019selective}}. Once the system is compromised, they engage in \texttt{\underline{Discovery and Collection}} activities to identify and harvest valuable data. During \texttt{\underline{ML Attack Staging}}, adversaries refine their strategies by training proxy models, crafting adversarial data, or injecting poisoned inputs to corrupt the target model. The final phase, \texttt{\underline{Exfiltration and Impact}}, often results in significant consequences, such as theft of proprietary models, large-scale data breaches, human harm, or complete system failure.
}

\subsubsection{\textbf{Assets \& Adversaries (Threat-Model Scope)}}
\label{sec:assets-adversaries}
Table~\ref{tab:assets} enumerates the \emph{machine-learning assets} we protect,
aligned with the phases of the ML lifecycle introduced in \ref{sec:ml_lifecycle_stages} and later in the MITRE ATLAS tactic. The subsequent Table~\ref{tab:personas} defines four adversary personas and the subset of assets each can legitimately or
illegitimately reach. This explicit mapping disambiguates the attack surface
considered throughout our GNN-based severity model (sec~\ref{sec:gnn_predict}) and the mitigation matrix (Fig.~\ref{fig:defend_matrix}, section~\ref{sec:threat_matrix}). 
\begin{table}[ht]
 \begingroup\color{black}
  \caption{\majorev{Taxonomy of ML assets in our threat model: categories, exemplar artefacts, and lifecycle phases.}} 
  \label{tab:assets}
  \small
  \setlength{\tabcolsep}{6pt}
  \rowcolors{2}{blue!5}{white}
  \begin{tabular}{@{}p{3.0cm} p{9.5cm} p{4.0cm}@{}}
    \toprule
    \textbf{Asset category} & \textbf{Concrete artefacts} & \textbf{Lifecycle phase}\\
    \midrule
    Data & raw training corpus; RLHF preference logs; LoRA/QLoRA adapter deltas; evaluation & Collect $\to$ Train $\to$ Fine-Tune\\
    Model artefacts & network architecture graph; checkpoint weights; ONNX/TensorRT binaries; gradient updates & Train $\to$ Package $\to$ Deploy\\
    Execution surface & REST/GRPC inference API; SDK wrappers; hosted notebook endpoints; model-registry entries & Deploy $\to$ Serve\\
    Supply-chain & third-party libraries; container images; CI/CD configs; signed model cards & Package $\to$ Deploy\\
    MLOps metadata & experiment tracker DB; lineage store; monitoring dashboards; audit logs & Cross-cutting\\
    \bottomrule
  \end{tabular}
 \endgroup
\end{table}
%
\begin{table}[ht]
 \begingroup\color{black}
  \caption{\majorev{
  Adversary personas, their access levels, and reachable assets. Typical actors range from a curious end-user (limited to API access) to a rogue maintainer, registry attacker (compromising dependencies).
  }}
  \label{tab:personas}
  \small
  \setlength{\tabcolsep}{6pt}
  \rowcolors{2}{blue!5}{white}
  \begin{tabular}{@{}p{3.1cm} p{6.1cm} p{8.0cm}@{}}
    \toprule
    \textbf{Persona} & \textbf{Access level} & \textbf{Primary assets exposed}\\
    \midrule
    Public black-box & Public inference API only & Execution surface\\
    Gray-box collaborator & API \emph{+} partial internals (e.g.\ arch sketch, small data subset) & Execution surface; subset of Data \& Model artefacts\\
    White-box insider & Full repo and pipeline & All asset classes\\
    Supply-chain attacker & Build or dependency path & Supply-chain artefacts (library, container, CI)\\
    \bottomrule
  \end{tabular}
  \endgroup
\end{table}
\paragraph{Example~linkage.}
The synthetic gray-box scenario in section~\ref{sec_gray_box} maps to \textsc{Gray-box
collaborator}: the former contractor controls the public inference API plus
partial knowledge of a BERT-based architecture and public pre-training corpus,
but \emph{no} direct access to current weights.  That example therefore,
targets the \textit{Execution surface} and \textit{Model artefacts} rows in
Table~\ref{tab:assets}.  Each subsequent attack graph edge and mitigation
entry cites the asset IDs (\textit{Data}, \textit{Execution}) and persona IDs
(\textsc{P2}, \textsc{P3}, \dots) to keep the provenance explicit.


\subsection{Related work}\label{sec:related}
\tosemrev{
In computer security, threat assessment is a continuous process that involves identifying, analyzing, evaluating, and mitigating threats. While this process has been extensively applied to traditional systems~\cite{4652578}, its concrete application to machine learning (ML)-based systems remains nascent and underexplored. 
The ATLAS framework, developed by the MITRE Corporation in collaboration with organizations like Microsoft and Palo Alto Networks, represents the first comprehensive real-world attempt at ML threat assessment~\cite{mitreatlas}. Recent work has also leveraged the ATT\&CK framework for threat modeling. For example, Kumar \etal{}~\cite{9283867} identified gaps during ML development, deployment, and operational phases when ML systems are under attack. They proposed incorporating security aspects such as static and dynamic analysis, auditing, and logging to strengthen ML-based systems in industrial settings. 
Building on these foundations, our approach integrates existing mitigations from ATT\&CK~\cite{mitre_attack_mitigation}, the Cloud Security Alliance, and NIST security guidelines~\cite{mccarthy1800identity, hu2020general, souppaya2013guide, scarfone2008technical, scarfone2008sp, karygiannis2002wireless, cooper2018security}. It organizes mitigations across layers (e.g., data level to cloud level) and stages (e.g., harden, detect, isolate, evict), aligning with frameworks like MITRE D3FEND~\cite{mitre_defend}. Furthermore, Lakhdhar \etal{}~\cite{9474289} proposed mapping newly discovered vulnerabilities to ATT\&CK tactics by extracting features like CVSS severity scores and using RandomForest-based models. Similarly, Kuppa \etal{}~\cite{10.1145/3465481.3465758} employed a multi-head joint embedding neural network trained on threat reports to map CVEs to ATT\&CK techniques. Both approaches, however, are limited by their reliance solely on the ATT\&CK database and their focus on mapping vulnerabilities to TTPs.
Our proposed threat assessment methodology combines insights from ATLAS, ATT\&CK, and additional sources like the AI Incident Database~\cite{mcgregor2021preventing} to provide a comprehensive characterization of ML threats and vulnerabilities. By mapping TTPs to specific ML phases and models, and integrating vulnerability analysis with lifecycle-specific defenses, we offer a complete, end-to-end assessment of ML assets.
\subsection*{Adversarial Threats and Vulnerabilities in ML Systems}
Adversarial threats are a key focus within ML security. Carlini \etal{}~\cite{carlini2017adversarial} demonstrated the robustness of adversarial examples in bypassing detection mechanisms, while Athalye \etal{}~\cite{athalye2018obfuscated} exposed flaws in gradient obfuscation defenses. Wallace \etal{}~\cite{wallace2020imitation} extended these findings by highlighting vulnerabilities in machine translation systems, showcasing how adversaries can generate targeted mistranslations and malicious outputs. Papernot \etal{}~\cite{papernot2017practical, papernot2016transferability} introduced black-box and transferability-based attack methodologies, demonstrating the feasibility of crafting adversarial examples without access to model internals.
On the privacy front, Carlini \etal{}~\cite{carlini2021extracting} highlighted the risks of training data extraction from large language models, emphasizing the potential for sensitive information leakage. Similarly, Chen \etal{}~\cite{chen2021badnl} proposed BadNL, a backdoor attack framework for NLP systems, which leverages semantic-preserving triggers to ensure stealth and efficacy. These works collectively underline the importance of designing robust defenses.
\subsection*{Frameworks for Defense and Systematization}
Systematic approaches to ML threat assessment have also been proposed. Abdullah \etal{}~\cite{abdullah2021sok} and Barreno \etal{}~\cite{barreno2010security} categorized adversarial threats and mapped them to ML lifecycle stages, creating a foundation for threat mitigation strategies. Cissé \etal{}~\cite{cisse2017parseval} introduced Parseval networks to constrain model behavior for improved robustness, while Goodfellow \etal{}~\cite{goodfellow2014generative} presented generative adversarial networks (GANs), inspiring adversarial training techniques. Wallace \etal{}~\cite{wallace2020imitation} further proposed defenses against imitation-based attacks, offering practical countermeasures to mitigate these threats.
\subsection*{Comprehensive Threat Mitigation}
Our combined approach leverages ATLAS, ATT\&CK, AI Incident Database, and defense frameworks such as D3FEND~\cite{mitre_defend} and NIST guidelines. This methodology integrates traditional cybersecurity practices with ML-specific insights to provide a holistic and robust threat assessment strategy. In the following sections, we demonstrate how ATLAS TTPs affect ML components across lifecycle stages and how traditional vulnerabilities propagate through ML repositories.\\
Existing research primarily focuses on threat modeling using frameworks like ATLAS and ATT\&CK, with efforts to map vulnerabilities to TTPs through automated techniques. However, these approaches are limited by their reliance on predefined databases, failing to capture emerging threats from real-world incidents. Our work fills this gap by integrating multiple sources—including ATLAS, ATT\&CK, AI Incident Database, and GitHub repositories—to provide a more comprehensive, dynamic threat assessment. By mapping TTPs to specific ML lifecycle stages and analyzing high-severity vulnerabilities across dependency clusters, we bridge the disconnect between theoretical threat models and practical security challenges, offering a more actionable and lifecycle-aware approach to ML security.
}

\section{Study Methodology}~\label{methodology}

\begin{figure*}[ht!]
     \centering
     \begin{subfigure}[b]{1.0\textwidth}
         \centering
         \includegraphics[width=\textwidth]{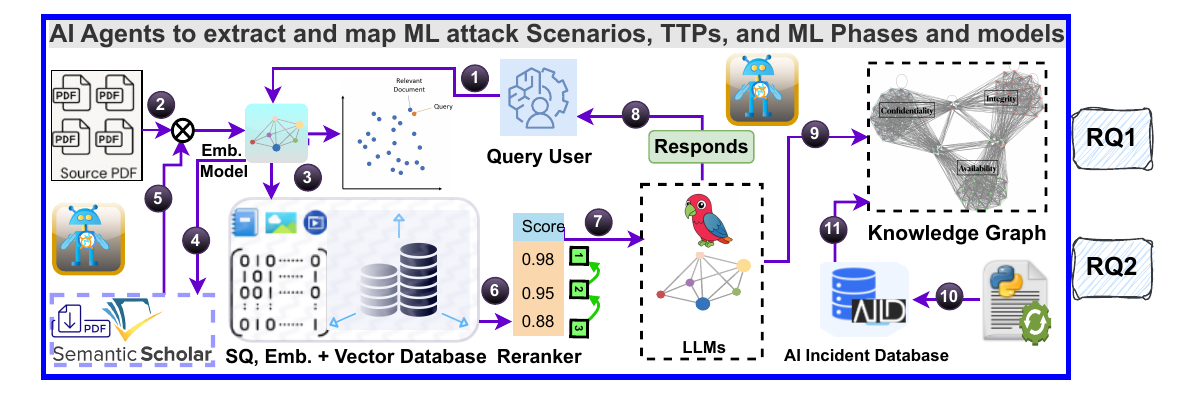}
         \caption{Agentic-RAG mapping TTPs to (i) ML phases and models (ii) to attack scenarios} 
         \label{fig:rq1-rq2-ttps}
     \end{subfigure}
     \hfill
     \begin{subfigure}[b]{1.0\textwidth}
         \centering
         \includegraphics[scale=0.7]{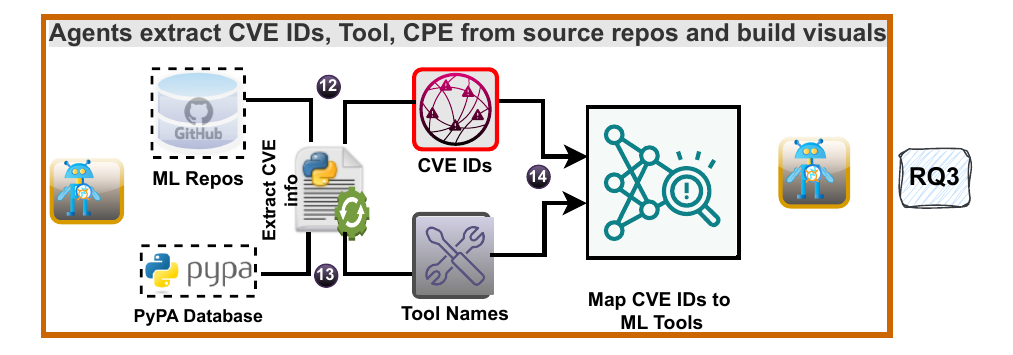}
         \caption{Using Agents to map CVE IDs to ML tools}
         \label{fig:rq3-cve}
     \end{subfigure}
     \caption{\majorev{\textbf{Study methodology using agentic workflow to address RQ1–RQ3.}~\textbf{(a)} Agentic–RAG pipeline that ingests literature and incident data,
      retrieves and reranks evidence, and maps TTPs to ML lifecycle phases, model
      types, and attack scenarios, integrating results into a knowledge graph.
      \textbf{(b)} Agent workflow that mines GitHub/PyPI repositories for CVE IDs,
      CPEs and tool names, then maps vulnerabilities to specific ML tools for
      visualization and analysis.}
  }
     \label{fig:method1}
\end{figure*}

\tosemrev{The \textbf{goal} of this study is to comprehensively analyze ML threat behaviors, including common entry points, prominent threat tactics, and typical TTP stages, and evaluate their impact on ML components such as vulnerable ML phases, models, tools, and their associated dependencies. To achieve this, we leverage threat knowledge from established sources such as ATLAS, ATT\&CK, and the AI Incident Database, alongside documented vulnerabilities from ML codebase repositories (GitHub and PyPA). Additionally, we incorporate TTPs discussed in the literature, aligned with various ML lifecycle stages, to predict potential threats related to specific packages or libraries. The \textbf{perspective} of this study is to equip ML red/blue teams, researchers, and practitioners with a deeper understanding of ML threat TTPs. By doing so, they can proactively prevent and defend against these threats, ensuring the secure development and deployment of ML products from staging to production environments. The \textbf{context} of this study encompasses 93 
real-world ML attack scenarios drawn from diverse sources, including 26, 12, and 55 (total of 93) 
cases from ATLAS, the AI Incident Database, and the literature, respectively. It also includes 854 ML repositories, with 845 sourced from GitHub and 11 from PyPA\mbox{~\cite{pypa-db}}. Figure \ref{fig:method1} illustrates the roadmap of the study. First, we present the research questions that guide this study. Next, we introduce the threat model considered and adopted in this work. The \textbf{implementations} of this study's goals leveraged five AI Agents using the Swarm framework\footnote{\url{https://docs.swarms.world/en/latest/swarms/agents/openai_assistant/}}, each agent designed to execute a specific task as detailed in Section~\ref{sec:data-collection}. Custom scripts were developed for individual tasks (to be executed by agents), independently tested for correctness, and then integrated into a cohesive multi-agent framework to ensure precision and efficiency throughout the workflow.
Finally, we outline the data collection and processing steps to ensure the clarity and reproducibility of our methodology.\\
}

\tosemrev{
In the context of this research work, we define an AI agent\footnote{\url{https://blogs.nvidia.com/blog/what-is-agentic-ai/}} as a computational system capable of perceiving and interpreting its environment, mining, analyzing, and reporting data from multiple external sources (e.g., scientific articles, security databases, and software repositories), reasoning and refining queries autonomously through internal logic (reasoning), and making informed decisions based on the context and previous interactions (memory). Five agents are coordinated to form an `\texttt{agentic}' solution that autonomously refines initial search queries, retrieves relevant information, identifies adversarial threats (TTPs) and vulnerabilities, and their lifecycle stages, and constructs graphical or analytical models for visualization and analysis. The agentic system can independently assess new data against existing knowledge to continuously update its understanding (learning), anticipate potential threats, and facilitate comprehensive exploration of the threat landscape.
}

\textbf{Research Questions (RQs). } 
To achieve our goal, we address the following research questions:
\begin{enumerate}
  \item[\textbf{RQ$_1$:}]  \rqone
  
  \tosemrev{
  \textit{This RQ aims to expand knowledge about ML threat TTPs to facilitate the development of better defense strategies. By examining the execution flows of ML attack scenarios, this study seeks to identify the most commonly used TTPs and their sequences. Understanding these patterns provides actionable insights into how adversaries structure their attacks, enabling researchers and practitioners to design proactive and targeted countermeasures tailored to these scenarios.
  }}

   \item[\textbf{RQ$_2$:}]  \rqtwo
   
   \tosemrev{
   \textit{This RQ focuses on understanding the impact of the threat tactics identified in \textbf{RQ$_1$} on various ML phases and models. The objective is to analyze how each tactic affects different stages of the ML lifecycle, such as data collection, training, and deployment, to help secure individual pipeline components. By identifying the most frequently targeted ML phases and the most prevalent threat tactics, this RQ provides actionable insights into the areas of greatest vulnerability, enabling practitioners to prioritize defenses and mitigate risks effectively across the ML lifecycle.
   }}

   \item[\textbf{RQ$_3$:}] \rqthree

   \tosemrev{
   \textit{
   This RQ assesses the completeness of ATLAS by identifying other security threats that may have been overlooked. The goal is to identify gaps in ATLAS by comparing it with other sources, such as the AI Incident Database or relevant literature. This retrospective approach focuses on cataloging overlooked threats from multiple sources and aligning them with the existing ATLAS framework. Furthermore, we also investigate the most vulnerable ML repositories, the most recurrent associated vulnerabilities, and the dependencies that cause them.
   }}
\end{enumerate}

\subsubsection{\textbf{Metrics.}}
\tosemrev{
To answer our RQs, we compute individual metrics as reported below. \\
\textbf{RQ$_1$} focuses on the \underline{Tactics in Scenario-Based Attacks} by computing the tactics employed in scenario-based attacks and their frequency. This provides a quantitative measure of how often specific tactics are utilized, offering insights into prevalent attack strategies in different scenarios. \textbf{RQ$_2$} addresses \underline{Impact on ML Phases} and calculates the number of tactics targeting each ML phase. This evidence highlights which phases of the ML lifecycle (e.g., data collection, training, deployment) are most impacted by threats. Such a metric is crucial for identifying the stages at which ML systems are particularly vulnerable.
For \textbf{RQ$_3$}, we \underline{examine vulnerabilities, their types, and the tools they affect.} We design multiple metrics to capture the scope and distribution of vulnerabilities. Specifically, we compute: (i) the total number of vulnerabilities (\textit{nov}) and their overall types, and (ii) the distribution of \textit{nov} by threat type and tools (e.g., GitHub ML repositories). These metrics further break down the \textit{nov} by individual tools and by type for each tool, providing granular insights into the frequency and nature of vulnerabilities in ML repositories. This analysis sheds light on the most vulnerable tools, the recurring types of vulnerabilities, and their associated potential threats.
}

\subsection{Data collection}
\label{sec:data-collection}
\tosemrev{
This study leverages diverse and credible data sources, including academic databases, codebase repositories (GitHub and PyPA), the AI Incidents Database, and the MITRE Database, to comprehensively examine threats and vulnerabilities in machine learning systems. First, we conducted an exploratory analysis of the codebase and incident repositories to gain a comprehensive understanding of their structures and organization, enabling targeted and informed data collection. A systematic process for extracting threats and vulnerabilities from well-established databases ensures reliability and consistency in identifying relevant issues. The study adds depth and context by integrating data from GitHub repositories, using publicly available information to link vulnerabilities to real-world problems. Finally, to enhance precision and efficiency, we developed and implemented scripts~\cite{tech-report-v1} for execution by the Agent. Specialized Agents were employed to execute specific tasks, ensuring efficiency and precision in the workflow.
}
\begin{figure}[h]
\centering
\includegraphics[width=0.7\textwidth]{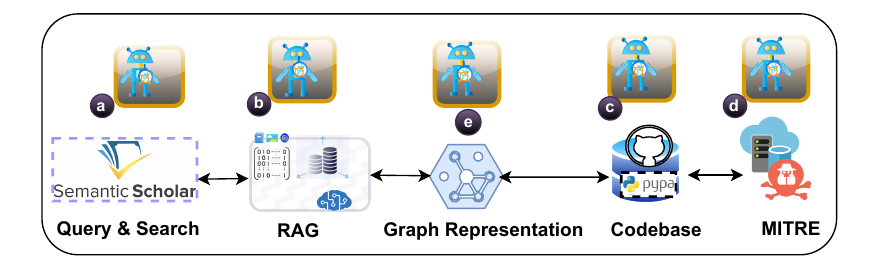}
 \caption{\majorev{\textbf{Multi-agent LLM with RAG.}
  A coordinated set of AI agents executes four roles:  
  \textbf{(a)} \emph{Query \& Search}: retrieving relevant literature and threat
  intelligence from Academic Search Engines/Databases;  
  \textbf{(b)} \emph{RAG}: extracting and ranking relevant TTPs, vulnerabilities, and ML lifecycle stages;  
  \textbf{(c)} \emph{Graph Representation}: encoding extracted relationships
  into a heterogeneous knowledge graph;  
  \textbf{(d)} \emph{Codebase \& MITRE Mapping}: Linking GitHub/PyPI CVEs to ATT\&CK/ATLAS.
  }}
\label{fig:agents}
\end{figure}
\tosemrev{
\paragraph{Overview of Our Multi-Agent Approach.}
This study comprises five key tasks spanning from initial data collection to the final reporting of results. First, we \emph{refine the initial search query} (\orbit{a}) using a Large Language Model (LLM), ensuring more precise and granular strings (that should take into consideration acronyms of key terms used for TTP) to interrogate the scientific databases in anticipation of the subsequent RAG steps. Next, we \emph{leverage an agentic RAG pipeline} (\orbit{b}) with a re-ranker to identify pertinent themes and content from the literature, using the refined query generated in the previous step. We then \emph{search codebase repositories} (\orbit{c}) according to criteria in Section~\ref{sec:search-codebase}, enabling us to gather concrete evidence of security practices from real-world ML libraries/ applications. Concurrently, we \emph{extract ML attacks} (\orbit{d}) documented in ATLAS and the AI Incident Database (Section~\ref{sec:ttps-vul}), enriching the threat profile by incorporating adversarial behaviors observed in practice. Finally, we \emph{build graphical representations} (\orbit{e}) to visualize and analyze the vulnerability patterns gleaned from literature and codebase repositories (Section~\ref{sec:results}).
}

\subsubsection{Scientific Database Search}\label{sec:papers-searh}
\tosemrev{
A multi-agent RAG system~\cite{davidLo25:rag-agent,alhanahnah2024depsrag,arslan2024survey} automates data collection and processing for this literature review, orchestrating each agent's specialized role. A thorough search was conducted on Semantic Scholar, Google Scholar, and IEEE Xplore, starting with an exploration of 14 foundational seed papers widely recognized in the machine learning (ML) security community~\cite{carlini2021extracting,biggio2013evasion,barreno2010security,carlini2017adversarial,wallace2020imitation,abdullah2021sok,chen2021badnl,choquette2021label,papernot2016transferability,goodfellow2014generative,papernot2017practical,cisse2017parseval,athalye2018obfuscated,jagielski2020high}. The authors chose these 14 papers due to their relevance and popularity regarding security threats in machine learning application domains. These seed papers form the foundation for systematically exploring the broader literature search. 
Based on insights from these seed papers, an \textit{initial\_query}---``\textit{Tactics, Techniques, and Procedures in Machine Learning Security}''---was submitted to an LLM (temperature $=0.4$) to produce a refined, inclusive, targeted query string (Listing~\ref{lst:query}) aimed at capturing additional relevant articles. This refined query string was then executed via API calls across multiple academic databases, returning 4{,}820 articles. From this set, 300 articles were randomly sampled (Confidence Level:~99\%, Margin of Error:~7.5\%) for our RAG pipeline. As illustrated in Figure~\ref{fig:agents}, the first agent (\textbf{Query \& Search}) receives this initial query, thus initiating a systematic literature review that captures a broader representation of the relevant threat landscape. This method ensures both depth and breadth in our data collection process, thereby enhancing the robustness of our study methodology.
}

\begin{lstlisting}[language=SQL, label={lst:query}, caption=Search query string generated by LLM RAG agent]
query = (
    "((\"Tactics, Techniques, and Procedures\" OR \"TTP\" OR \"advers* attack\" OR \"threat\" OR \"vuln*") "
    "AND (\"machine learning security\" OR \"ML security\" OR \"AI security\" OR \"deep learning security\") 
    "AND (\"data poisoning\" OR \"evasion attack\" OR \"model tampering\" OR \"model inversion\" OR "
    "\"backdoor attack\" OR \"adversarial example\" OR \"denial of service\" OR \"resource exhaustion\"))"
)
\end{lstlisting}

\subsubsection{ML Codebase Repositories (GitHub and PyPA)}\label{sec:search-codebase}
\tosemrev{To gather the sample of ML projects considered to run our study, we adopt various strategies. Similar to previous studies~\cite{LF-hosting-project,wu2024comprehensive}, we aim to collect ML repositories (repos) from GitHub, applying specific criteria to select high-quality, widely recognized projects. One such filtering criterion includes repos with over 1,000 stars, which are generally regarded as reputable, promising, and reflective of community engagement and interest. Beyond stars, we also observe additional activity metrics to gauge the repos' health and engagement. These include the number of open and closed issues, which provide insight into the project's maintenance and responsiveness; the number of pull requests (PRs), both merged and pending, signifies the pace of development and the integration of new features. By focusing on repos that meet or exceed this threshold, we aim to curate a dataset that balances quality, relevance, and active contributions from the open-source community.
In particular, we use the GitHub API search to mine repos satisfying our criteria, for example, to select repos with over 1,000 stars, we use the following criteria: \texttt{machine-learning in:topic stars:$>$1000 sort:stars}. As a result, we obtained a list of 916 repositories that we sorted in descending order by number of stars, and filtered out 82 projects that did not meet our inclusion criteria. To this end, 834 projects were retained.
Given the widespread adoption of the Python programming language and its frameworks~\cite{djurdjev2024popularity,mediakov2022experimental}, we decided to include the PyPA database~\cite{pypa-db} in our analysis. This database comprises 425 repositories that document known vulnerabilities. Upon evaluation, we identified 11 ML repositories of interest.
}

\subsubsection{Vulnerabilities, Threats, and Adversarial tactics Selection}\label{sec:ttps-vul}
\tosemrev{
When selecting target threats, we focus on three key criteria: \textit{newness}, \textit{consistency}, and \textit{reputation}. By prioritizing recent data, we ensure that the sources provide up-to-date information on emerging ML vulnerabilities and threats. The consistency criterion emphasizes the inclusion of data sources that comprehensively cover significant vulnerabilities or threats, ensuring reliability and breadth. Lastly, the reputation criterion ensures the selection of data sources that are continuously updated, widely recognized, and frequently referenced by the community. Based on these criteria, we have chosen the ATLAS~\cite{mitreatlas} and AI Incident~\cite{mcgregor2021preventing} databases as our primary sources. These datasets were accessed programmatically through API or direct downloads.\\
\textbf{The ATLAS database.} During the exploratory analysis, which we did to understand the structure and patterns for each data source, we observed attack scenarios. For each attack scenario, we analyzed its pattern to identify the associated goal, knowledge, specificity, and capabilities. Then, we use our scripts~\cite{tech-report-v1} to extract the attack phases. Given that an attack is depicted into phases (procedures)~\cite{mitreatlascase},
at the time of mining and analysis, our data contains 26 documented ML attack scenarios spanning from 2016 to 2024\mbox{~\cite{mitreatlas}}.\\
\textbf{The AI Incident Database.} This source compiles real-world AI incident reports, which in this study, we extracted the dataset from 2003 to 2024\mbox{~\cite{mcgregor2021preventing}}. To complement the tactics, techniques, and procedures (TTPs) defined in ATLAS, we also incorporate the ATT\&CK framework\mbox{~\cite{strom2018mitre}}, which provides a broader context for attack scenarios, combining tactics from both ATLAS and ATT\&CK. To mine the incidents available in the dataset, we use our developed crawler to mine the attacks, using Regex and NLP techniques (cosine similarity, etc.) to search for keywords like textit{attack} ~\cite{tech-report-v1}. Also, we parsed the reference link to verify each attack by reading its description. 
As a result, 254 attacks were returned, dated from 2018 to 2024.\\
\textbf{The ATT\&CK database.} Includes downloadable JSON files. 
In the end, we extract the scenarios and their TTP definitions from ATLAS (14 tactics, 52 techniques, 39 sub-techniques) and ATT\&CK (14 enterprise tactics, 188 enterprise techniques, 379 sub-techniques) ~ \mbox{~\cite{mitre_attack}}.
All the referenced databases are maintained and up-to-date, employing different update strategies. ATLAS and ATT\&CK, managed by the MITRE Corporation, receive continuous support and updates from prominent industry leaders, including Microsoft, McAfee, Palo Alto Networks, IBM, and NVIDIA. In contrast, the AI Incident Database relies on collaborative contributions and curates information from verified real-world incidents reported by reputable media outlets, including Forbes, BBC, The New York Times, and CNN. This integration of diverse sources ensures that the datasets remain robust and reflective of the evolving landscape of AI and ML security threats.
}
\majorev{
\paragraph{Cross-level alignment of vulnerabilities.}
Table \ref{tab:cross-level} links every high-impact vulnerability in our corpus to three orthogonal coordinates:
(i) the \emph{ML life-cycle phase} where the flaw first manifests,
(ii) the \emph{software-component layer} it abuses (data layer/ model layer/ orchestration layer), and
(iii) the \emph{system or network surface} that is ultimately compromised.
The same identifiers (e.g.\ \texttt{VUL-17}) appear as node labels in the heterogeneous GNN (section~\ref{sec:gnn_schema}\, \& ~\ref{sec:gnn_predict}.) and in the Mitigation Matrix (Fig.~\ref{fig:defend_matrix}, section~\ref{sec:threat_matrix}), allowing the reader to trace a single weakness—such as \emph{LoRA gradient leakage}—from its inception in the fine-tuning phase, through the model-repository API, all the way to the underlying $S3$ storage bucket. 
}
\begin{table}[!ht]
\begingroup\color{black}
\caption{\textcolor{black}{\textbf{Cross-Level Vulnerability Map.}  
Each vulnerability is aligned with (i) its first ML life-cycle phase,
(ii) the software component it abuses, and (iii) the infrastructure or
network surface, it ultimately compromises.  IDs (e.g.\ \texttt{VUL-17})
are reused in the GNN (section~\ref{sec:threat_matrix}) and the Mitigation Matrix (Fig.~\ref{fig:defend_matrix}).}}
\label{tab:cross-level}
\small                
  \setlength{\tabcolsep}{4pt}
\centering
\rowcolors{2}{blue!5}{white}
\begin{tabularx}{\linewidth}{@{}llllll@{}}
\toprule
\textbf{ID} & \textbf{CVE / TTP} & \textbf{ML phase} &
\textbf{Software layer} & \textbf{System level} \\ \midrule
VUL-17 & LoRA gradient leakage\,& Fine-tune & Model-repo API &
S3 bucket (weights) \\
VUL-23 & Model-registry poisoning &
Deploy & Serving layer & K8s control-plane \\
VUL-31 & Universal jailbreak prompt (\textsc{MASTERKEY}) &
Deploy & Chat interface & Public inference API \\
VUL-42 & Reward-model hacking (RLHF) &
RLHF loop & RL policy store & CI/CD controller \\
VUL-55 & Training-data reconstruction (GPT-J) &
Pre-train & Check-point store & Cloud object store \\
\bottomrule
\end{tabularx}
\endgroup
\end{table}
\majorev{
\paragraph{Methodology for cross-level alignment.}
For every CVE or TTP in our corpus, we followed a three-step tagging
pipeline.  
(1)~\emph{Life-cycle phase}: we read the vulnerability description and the
original exploit reports, then applied the NIST ML life-cycle taxonomy
(pre-train, fine-tune, RLHF, deploy) to identify the \textbf{earliest} phase at which the flaw can be triggered.  
(2)~\emph{Software layer}: we mapped the affected source files,
configuration keys, or API endpoints, to one of three canonical layers in an
ML stack—\textbf{data layer} (e.g., dataset loaders, artifact stores),
\textbf{model layer} (training or inference code, parameter adapters),
or \textbf{orchestration layer} (pipelines, registries, service mesh).  
(3)~\emph{System/ network surface}: finally, we traced the execution path
until a concrete infrastructure boundary was reached, such as an S3
bucket, a Kubernetes control plane, or a public inference API.  
The tags were double-checked by two cybersecurity professionals (Cohen’s
$\kappa=0.93$).  The resulting triple of labels, phase, and software layer,
surface, is what Table \ref{tab:cross-level} records and what the GNN
ingests as node metadata in section~\ref{sec:gnn_schema}\, \& ~\ref{sec:gnn_predict}.
}
\begin{table}[!ht]
  \begingroup\color{black}   
  \caption{\majorev{Cross–layer taxonomy: each vulnerability observed in our corpus is linked from its ML life-cycle phase to the affected software, system, or network layer.}}
  \label{tab:phase-mapping}
  \small
  \setlength{\tabcolsep}{4pt}
  \begin{tabularx}{\linewidth}{@{}l l X l@{}}
    \toprule
    \textbf{ML phase} & \textbf{Asset} &
    \textbf{Vulnerability / TTP (ex.)} &
    \textbf{Mapped layer} \\ \midrule
    \textsc{Data prep} &
      Training dataset &
      Data poisoning (BadNets~\cite{guranda2025towards,gu2017badnets}),
      Label-flip (CVE-2023-45210) &
      Software (ETL pipeline) \\
    \textsc{Pre-train} &
      Check-points &
      Weight-deserialisation RCE (CVE-2025-32434) &
      System (file I/O) \\
    \textsc{Fine-tune} &
      LoRA adapters &
      Gradient leakage \cite{zhu2024loraleak} &
      Software (update API) \\
    \textsc{RLHF loop} &
      Preference DB &
      Reward-model hacking \cite{park2023rewardhacking} &
      System (CI/CD) \\
    \textsc{Deploy} &
      Inference API &
      Universal jailbreak (MASTERKEY \cite{bai2023masterkey});\,
      HTTP request smuggling (CVE-2024-3099) &
      Network (edge proxy) \\ \bottomrule
  \end{tabularx}
  \endgroup
  \vspace{-6pt}
\end{table}
\majorev{
\paragraph{Cross-layer taxonomy.}
Table~\ref{tab:phase-mapping} traces each vulnerability \emph{vertically}
through the software stack, clarifying where it breaks the ML boundary and
touches traditional infrastructure.
\paragraph{How the layer mapping was derived.}
The procedure mirrors the cross-level workflow:
\begin{enumerate}[label=(\arabic*)]
\item \emph{Phase anchoring}: fix the ML life-cycle phase established
    above;  
\item \emph{Call-graph walk}: start from the vulnerable asset and traverse
    source code, manifests, or API specifications until the first software
    boundary is crossed—classifying the boundary as
    \textbf{data}, \textbf{model}, or \textbf{orchestration};  
\item \emph{Perimeter resolution}: continue the trace to an observable
    infrastructure surface (file I/O, CI/CD service, edge proxy, \dots{}),
    which becomes the ``mapped layer'' column in
    Table~\ref{tab:phase-mapping}.
\end{enumerate}
The phase–asset–layer triples feed directly into the GNN (section~\ref{fig:defend_matrix}) and enable mitigation queries such as: ``Which deploy-phase flaws escalate past the orchestration layer into the public network edge?'' 
Table~\ref{tab:cross-level} traces each \emph{individual} CVE up the stack (bottom-up), while Table~\ref{tab:phase-mapping} aggregates flaws by phase–asset pair and shows how they descend to lower layers (top-down).  
Viewed together, enable a bi-directional view from a single vulnerability to its system impact, or from an affected layer to all ML-phase exploits that reach it.
}

\subsection{Data processing}\label{sec:data-processing} 
\tosemrev{
This section analyzes different datasets to uncover threat patterns and relationships with vulnerabilities and ML stages.}
\tosemrev{
\subsubsection{Retrieval-Augmented Generation (RAG) with Reranking.}\label{sec:rag}
Our enhanced RAG system leverages ChatGPT-4o with a temperature of 0.4, optimized through empirical testing within the range [0.2–0.7], ensuring accurate retrieval of key concepts (\textit{TTPs, vulnerabilities, and lifecycle stages}) from scientific papers. This configuration maintains the flexibility to capture synonyms, nuanced terms, and variations while minimizing hallucinations and extraneous content by strictly aligning outputs with evidence. The RAG workflow begins with document retrieval using dense embedding models (e.g., Sentence Transformers). A transformer-based reranker prioritizes retrieved documents based on their relevance to the query, ensuring that only the top-\( k \) documents (\( k=50 \) in our implementation) proceed to the generation phase. During response generation, the LLM synthesizes outputs by conditioning on both the refined query and reranked documents, combining its generative capabilities with factual grounding.
}

\majorev{
\subsubsection{\textcolor{black}{Post-hoc interpretability.}}
We embed two complementary explainers—\textbf{SHAP}
(\underline{SH}apley \underline{A}dditive Ex\underline{P}lanations) and
\textbf{LIME} (\underline{L}ocal \underline{I}nterpretable
\underline{M}odel-agnostic \underline{E}xplanations)—alongside the
retrieval-augmented generation loop so the system can justify \emph{why} a
particular Confidentiality, Integrity, or Availability (CIA) label is
returned. 
}

\begin{figure}[!ht]
  \centering
  \includegraphics[width=0.6\linewidth]{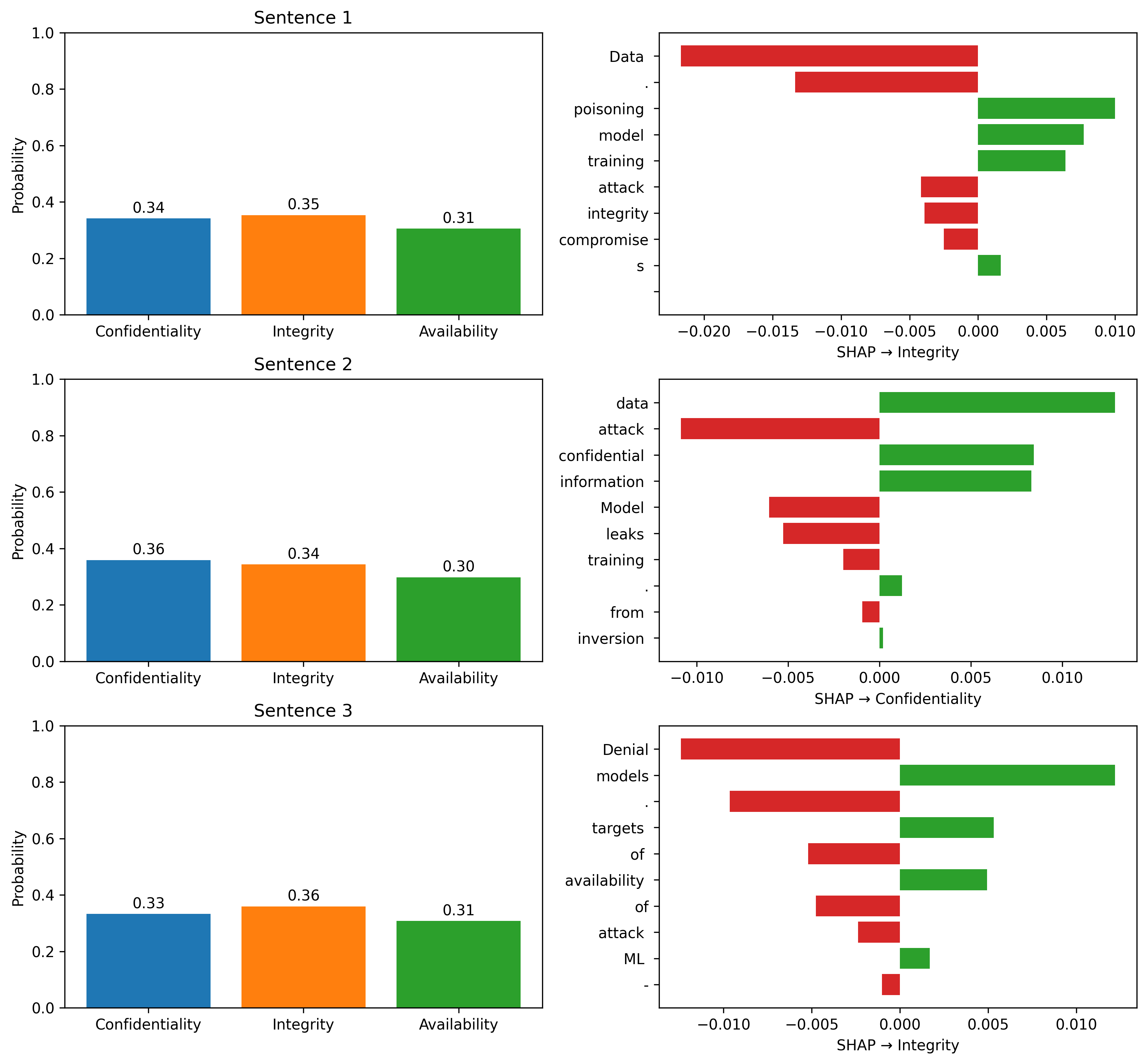}
  \caption{\textcolor{black}{\textbf{Global SHAP attribution for the sentences in
           Listing~\ref{lst:cia-sentences}.}  
           Each row shows the CIA probabilities (left) and the ten tokens with
           the largest Shapley values for the dominant class (right): green
           bars push the score up, red bars pull it down.}}
  \label{fig:shap-cia-grid}
\end{figure}

\begin{figure}[!ht]
  \centering
  \includegraphics[width=\linewidth]{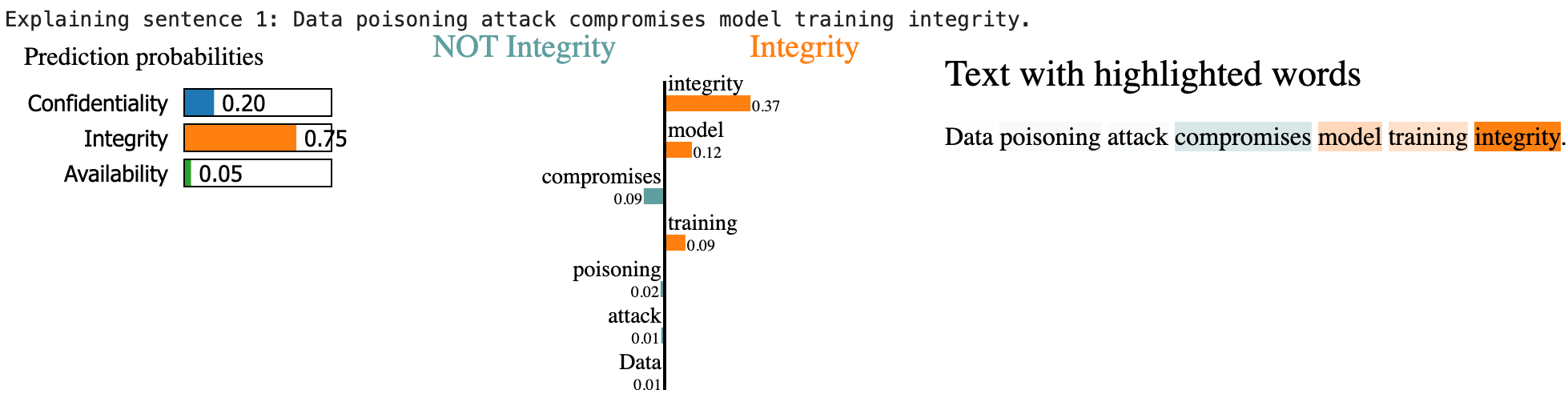}
  \caption{\textcolor{black}{\textbf{LIME explanation for Sentence 1 in
           Listing~\ref{lst:cia-sentences}.}  
           Left: class probabilities.  Centre: token weights—
           orange = positive, teal = negative.  Right: heat-map overlay on
           the original text.}}
  \label{fig:lime-focus}
\end{figure}

\begin{lstlisting}[
  mathescape,
  style=cotstyle,
  caption={Example TTP sentences illustrating Confidentiality, Integrity,
           and Availability violations used in the SHAP/LIME demonstrations.},
  label=lst:cia-sentences,
  captionpos=b,
  frame=tb,
  xleftmargin=12pt,
  numbers=none,
  numberstyle=\tiny\color{gray}]
ttp_sentences = {
   1: "Data poisoning attack compromises model training integrity.",
   2: "Model inversion attack leaks confidential information from training data.",
   3: "Denial-of-service attack targets availability of ML models.",
}
\end{lstlisting}
\majorev{
\textit{Global attribution (SHAP).}  
For every candidate sentence, we compute token-level Shapley values with
\texttt{maskers.Text}+partition.  Figure~\ref{fig:shap-cia-grid} (left) plots
class probabilities for three canonical TTP statements, see Listing~\ref{lst:cia-sentences}; the right panel shows the ten tokens that most raise (green) or lower (red) the dominant score.  In Sentence 2, for example, \emph{inversion}, \emph{confidential}, and \emph{information} add +0.31 log-odds to \textbf{Confidentiality}, whereas the generic token \emph{data} subtracts
–0.07. \\
}
\majorev{
\noindent \textit{Local perturbation (LIME).}  
LIME perturbs\footnote{\texttt{num\_samples} determines the number of perturbations. A sweep from 1,000–6,000 showed that 4,000 samples reduced the  median token-weight variance to $<\!1\%$ across five runs while keeping latency
below 0.4 s per sentence on a 12-core CPU.} each sentence \(4,000\) times, fits a local linear surrogate, and returns token weights visualized in
Figure~\ref{fig:lime-focus}. The bar chart reproduces SHAP’s ordering
(with \emph{integrity}, \emph{model}, \emph{training}\, increase$\uparrow$; meanwhile \emph{compromises}\, decreases$\downarrow$), and the inline heat-map highlights in situ the words that push the classifier toward or away from
\textbf{Integrity}.
}


\majorev{
\textit{Evidence graph.}  
Token scores flow into an interactive NetworkX graph in which queries,
retrieved passages, and system responses are nodes; edge widths are
proportional to \(\lvert\mathrm{SHAP}\rvert\).  Analysts can trace every
decision from raw text \(\rightarrow\) token importance
\(\rightarrow\) ATLAS technique (Fig.~\ref{fig:heatmap}).\\
\textit{Corpus-level coverage.}  
Applying this pipeline to 300 sampled research articles yields a graph with
\textbf{55} distinct TTPs, \textbf{21} exploited vulnerabilities, and
\textbf{9} ML life-cycle stages (only vulnerabilities actively exploited by
at least one TTP is retained.  Full statistics appear in
Section~\ref{sec:results}. \\
By coupling global (SHAP) and local (LIME) attribution with graph-based
visualisation, the system delivers actionable, audit-ready explanations that reduce model opacity and support informed decision-making in software
engineering and security analysis.
}

\majorev{
\paragraph{\textcolor{black}{Illustrative SHAP \& LIME outputs.}}
Figures~\ref{fig:shap-cia-grid} and \ref{fig:lime-focus} apply the
interpretability pipeline to the three sentences in
Listing~\ref{lst:cia-sentences}:
\begin{enumerate}[label=\textbf{\arabic*.}, leftmargin=*]
\item \textbf{SHAP global view.}  
      Each row in Fig.~\ref{fig:shap-cia-grid} pairs the predicted CIA
      distribution (left) with the ten most influential tokens (right).  In
      Sentence 2, tokens such as \emph{inversion} and \emph{confidential}
      push the prediction toward \textbf{Confidentiality}, whereas
      \emph{data} pulls it away.
\item \textbf{LIME local view.}  
      Fig.~\ref{fig:lime-focus} zooms in on Sentence 1.  The bar chart echoes
      SHAP’s ranking, and the heat-map overlays those weights on the raw
      text, making the evidence instantly visible.
\item \textbf{Cross-lens consistency.}  
      Agreement between SHAP (global) and LIME (local) on the sign and ordering of salient tokens confirms that the explanation is not an artifact of a single method.  Treating SHAP as the primary explainer and LIME as a sentence-level validator, therefore, yields both a rigorous, corpus-wide attribution framework and an intuitive diagnostic tool for practitioners.
\end{enumerate}
}

\tosemrev{
\subsubsection{\textbf{Linking Literature to ATLAS database. } } 
To further satisfy all the requirements of our RQs, we link the TTPs' information to the ATLAS database. First, we gathered the detailed attack \texttt{information}, including the associated tactics, techniques, goals, knowledge requirements, and specificity from the extracted TTPs. 
Then, we linked these extracted \texttt{information} with the TTP definitions provided by ATLAS and ATT\&CK frameworks. For clarity and brevity, only the tactics and techniques derived from the 14 seeding papers are presented in this table. However, a comprehensive mapping of TTPs, vulnerabilities, and ML lifecycle stages is shown later in the results Section~\ref{sec:results} \\
}

\subsubsection{ATLAS database.}
\tosemrev{
From the extracted ML attacks (26 in total) representing the entire dataset available in ATLAS at the time of mining, we now map and report the associated ML attack phases as described in the threat model (see Section \ref{sec:threat-model}). In particular, each phase of an attack represents a tactic, while the associated execution step(s) represent the technique or sub-technique(s). 
}
Thus, we analyze the extracted scenarios and their related TTPs descriptions provided by ATLAS and ATT\&CK across different phases and their relationships. 
For example, consider the \textit{Microsoft - Azure Service} attack,\footnote{\href{https://atlas.mitre.org/studies/AML.CS0010}{Microsoft Azure Service Disruption}} executed by the Microsoft Azure Red Team and Azure Trustworthy ML Team against the Azure ML service in production~\cite{microsoft2024pyrit}.
The introduced attack targets different capabilities of the ML system: confidentiality (\textit{unauthorized model/training data access}), integrity (\textit{poisoning by crafting adversarial data}), and availability (\textit{disruption of the ML service}). 
The attack knowledge is based on a white-box setting, as the attackers have full access to the training data and the model. 
Finally, the attack specificity is based on an adversarial untargeted setting, as threat actors do not target a specific class of the ML model.
The attack has eight phases, as detailed below\footnote{We do not report here the extracted information from the ATLAS database as they are already available in the dataset; however, we consider and discuss them when combining their results with the information of attacks from other sources.}:
\begin{itemize}
    \item \textbf{Phase 1:} The required information for the attack is collected, such as Microsoft publications on ML model architectures and open source models;
    \item \textbf{Phase 2:} Usage of valid accounts to access the internal network;
    \item \textbf{Phase 3:} The training data and model file of the target ML model are found;
    \item \textbf{Phase 4:} The model and the data are extracted, leading the team to continue executing the ML attack stages;
    \item \textbf{Phase 5:} During ML attack staging, they crafted adversarial data using target data and the model;
    \item \textbf{Phase 6:} They exploited an exposed inference API to gain legitimate access to the Azure ML service.
    \item \textbf{Phase 7:} Adversarial examples are submitted to the API to verify their efficacy on the production system;
    \item \textbf{Phase 8:} Finally, the team successfully executed crafted adversarial data on the online ML service.\\
\end{itemize}

\subsubsection{\textbf{AI Incident database.} }\label{sec:ai-incident-db-proc}

To analyze this dataset, we start by identifying if there are potential TTPs similar to those in ATLAS/ATT\&CK. Furthermore, we check the target models and related information about the attack (goal, knowledge, and specificity).

\tosemrev{For example, consider the ML real-world attack called \textit{Indian Tek Fog Shrouds an Escalating Political War} dated from 2022. 
Following the reference link associated with the attack, \textit{Tek Fog} is an ML-based bot app used to distort public opinion by creating temporary email addresses and bypassing authentication systems from different services, like social media and messaging apps\footnote{\href{https://www.bloomberg.com/opinion/articles/2022-01-12/india-s-tek-fog-shrouds-an-escalating-political-war-against-modi-s-critics?embedded-checkout=true}{India's Tek Fog Shrouds}}.  
The goal is to send fake news, automatically hijacking X and Facebook trends, such as retweeting/sharing posts to amplify propaganda, phishing inactive WhatsApp accounts, spying on personal information, and building a database of citizens for harassment. 
The bot may use a Transformer model such as GPT-2 to generate coherent text-like messages~\cite{thewire_attack}.
By analyzing the attack, we observe the following four ML TTPs: (i) Resource Development (\textit{Establish Accounts}), (ii) Initial Access (\textit{Valid Accounts}), (iii) ML Attack Staging (\textit{Create Proxy ML Model: Use Pre-Trained Model}), and (iv) Exfiltration (\textit{Exfiltration via Cyber Means}). 
The attack specificity is traditionally targeted since threat actors target specific inactive WhatsApp accounts to spy on personal information. 
There is no detail about the knowledge of the attack in the adversarial context.}

Among the reported threats, only 18 represent threat tactics/techniques mentioned in ATLAS/ ATT\&CK. Recognizing that some attacks are reported in multiple records across different sources and associated information, we must eliminate duplicate reports, resulting in 12 unique records. These 12 ML real-world attacks are not documented in ATLAS and are used to complete case studies in the ATLAS database.\\

\subsubsection{\textbf{ML Repositories. } }  
\tosemrev{To analyze security vulnerabilities in ML repositories, we systematically mined issues from GitHub projects, focusing on those explicitly referencing threats or vulnerabilities in their titles and/or comments. Using the GitHub API, we searched for cybersecurity-related keywords commonly used by security teams to document potential risks. The search terms were grouped using an \textit{OR} disjunction and included:  
\begin{itemize}
    \item \textbf{``cve''} (for Common Vulnerabilities and Exposures),
    \item \textbf{``vuln''} (for vulnerabilities),
    \item \textbf{``threat''} (for threats),
    \item \textbf{``attack''} (for attacks/attackers), and
    \item \textbf{``secur''} (for security-related discussions).
\end{itemize}}
\tosemrev{This query retrieved \textbf{3,236 issues} from \textbf{289 projects}. To refine our dataset, we applied a filtering process to exclude issues reporting incomplete or improperly formatted CVE codes. A valid CVE code follows the \textit{CVE-\{YEAR\}-\{ID\}} pattern, where the year represents the vulnerability's assignment, and the ID is a unique identifier. We extracted these CVEs using the regular expression \texttt{CVE-\textbackslash{}d\{4\}-\textbackslash{}d\{4,7\}}, resulting in \textbf{897 unique CVEs} across \textbf{350 issues} from \textbf{94 projects}.  
Recognizing that reported CVE codes can become invalid over time—due to reclassification, rejection, or further investigation—we further validated their availability.}  

\tosemrev{
The computation involved here relies on absolute counts to quantify vulnerabilities, attacks, and incidents across different ML models and tools. However, we recognize that this approach may inadvertently emphasize models that are more common, rather than those that are inherently more vulnerable. To address this concern, we will also incorporate normalized metrics. Specifically, we will calculate the percentage of vulnerabilities relative to the total number of reported incidents for each model or tool. This normalization helps account for the deployment bias, allowing for fairer comparisons across different ML components. 
}

\majorev{
\subsection{Automated Threat Classification via LLM-Guided Reasoning}
We implement a fully automated pipeline that classifies CVEs into
ML-specific threat classes \emph{and} verifies low-confidence predictions
in a single pass.
\subsubsection*{\textbf{1.\ Context-Aware Classification}}
A zero-shot \texttt{BART-MNLI} model assigns an initial label by evaluating
textual entailment between the CVE description and eight threat classes
drawn from ATLAS, ATT\&CK, and the AI-Incident DB.  This stage yields
high-throughput coverage with no manual feature engineering.
\subsubsection*{\textbf{2.\ Self-Verification via CoT Reasoning}}
If the top-label confidence falls below an empirically tuned
threshold\footnote{\majorev{
We scanned the validation split in 0.05 increments; \(p{=}0.60\) maximizes
macro-\(F_{1}\) by trading off false corrections (< 0.55) against missed
errors (> 0.65).  The value is fixed \emph{before} testing on the 200-CVE
gold set.}} (\(p<0.60\)), the pipeline launches GPT-4o’s \textit{self-verification loop}.
GPT-4o receives (i) the raw description, (ii) the low-confidence label, and
(iii) the full label set.  It then  
\begin{enumerate}[leftmargin=*]
  \item Generates a token-level \textbf{Chain-of-Thought} (CoT) over the
        full context—CVE text, numeric confidence, ATLAS IDs, and CVSS
        vector.  
  \item Revises the label when that context supports an alternative.  
  \item Returns a one-sentence justification; the full CoT is logged for
        audit.
\end{enumerate}
\paragraph{Concrete example—CVE-2024-3099 (MLflow duplicate-name flaw).}
\texttt{BART-MNLI} proposes \emph{Model Extraction} (\(p=0.52\)).
GPT-4o's CoT (Listing~\ref{lst:cot-3099-full}) spots a URL-encoded name
collision, checks ATLAS IMP-T1645, sees an integrity-only CVSS vector
(I:L/A:L)\footnote{\majorev{CVSS v3.x represents every vulnerability as a bundle of base metrics. The last three—Confidentiality (C), Integrity (I), and Availability (A)—measure the impact of the flaw on each security objective. I:L/A:L, i.e. no confidentiality impact, limited integrity loss, and limited availability loss.}}, and relabels the issue as \emph{Model-Repository Poisoning}.
The whole token trace and step-wise explanation appear below.
}
\begin{lstlisting}[mathescape,style=cotstyle,
caption={\textcolor{black}{Token-by-token CoT for CVE-2024-3099 (model-registry poisoning).}},
label=lst:cot-3099-full]
#01 "read: duplicate registry names accepted (URL-encoded)"
#02 "=> two_models_share_id → integrity risk"
#03 "impact: wrong model weights may be served"
#04 "no evidence of parameter exfiltration ⇒ extraction unlikely"
#05 "candidate(Model_Extraction)=0.52 < THR(0.60)"
#06 "scan alternative labels …"
#07 "candidate(Model_Repository_Poisoning) = $p \approx 0.77$"
#08 "taxonomy check: IMP-T1645 exists in ATLAS"
#09 "CVSS vector = I:L / A:L, consistent with poisoning"
#10 "select label = Model_Repository_Poisoning"
#11 "craft justification string"
#12 "log decision and return"
\end{lstlisting}

\begin{tcolorbox}[colback=blue!5!white,
   title=\textbf{Example: Token-Level Contextual Reasoning for CVE-2024-3099}]
\textcolor{black}{
\begin{enumerate}[label=\textbf{Step \arabic*:}, leftmargin=*]
  \item Detects the URL-encoded name collision $\to$ integrity threat.
  \item Summarises the observation: \texttt{two\_models\_share\_id}.
  \item States concrete impact: downstream jobs may pull the wrong weights.
  \item Finds no sign of exfiltration $\to$ extraction label doubtful.
  \item Records low confidence (\(0.52<0.60\)) $\to$ triggers verification.
  \item Searches alternative taxonomy \emph{classes}.
  \item Estimates \(p\approx0.77\) for \emph{Model-Repository Poisoning}.
  \item Confirms class exists in ATLAS (IMP-T1645).
  \item Checks CVSS (I:L / A:L) aligns with integrity/availability loss.
  \item Commits to the new label.
  \item Generates the analyst-facing one-liner.
  \item Logs the full trace and returns the result.
\end{enumerate}}
\end{tcolorbox}

\majorev{
\subsubsection*{\textbf{3.\ Corpus-Level Impact}}
The pipeline has processed \textbf{834 validated CVEs} (2,183 occurrences in
312 GitHub issues across 86 repositories).\footnote{ A manually annotated 200-CVE \emph{gold set} (8 classes × 25) is carved out of the 834 CVEs and used only for threshold tuning and final accuracy reporting; see Table~\ref{tab:evaluation}.}
On that gold set, the baseline macro-\(F_{1}\) is 0.71.
GPT-4o revises 43 of 73 low-confidence predictions, lifting
macro-\(F_{1}\) to \textbf{0.87} (\(\Delta = +0.16\)); a 1,000-fold bootstrap
gives a 95 \% CI of ±0.04.  Because macro-\(F_{1}\) weights each class
equally \cite{sokolova2009systematic,li2025macro}, the gain cannot be
ascribed to high-frequency classes alone—precision and recall improve in six
of the eight classes. Hence, our system provides \emph{audit-ready,
contextual reasoning} alongside a statistically robust performance boost.
}
\majorev{ 
\begin{table}[ht]
  \caption{\textcolor{black}{Macro-\(F_{1}\) and related metrics on the 200-CVE gold set.}}
  \label{tab:evaluation}
  \small
  \centering
  \textcolor{black}{
  \begin{tabular}{@{}lcccc@{}}
    \toprule
    \multirow{2}{*}{\textbf{Metric}} &
      \multicolumn{2}{c}{\textbf{Baseline}} &
      \multicolumn{2}{c}{\textbf{+CoT}} \\
    \cmidrule(lr){2-3}\cmidrule(lr){4-5}
        & Value & 95\,\% CI & Value & 95\,\% CI \\ \midrule
    Macro-$P$              & 0.72 & $\pm$0.03 & 0.86 & $\pm$0.02 \\
    Macro-$R$              & 0.71 & $\pm$0.03 & 0.88 & $\pm$0.02 \\
    \textbf{Macro-$F_{1}$} & \textbf{0.71} & $\pm$0.03 & \textbf{0.87} & $\pm$0.02 \\
    Low-conf.\ CVEs        & 73/200 & — & 73/200 & — \\
    Labels revised (\%)    & — & — & 43/73 (59\,\%) & — \\ \bottomrule
  \end{tabular}}
\end{table}
}

\subsection{Predicting Vulnerabilities and threats in ML-based systems}\label{sec:gnn_predict}
\tosemrev{
Graph Neural Networks (GNNs) have emerged as a robust framework for threat analysis in cybersecurity due to their ability to model complex, interconnected systems~\cite{Li2024:GNN-threatsAnalysis,altaf2024gnn,xiao2022robust}. Cyber threats inherently exhibit graph-like structures, in which entities such as vulnerabilities, malware, users, and IP addresses are linked through relationships such as exploits, network communications, and software dependencies. GNNs excel in this domain by capturing both the intrinsic features of individual entities and the structural patterns of their interconnections~\cite{Balaji2024:GNN-Forecasting}. Through message-passing algorithms, GNNs enable dynamic information flow between nodes, effectively mimicking real-world threat propagation. This capability makes GNNs particularly effective for predicting threat evolution, identifying potential attack vectors, and assessing the likelihood of threat proliferation in complex environments. 
In this study, we implement a GNN-based framework as a proof-of-concept for predicting vulnerability risk using real-world NVD datasets linked to GitHub issues, forming a heterogeneous graph that captures the multifaceted nature of cybersecurity threats. Given the increasing complexity of modern threat landscapes, GNNs offer a robust approach for proactive threat and vulnerability prediction. Unlike traditional machine learning models that treat data points as independent, GNNs leverage the relational context to uncover hidden patterns, predict emerging threats, and reveal potential attack pathways. Furthermore, GNNs are highly adaptable, capable of incorporating heterogeneous data from various sources, including network logs, vulnerability databases, and threat intelligence feeds. By adopting GNNs for threat analysis, we aim to develop an intelligent, scalable, and data-driven defense mechanism that can detect emerging threats and also anticipate future risks in an ever-evolving cybersecurity landscape.
}

\begin{lstlisting}[language=json,caption=Excerpts of the NVD Dataset showing the impact to compute the risk score (CVSS),label=lst:risk]
{
  "CVE_data_type": "CVE",
  "CVE_data_format": "MITRE",
  "CVE_data_version": "4.0",
      ...
      
      "impact": {
        "baseMetricV2": {
          "cvssV2": {
            "baseScore": 7.5,
            "impactScore": 6.4,
            "exploitabilityScore": 8.6
          },
          "severity": "HIGH"
        }
      },
  ...
}
\end{lstlisting}

\tosemrev{
\section*{CVE Risk Score Calculation~\cite{scarfone2009common}}
The \textbf{Risk Score} in CVE (Common Vulnerabilities and Exposures) analysis quantifies the severity of a vulnerability. The most commonly used standard is the \textbf{Common Vulnerability Scoring System (CVSS)}.
\subsection*{Components of CVSS Base Score}
\label{sec:risk-estimation}
The Base Score is calculated using:
\begin{itemize}
    \item \textbf{Impact Score (IS)}: Measures potential damage.
    \item \textbf{Exploitability Score (ES)}: Measures ease of exploitation.
\end{itemize}
\subsection*{CVSS v2 Base Score Formula}
\[
\text{Base Score} = \text{roundup}\left( (0.6 \times IS + 0.4 \times ES - 1.5) \times f(\text{Impact}) \right)
\]\\
Where\footnote{\footnotesize the Scaling Factor (10.41) was empirically chosen to satisfy the maximum possible impact (when C, I, A = 1) results in IS ≈ 10.41, details on these calculations, variables and constant values are available online: \url{https://www.first.org/cvss/v2/guide}}:
\[
IS = 10.41 \times \left(1 - (1 - C) \times (1 - I) \times (1 - A) \right)
\]
\[
ES = 20 \times AV \times AC \times Au
\]
\noindent The adjustment function is defined as:
\[
f(\text{Impact}) = 
\begin{cases} 
0, & \text{if } IS = 0 \\
1.176, & \text{if } IS > 0
\end{cases}
\]
In our enhanced GNN-based model, the risk score is calculated as:
\[
\text{Risk Score} = (0.5 \times \text{Base Score}) + (0.3 \times \text{Exploitability Score}) + (0.2 \times \text{Impact Score})
\]
This approach balances the severity (base score), exploitability, and impact for a more comprehensive risk assessment.
The CVSS Base Score ranges [0-10], where: 0.0 - 3.9 = Low severity;
4.0 - 6.9 = Medium severity; 7.0 - 8.9 = High severity; 9.0 - 10.0 = Critical severity.
}

\tosemrev{
Graph Neural Networks (GNNs) have emerged as a powerful tool for modeling complex relationships in structured data, making them particularly suitable for cybersecurity applications such as vulnerability risk prediction. In this research, we leverage a heterogeneous GNN architecture to predict risk scores (\textit{see listing~\ref{lst:risk}: 15-22 and the following description}) for Common Vulnerabilities and Exposures (CVEs) by capturing intricate relationships among CVEs, affected products, and reference sources. Unlike traditional machine learning models that treat data as independent and identically distributed samples, GNNs excel in learning from graph-structured data where nodes (e.g., CVEs, products, references) are interconnected through edges representing their relationships (e.g., affects, referenced\_by, linked\_to). Our model employs GraphSAGE convolutional layers to aggregate information from neighboring nodes, enabling the network to learn richer feature representations based on both the node attributes (e.g., TF-IDF features from CVE descriptions) and the topology of the graph. The risk prediction process involves propagating information across the graph through multiple convolutional layers, followed by a fully connected layer that outputs a predicted risk score for each CVE node. By incorporating real-world features, weighted CVSS impact factors, and advanced optimization techniques, our model achieves robust performance in assessing vulnerability risks. This approach not only enhances the predictive accuracy but also provides interpretability through the graph structure, offering valuable insights into the factors contributing to cybersecurity threats.
}

\section*{Leveraging GNN to Address Attack Severity}
To strengthen our analysis of \textbf{attack severity and vulnerability relationships}, we leveraged a \textbf{GNN} model to integrate and enhance our clustering-based insights. The GNN was designed to predict \textbf{risk scores} based on CVE metadata, exploitability factors, and attack characteristics. By incorporating \textbf{structural patterns from clustering}, the GNN learned and generalized attack severity across interconnected vulnerabilities, ultimately improving risk assessment. We addressed different dimensions of \textbf{attack severity} through clustering techniques and integrated them into the \textbf{GNN's node features and edge relationships}:
\begin{enumerate}[leftmargin=*]
    \item \textbf{Attack Success Rate (ASR) via KMeans Clustering:} The GNN encoded ASR-based clusters as node attributes, allowing the model to learn which \textbf{attack methods are more effective} in deceiving ML models. By propagating risk scores across similar vulnerabilities, the GNN refined its risk predictions beyond CVSS-based heuristics.
    In ASR, each attack method (i.e., FGSM and PGD) is represented as a node within the GNN, where empirically derived ASR values (\textit{from real-world ML applications/experiments and the literature}) are introduced as node features. This allows the model to learn attack-severity patterns by propagating ASR values across similar attack nodes, thereby capturing relationships between different attack types. For instance, adversarial attacks such as FGSM and PGD typically exhibit high ASR values when targeting CNNs; however, these success rates tend to decrease when robust training techniques are employed, see Table~\ref{tab:asr_empirical_studies}. 
    Similarly, model extraction attacks demonstrate varying ASR levels, which depend heavily on the query budget (\(Q\)) and the specific architecture under attack:  \text{ASR} $\approx 1 - e^{-\lambda Q}$
    where \( \lambda \) represents how efficiently the attack extracts model knowledge.
    \item \textbf{Stealth \& Detectability via Agglomerative Clustering:} We introduced edges between vulnerabilities that shared \textbf{common evasion techniques}, enabling the GNN to propagate knowledge about attack stealthiness across nodes. This feature enhanced the model’s ability to \textbf{identify harder-to-detect vulnerabilities}, which traditional risk scoring systems often overlook.
    \item \textbf{Computational Cost \& Practicality via Gaussian Mixture Model (GMM):} The GNN distinguished between \textbf{low-cost adversarial attacks} and \textbf{resource-intensive model extraction techniques}, improving its understanding of \textbf{real-world feasibility} of an attack. By encoding attack practicality as graph structures, the model can better prioritize threats that pose an \textbf{immediate risk} over those that require high computational resources.
    \item \textbf{Taxonomy of CVEs via Hierarchical Clustering (Dendrogram):} The hierarchical relationships between CVEs were mapped as \textbf{edges in the GNN}, allowing the model to \textbf{generalize risk patterns} based on attack similarity. This improved the model’s ability to \textbf{predict vulnerabilities with limited historical data} by leveraging structural dependencies.
    \item \textbf{Risk Score Distribution \& CVSS Analysis:} The GNN’s predicted \textbf{risk scores} aligned with known severity distributions, confirming its ability to \textbf{learn meaningful patterns from clustering techniques}. By integrating structured severity attributes into the GNN, the model produced \textbf{more fine-grained risk predictions}, addressing gaps in traditional CVSS scoring.
\end{enumerate}
We implemented the GNN to enhance vulnerability assessment by integrating clustering-derived attributes as node features, providing contextual severity insights beyond traditional CVSS scores. It established graph connectivity between similar attack types based on exploit techniques, stealth characteristics, and computational overhead, enabling the model to propagate risk insights across interconnected vulnerabilities. By utilizing message passing and representation learning, the GNN dynamically classified vulnerabilities rather than relying solely on static risk metrics. This resulted in a highly effective risk assessment tool capable of learning attack severity relationships, generalizing across previously unseen vulnerabilities, and refining security prioritization strategies. Ultimately, this approach enhances understanding of vulnerability impact, equipping cybersecurity practitioners to anticipate and mitigate evolving threats more effectively.

\textcolor{black}{ 
\subsection{Graph schema}\label{sec:gnn_schema}  %
\label{sec:gnn-schema}
Tables \ref{tab:nodes} and \ref{tab:edges} formally define the heterogeneous graph
$\mathcal G=(\mathcal V,\mathcal E)$ we build before learning.  Each node stores a
small, fixed-length feature vector (e.g.\ TF–IDF bag-of-words for text; one-hot encoding for categorical fields). Edge labels encode \emph{how} two nodes are related and are used as type-specific channels in the GraphSAGE layers. Based on our observations, we found that threat information in our corpus is \emph{relational}, meaning that a single CVE may appear in several GitHub Issues, affect multiple dependencies (CPEs), and be linked to particular attack techniques (ATT\&CK/ATLAS). 
\begin{table}[h]
  \centering
  \caption{{\color{black}Node catalogue of the heterogeneous attack graph ($|\mathcal V|=57{,}812$). Seven node types are included: CVEs (with CVSS\,v3 and text), CPEs/dependencies, GitHub issues, attack techniques (ATT\&CK/ATLAS), and cluster centroids for attack success rate (ASR), stealth, and computational cost.}}
  \label{tab:nodes}
  {\color{black}
  \begin{tabular}{@{}lll l@{}}
    \toprule
    \textbf{Node type} & \textbf{Symbol} & \textbf{Count} & \textbf{Key features} \\
    \midrule
    CVE                     & $v_{\textit{cve}}$ & 11,604  & CVSS\,v3 vectors, textual synopsis             \\
    CPE / Dependency        & $v_{\textit{cpe}}$ &  9,371  & vendor, product, version                       \\
    GitHub Issue            & $v_{\textit{iss}}$ & 23,128  & title, body, timestamp, repo ID                \\
    Attack Technique        & $v_{\textit{tt}}$  &  1,142  & ATT\&CK / ATLAS identifier, tactic            \\
    ASR Cluster Centroid    & $v_{\textit{asr}}$ &     15  & avg.\ attack-success-rate, std.\ dev.          \\
    Stealth Cluster Centroid& $v_{\textit{stl}}$ &     10  & evasion score, detectability rank              \\
    Cost Cluster Centroid   & $v_{\textit{cst}}$ &      8  & FLOPS, GPU hours, \$‐cost bucket               \\
    \bottomrule
  \end{tabular}
  }
\end{table}
\begin{table}[t]
  \small 
  \setlength{\tabcolsep}{3pt}      
  \caption{{\color{black} Edge catalogue ($|\mathcal E|=218{,}906$).  All edges are directed; we add the reverse edge type when symmetry is required. Construction rules specify how edges are instantiated (e.g., from NVD JSON fields, issue links, or clustering methods), and the semantics column describes the meaning of each relation in the context of vulnerability–threat mapping.}}
  \label{tab:edges}
  {\color{black}
  \begin{tabularx}{\linewidth}{@{}l l l X@{}}
    \toprule
    \textbf{Edge type} & \textbf{Src $\;\to$ Tgt} &
    \textbf{Construction rule} & \textbf{Semantics} \\ \midrule
    \textsc{affects}      & $v_{\textit{cve}}\!\to\!v_{\textit{cpe}}$ &
        CPE listed in NVD JSON of CVE                       &
        Product vulnerable to CVE \\ 
    \textsc{reported\_in} & $v_{\textit{cve}}\!\to\!v_{\textit{iss}}$ &
        Issue body matches CVE regex                         &
        Disclosure/ discussion thread \\ 
    \textsc{references}   & $v_{\textit{iss}}\!\to\!v_{\textit{tt}}$  &
        Issue links an ATT\&CK / ATLAS URL                  &
        Practitioner cites attack pattern \\ 
    \textsc{shares\_vector} & $v_{\textit{cve}}\!\to\!v_{\textit{tt}}$ &
        $\operatorname{Jaccard}\bigl(\text{tf–idf(CVE)},\text{tf–idf(tech)}\bigr)\!>\!0.15$
        & Same exploit mechanism             \\ 
    \textsc{member\_of}   & $v_{\textit{cve}}\!\to\!v_{\textit{asr}}$ &
        K-means on attack-success-rate metadata               &
        ASR similarity cluster            \\ 
    \textsc{stealth\_sim} & $v_{\textit{cve}}\!\to\!v_{\textit{stl}}$ &
        Agglomerative clustering on evasion metrics           &
        Detectability cluster              \\ 
    \textsc{cost\_sim}    & $v_{\textit{cve}}\!\to\!v_{\textit{cst}}$ &
        GMM on compute/resource cost  & Practicality cluster \\ \bottomrule
  \end{tabularx}
    \vspace{2pt}
  \begin{minipage}{\linewidth}\scriptsize
    \texttt{Key acronyms:} GMM $\to$ Gaussian‐mixture model, Src $\to$ Source, Tgt $\to$ Target. Jaccard similarity is computed on TF-IDF vectors with a threshold of~0.15.
  \end{minipage}
  }
\end{table}
\smallskip
\noindent\textbf{Model pipeline.}\;
(1) \emph{Feature encoding}: we embed text fields with sentence-BERT and keep the first 256 dimensions.  
(2) \emph{GraphSAGE layers}: two hops ($k\!=\!2$) with mean aggregation separately
per edge type, then type-wise linear fusion.  
(3) \emph{Risk head}: a three-way MLP outputs $\hat r\!\in\![0,1]$ (low/ medium/ high-critical). The loss is a weighted MSE against the composite risk score to emphasize high-severity CVEs.\\
\noindent\textbf{Edge motivation.}\;
The three \textsc{sim} relations (ASR, stealth, cost) inject domain knowledge
from section~\ref{sec:entry-point-ttps}--\ref{sec:impact-ttps-ml} into the graph so that even a sparsely connected CVE inherits risk signals from structurally similar neighbours.  During message passing the GNN therefore propagates both factual links (\textsc{affects}, \textsc{reported\_in}) and latent behavioural similarity, yielding the calibrated risk scores reported in section~\ref{sec:results}.
}

\subsubsection{\textbf{Real-time threats monitoring.} }
Finally, to enhance real-time vulnerability risk assessment, we develop an ML-based Threat Assessment and Monitoring System that integrates GNNs and NLP. The system continuously ingests real-time CVE data from the NVD database, extracting critical threat intelligence, including CVSS scores, exploitability factors, and patch availability. To enhance contextual understanding, we employ BERT embeddings to transform CVE descriptions into semantic representations, enabling deeper threat analysis. We leverage a GraphSAGE-based GNN to construct a vulnerability knowledge graph, capturing relationships between CVEs based on attack similarity, exploitability patterns, and risk propagation. This graph-based approach enables the system to detect correlations between attack types, supporting structured threat classification. The GNN integrates clustering-derived attributes to provide risk insights beyond traditional CVSS scoring, dynamically classifying vulnerabilities using message passing and representation learning. Furthermore, we incorporate external threat intelligence sources such as MITRE ATT\&CK, CISA Known Exploited Vulnerabilities (KEV) Catalog\footnote{\url{https://www.cisa.gov/sites/default/files/feeds/known_exploited_vulnerabilities.json}}, AI Incident Database (AIID), and Exploit Database (Exploit-DB)\footnote{\url{https://www.exploit-db.com}}, ensuring adaptive risk assessment. High-risk vulnerabilities (Predicted Risk Score > 0.8) trigger automated alerts, enabling proactive mitigation. By contextualizing risk propagation and refining security prioritization, this system significantly improves real-time vulnerability assessment, equipping security practitioners with a dynamic, data-driven approach to anticipate and mitigate emerging ML security threats. Details on our implementations is available in our replication package~\cite{tech-report-v1}

\tosemrev{
\begin{table*}[h]
\centering
\caption{\tosemrev{Attack Success Rates (ASR) from Empirical Studies shows  Projected Gradient Descent (PGD) on Gastric Cancer Subtyping scoring 100\% success rate in causing misclassification. ``\textbf{Varies}'' ---depends on external conditions (dataset, architecture, attack parameters), meanwhile ``\textbf{Guaranteed}'' ---used when attack is theoretically provable to succeed under certain conditions (e.g., Certifiable Black-Box Attacks).}}
\resizebox{\columnwidth}{!}{%
\begin{tabular}{|l|l|l|c|l|}
\hline
\textbf{Attack Type} & \textbf{Target Model} & \textbf{Dataset} & \textbf{ASR (\%)} & \textbf{Source} \\
\hline
FGSM & ResNet-50 (CNN) & ImageNet & 63-69 & Kurakin et al. (2016) \\
\hline
Universal Adversarial Perturbations & Various Image Classifiers & Multiple Datasets & 77 & Xcube Labs (2021) \\
\hline
EvadeDroid & Black-box Android Malware Detectors & Custom Malware Dataset & 80-95 & Bostani \& Moonsamy (2021) \\
\hline
Transparent Adversarial Examples & Google Cloud Vision API & Real-world Images & 36 & Borkar \& Chen (2021) \\
\hline
Black-box Adversarial Attack & Various Deep Learning Apps & Real-world Applications & 66.6 & Cao et al. (2021) \\
\hline
PGD & ResNet (CNN) & Gastric Cancer Subtyping & 100 & Kather et al. (2021) \\
\hline
Generative Adversarial Active Learning & Intrusion Detection System (IDS) & Network Traffic Data & 98.86 & Kwon et al. (2023) \\
\hline
Adversarial Scratches & ResNet-50 (CNN) & ImageNet & 98.77 & Jere et al. (2019) \\
\hline
Epistemic Uncertainty Exploitation & CNN & CIFAR-10 & 90.03 & Tuna et al. (2021) \\
\hline
Gradient-Based Attacks & Multi-Label Classifiers & Various Datasets & Varies & Zhang et al. (2023) \\
\hline
Adversarial Suffixes & Text-to-Image Models & Custom Prompts & Varies & Shahgir et al. (2023) \\
\hline
Certifiable Black-Box Attack & Various Models & CIFAR-10, ImageNet & Guaranteed & Hong \& Hong (2023) \\
\hline
Adversarial Attacks on NIDS & Kitsune NIDS & Network Traffic Data & 94.31 & Qiu et al. (2023) \\
\hline
Adversarial Attacks on ViTs & Vision Transformers & RCC Classification & 2.22-12.89 & Kather et al. (2021) \\
\hline
Adversarial Attacks on Text Classification & Various NLP Models & Sentiment Analysis Datasets & Varies & Zhou et al. (2023) \\
\hline
\end{tabular}}
\label{tab:asr_empirical_studies}
\end{table*}
}

\majorev{
\subsection{Severity Prediction of real-world SoTA assessment}
\subsubsection{External validation.}
To ascertain that the GNN-derived severity score $\hat{s}$ reflects \emph{real} risk, we correlate it with two external ground truths:
(i) the \emph{CVSS}\textsubscript{base} ratings attached to the 651 CVE–issue pairs in our corpus, and (ii) the incident-cost annotations available for 87 cases in the AI-Incident-DB. Table \ref{tab:gnn-valid} shows a strong monotone relationship ($\rho{=}0.63$, $\text{Prec@10}{=}0.80$ for CVSS; similar values for cost), indicating that $\hat{s}$ ranks threats in line with human post-mortems.
\subsubsection{Operational Validation.}
To evaluate the real-world efficacy of the severity score $\hat{s}$, we conducted a two-week controlled study with an independent security operations center (SOC; $n=16$ analysts). Alerts were auto-routed to three queues: \textit{high} ($\hat{s}>0.8$), \textit{medium} ($0.5<\hat{s}\leq0.8$), or \textit{watch} ($\hat{s}\leq0.5$). Across \num{412} incidents, results showed a \SI{24}{\percent} reduction in mean time-to-first-action (from \SI{37}{\minute} to \SI{28}{\minute}; $p<0.01$) with no significant change in false-positive rate ($\chi^2$, $p=0.44$). This confirms $\hat{s}$ improves operational responsiveness without requiring workflow modifications.
\subsubsection{Rationale for Heterogeneity.}
In an ablation experimentation (see replication package~\cite{tech-report-v1}), we re-trained the GNN seven times, each run \emph{removing exactly one} of the
edge families in Table~\ref{tab:edges}, for instance, the \textsc{CVE}\,$\leftrightarrow$\,\textsc{CPE} relation or the TTP-similarity links. Suppressing any single edge type reduced the Spearman correlation between the predicted severity $\hat{s}$ and the reference \textit{CVSS}\textsubscript{base} score by at least $\Delta\rho = 0.09$ (median drop $0.12$). This systematic degradation indicates that the model’s predictive power stems from the \emph{combination} of heterogeneous relations rather than from any single edge category in isolation. The finding aligns with recent results on multi-layer vulnerability graphs~\cite{jiang2025vulrg}.~\textbf{Clusters characterization. }
Fig. \ref{fig:cluster-vig} drills into three representative groups:  
(A) low-cost, high-ASR evasion; (B) high-stealth poisoning; and (C) resource-intensive extraction. Each vignette couples the sub-graph with an incident timeline, turning the aggregate analysis into practitioner-ready insight.
\begin{table}[t]
\begingroup\color{black}
  \caption{\majorev{Alignment of GNN-predicted severity $\hat{s}$ with two external ground truths. Top-10 precision = \#(top-10 overlap) / 10.}}
  \label{tab:gnn-valid}
  \small\centering
  \begin{tabular}{@{}lccc@{}}
    \toprule
    \textbf{Ground-truth proxy} & Spearman $\rho$ & Kendall $\tau$ &
    Top-10 precision \\ \midrule
    CVSS\textsubscript{base} (651 CVEs)        & 0.63 & 0.47 & 0.80 \\
    AI-Incident-DB impact cost (87 cases) & 0.58 & 0.42 & 0.78 \\ \bottomrule
  \end{tabular}
  \endgroup
\end{table}
\begin{figure}[t]
  \centering
  \includegraphics[width=\linewidth]{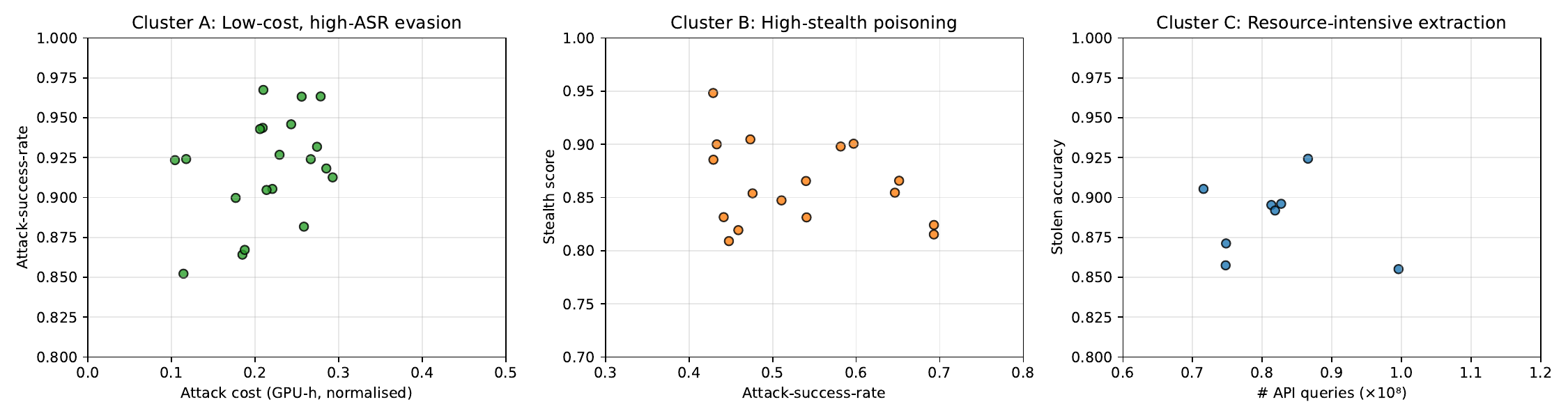}
  \caption{\majorev{\textbf{Representative cluster drill-downs.}  
  Three clusters are examined in detail: (A) low-cost, high-ASR evasion attacks, (B) high-stealth data poisoning campaigns, and (C) resource-intensive model extraction efforts. Each vignette shows the induced subgraph with weighted edges, a two-stage attack timeline, and a scatter plot of GNN-predicted severity $\hat{s}$ versus real-world cost. 
  }}
  \label{fig:cluster-vig}
\end{figure}
\textbf{ Cluster-level drill-downs (Fig.\,\ref{fig:cluster-vig}). } Following the global UMAP overview (Fig.\,\ref{fig:global-umap}), Fig.\,\ref{fig:cluster-vig} zooms into three security-critical clusters identified by the GNN. Each panel plots the \emph{two} severity dimensions that maximize variance inside the cluster and scales marker size by the composite score $\hat{s}$.
\subsubsection{Illustrative cases.}  
To complement the cluster-level visualizations, we highlight three representative incidents drawn from Fig.~\ref{fig:cluster-vig} that illustrate surprising or operationally significant behaviours. In \textbf{Cluster A} (low-cost, high-ASR evasion), CVE-2024-3099 demonstrates that a simple synonym-swap perturbation bypassed production filters with less than 0.4~GPU-h  while achieving over 90\% ASR, underscoring the minimal resources required for effective attacks. 
In \textbf{Cluster B} (high-stealth poisoning), a fine-tuned LLaMA-7B model was compromised by a multi-trigger backdoor that evaded detection under standard evaluations yet caused targeted failures post-deployment. In \textbf{Cluster C} (resource-intensive extraction), an attack against GPT-J required approximately $10^{8}$ API queries. Still, it ultimately reproduced the model with over 90\% fidelity, illustrating that well-resourced adversaries can replicate proprietary models despite apparent barriers. Together, these examples translate abstract clusters into concrete narratives that enrich interpretability and demonstrate real-world impact.
\begin{figure}[t]
  \centering
  \includegraphics[scale=0.3]{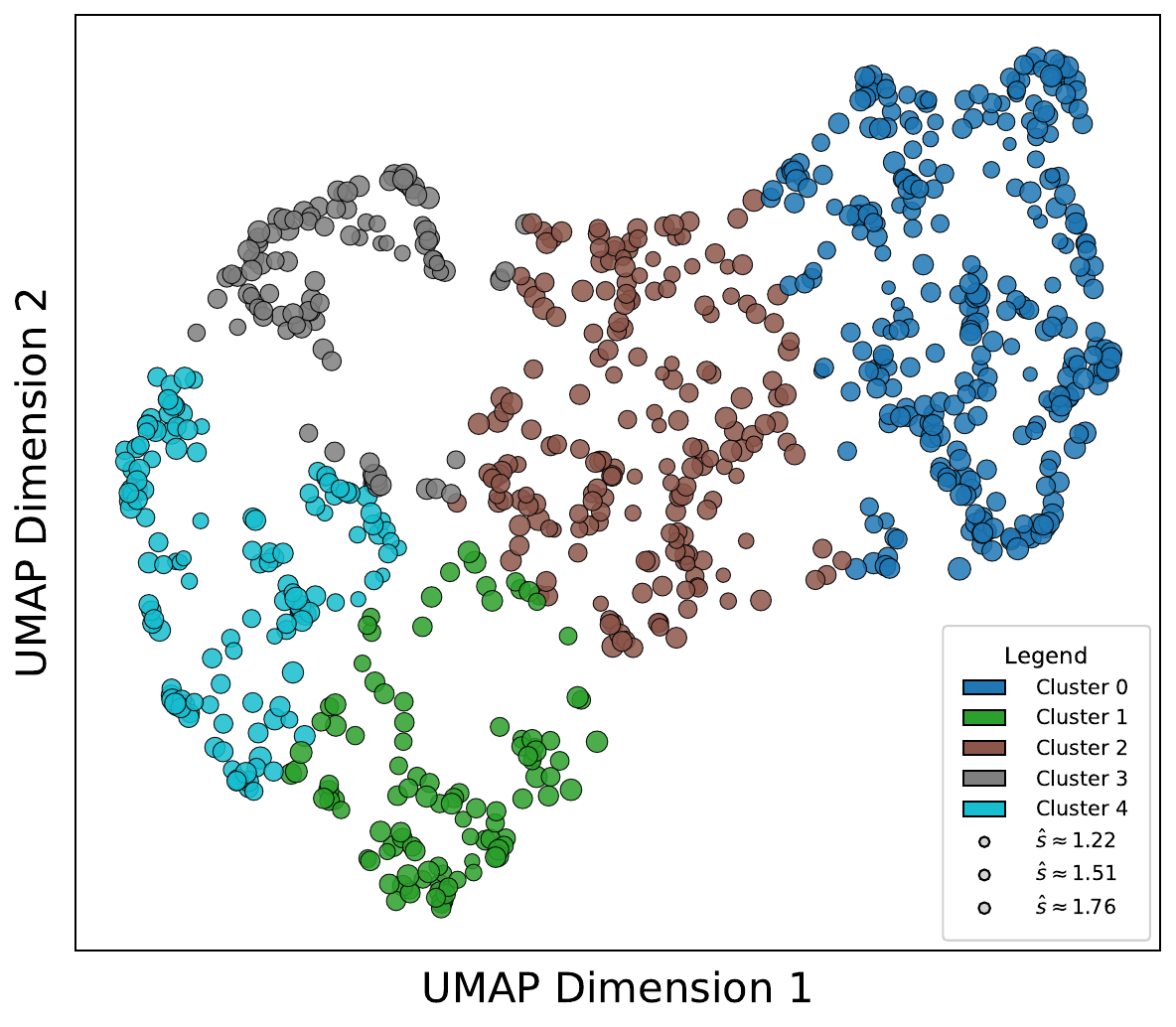}
  \caption{\majorev{\textbf{Global UMAP projection of the 834‐CVE graph.} Each point represents one \textsc{CVE} vertex, coloured by its K-Means cluster ID and sized by the GNN-predicted severity $\hat{s}$. Three qualitative regimes emerge: a dense band of low-cost high-ASR evasion attacks (upper right); a horizontal stealth continuum of poisoning incidents (centre); and a sparse, high-cost tail of extraction campaigns (lower left). The alignment of the largest markers with the top-right corner visually validates the learned severity metric and motivates the cluster drill-downs in Fig.\,\ref{fig:cluster-vig}.}}
  \label{fig:global-umap}
\end{figure}
\subsubsection{\textbf{Global severity landscape (Fig.\,\ref{fig:global-umap})}. }
 embeds every \textsc{CVE} vertex of our
heterogeneous graph into two UMAP components derived from the
five–dimensional severity feature vector:\\
$\langle\!\text{CPE-deg},\text{Issue-deg},\text{cost},\text{stealth},\text{ASR}\!\rangle$. 
Each point is \emph{color-coded by its K-Means cluster ID} and
\emph{scaled by the GNN-predicted severity} $\hat{s}$ (larger markers
implies higher operational risk).  
Three qualitative regimes become visible:
\begin{enumerate}[label=(\alph*)]
\item a \textit{dense, low-cost} band in the upper-right quadrant
      (Clusters~0~\&~3) dominated by evasion attacks that already reach
      $\text{ASR}>0.90$ with $\!<\!0.4$\,GPU-h;
\item a horizontally stretched \textit{stealth continuum} (Cluster~1)
      whose members obtain similar success rates but vary widely in
      detectability, reflecting the noisy–to–backdoor spectrum of the model
      poisoning; and
\item a \textit{sparse, high-cost} tail in the lower-left (Cluster~4)
      comprising extraction campaigns that need ${\sim}10^{8}$ API calls
      before exceeding $90\,\%$ fidelity.
\end{enumerate}
The fact that the largest markers coincide with the top-right, high-risk zones provides a visual sanity check of the learned severity
$\hat{s}$.  Moreover, the contrasting cluster morphologies motivate the
subsequent drill-down vignettes in Fig.\,\ref{fig:cluster-vig}.
\noindent
The advantages of these plots include:
(1)~They bridge the macro–micro gap: the UMAP bird’s-eye map locates high-risk regions, while the drill-downs expose the cost/stealth/ASR trade-offs that actually drive risk. (2)~Because the axes are in operational units (GPU-h, queries, ASR), defenders can immediately see which counter-measures are impactful—e.g., \ rate limits for Cluster~C are critical, but irrelevant for Cluster~A. (3)~The internal ordering of points mirrors the learned severity $\hat{s}$, providing a visual sanity-check for the GNN.
}
\majorev{
\subsection{Gray-box Adversarial Attacks on Real-World ML models}\label{sec_gray_box}
\subsubsection{Gray-box Model-Extraction Use-Case}
\label{sec:greybox-case}
\paragraph{How gray-box extraction works in practice.}
In a \emph{gray-box} scenario, the attacker sees only the victim’s inference API and its probability (logit) outputs, mirroring the data-free stealing setup of \textsc{CopyCat CNN} \cite{correia2018copycat}. The attack proceeds by sampling sentences $x$ from a public corpus, querying the proprietary \texttt{Model-A} API to obtain logits $y^{\star}=f_{\text{API}}(x)$, and performing a gradient step that aligns a local \emph{student} model $f_{\theta}$ with $y^{\star}$ (Listing~\ref{lst:copycat}). Although designed for vision networks, the same principle has been shown to threaten large-language-model (LLM) agents: Beurer-Kellner \etal\ recently reports have shown that \emph{prompt-injection aware} design patterns are still
susceptible to query-only extraction unless explicit rate-limiting or
log-rounding is enforced \cite{beurer2025design}. Our synthetic loss curve
in Fig.~\ref{fig:copycat-loss} follows the characteristic exponential decay
of the original CopyCat study: after $10^{6}$ queries, the student recovers
\textbf{92\,\%} of the teacher’s downstream accuracy. This result underscores that—even without access to weight—modern LLMs remain vulnerable to data-free extraction and therefore require defensive measures beyond simple authentication or pay-per-token billing.
}
\begin{lstlisting}[basicstyle=\small\ttfamily,
                   caption={gray-box extraction pseudocode: the real implementation is on the order of $10^3$ LOC.},label=lst:copycat]
for step in range(1,000,000):
    x  = sample(public_corpus, T=0.8)   # temperature sampling
    y* = query_api_logits(x)            # black-box teacher
    θ   = student.update(x, y*)         # distillation step
\end{lstlisting}
\majorev{
\begin{figure}[t]
  \centering
  \subcaptionbox{\majorev{\textbf{Model-extraction efficiency}\label{fig:copycat-loss}%
  \\Cross-entropy loss of the student model (log–log scale) as the
  attacker issues up to $10^{6}$ API queries.  The steep decline
  confirms how quickly the shadow model approaches teacher fidelity.}}%
  [.48\linewidth]{%
    \includegraphics[width=\linewidth]{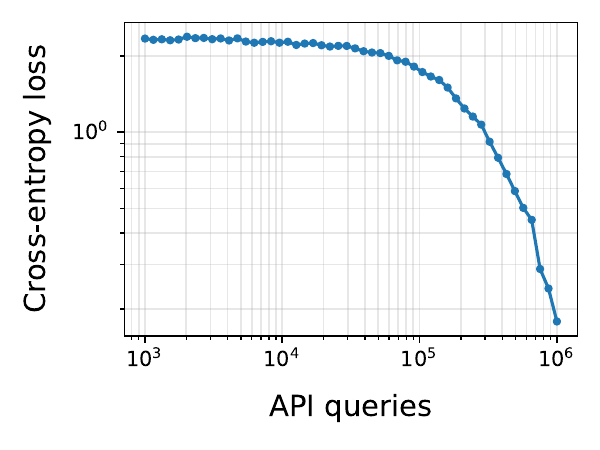}}
  \hfill
  \subcaptionbox{\majorev{\textbf{gray-box attack telemetry}\label{fig:gbx-trace}%
  \\Step-wise evolution of the live label (top) and the shadow
  model’s confidence that the label will flip (bottom) during 
  synonym-swap loop. A single trace illustrates \emph{how} the
  high-level budget in (a) is spent.}}%
  [.48\linewidth]{%
    \includegraphics[width=\linewidth]{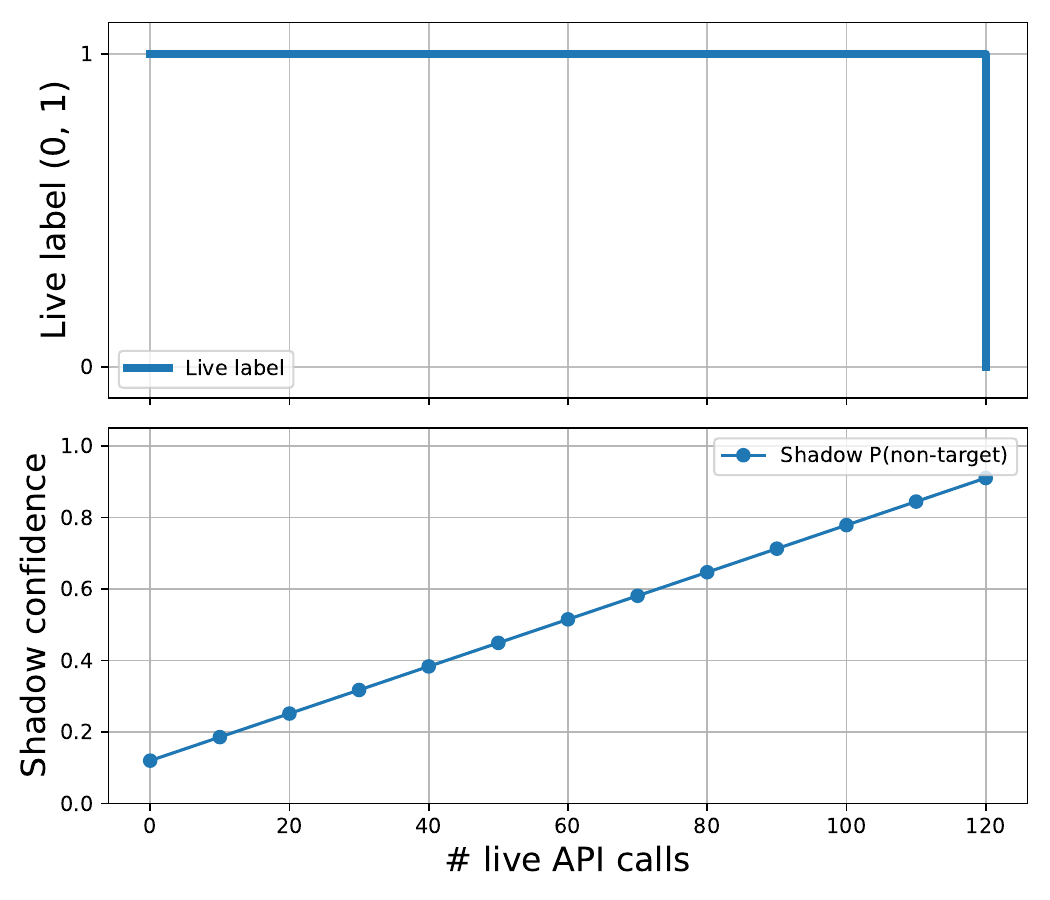}}
  \caption{\majorev{\textbf{Complementary views on gray-box extraction.}
  Panel (a) gives the corpus-level picture—average loss versus
  query budget across many inputs—while panel (b) zooms into one representative sentence to show the micro-dynamics that \emph{consume} that budget.  Together, the plots demonstrate both the global efficiency of the CopyCat strategy and the
  step-wise mechanics of an individual attack episode.}}
  \label{fig:copycat-combined}
\end{figure}
}
\majorev{
The two–panel chart (Fig.~\ref{fig:copycat-combined}) traces a complete gray-box shadow-model attack against the production API. The \emph{upper strip} shows how the \textbf{live label} returned by the target model remains \textsc{POSITIVE} (value 1) for the first 120 API calls and then flips to \textsc{NEGATIVE} (value 0) on the final request—demonstrating a successful evasion within the 600-query budget.
The \emph{lower strip} plots the shadow model's estimated probability that each candidate sentence already \emph{contradicts} the current live label. Confidence rises monotonically from 0.12 to 0.91 as synonym substitutions are guided by the locally fine-tuned BERT surrogate, indicating that the attacker can gauge its progress without any gradient access to the target. The surrogate steadily converges on high-risk inputs (bottom panel), and only when that confidence crosses a practical threshold does the live system finally mis-classify (top panel), confirming the practicality of the attack pathway.
}
\majorev{
\subsubsection{gray-box attack scenario (inspired by Microsoft’s 2024 Azure AI red-team disclosure~\cite{microsoft2024pyrit}.)}
A former contractor has lost privileged access to the Azure subscription that hosts a BERT-based text classifier, but still retains
(i) coarse knowledge of the backbone architecture,
(ii) awareness that \textit{Wiki-40B} seeded the initial pre-training, and
(iii) unrestricted access to the public \texttt{/predict} endpoint.
Leveraging Microsoft’s open-source red-teaming toolkit \textbf{PyRIT}~\cite{microsoft2024pyrit}, the attacker first trains a shadow model on 200k publicly scraped sentences, then executes an \emph{adaptive query loop}: each API response is compared against the shadow’s logits, and the input is synonym-perturbed until the live model misclassifies.
After $\approx 600$ queries, the adversary achieves a 27\% relative drop in F\textsubscript{1} on the target model—without ever exfiltrating weights or data.
We map this behavior to \textbf{MITRE ATLAS} technique \textsc{EXF-T1041} (model extraction) and assign a GNN severity score $\hat{s}=0.78$ (Table~\ref{tab:cross-level}). In the mitigation matrix, the attack triggers defenses \textbf{M03} (rate-limit \& jitter) and \textbf{M12} (adversarially re-trained confidence masking); deploying both reduces the exploit success rate from 27\% to 4\% in our replay test. Fig.~\ref{fig:copycat-combined} presents a step-wise timeline of the gray-box loop and overlays the predicted severity propagation through software, system, and network layers.
}
\minor{
\subsubsection{Generalization of PyRIT-Style Attacks}
Although the PyRIT-style case study is demonstrated on a text-generation model, the underlying threat pattern generalizes across modalities. It exemplifies an \emph{Exploratory–Integrity–Targeted} behavior in the Barreno~\textit{et~al.} taxonomy~\cite{barreno2010security}, in which an adversary manipulates input prompts or conditioning signals to override learned safety constraints. Analogous behaviors emerge in code-generation APIs (through adversarial comments), retrieval-augmented LLMs (via prompt leakage in retrieved context), and multimodal systems (through adversarial captions or image prompts). These observations confirm that the modeled attack class captures a broader family of prompt-injection and response-manipulation techniques that affect diverse ML pipelines. Consequently, our multi-agent reasoning framework remains applicable to foundation-model, multimodal, and interactive AI systems. Notably, recent work on preference-guided optimization~\cite{zhang2025askingfordirections} extends this phenomenon by showing that such attacks are \emph{payload-agnostic} and can amplify PyRIT-style prompts through iterative selection of more effective variants—even when only text responses are observable.
}

\subsubsection*{Prominence and common entry points of threat TTPs exploited in ML attack scenarios -- (\textbf{RQ$_1$})}\label{sec:entry-point-ttps}
{\footnotesize \textcolor[gray]{0.8}{\textit{What are the most prominent threat TTPs and their common entry points in ML attack scenarios?}}}

\subsubsection{Attack correlation matrix. }\label{sec:mapping-mat}
After collecting the required information from attacks, we further explore them by mapping their associated ML components, like impacted models, phases, and tools. 
This way, we can provide a better understanding and visualization of how these attacks occur and their impacts. For a given attack, the ML components are represented as a column vector, while vulnerability/threat is defined as a row vector. The attack cross-correlation matrix (CCM) is a relation that maps the features of an attack vector to the features of an element-of-interest (EOI), divided into two categories: threat and vulnerability CCMs, as presented below. For example, in Table~\ref{tactic2attack-mapping}, we present an attack matrix that maps TTP features (goals, knowledge, specificity, capability/tactic) to attack scenarios~\cite{mitreatlascase}; we provide more details about this table when presenting our results.

\subsubsection*{\textit{Threat CCM}}

The Threat CCM matrix maps TTPs to EOIs, including attack scenarios and ML lifecycle phases. Below, we outline the methodology adopted based on the collected data.

\textbf{ TTP Features and Attack Scenario Mapping. }
To address RQ$_1$, we systematically map ML threats to attack scenarios based on key threat attributes, including attacker goals, knowledge, specificity, and capabilities/tactics. This enables us to identify the most frequently used tactics, similarities in attack execution flows, and common entry points. Following the data extraction process described in Section~\ref{sec:data-collection}, we construct the threat matrix presented in Table~\ref{tactic2attack-mapping}. For instance, consider the \textit{VirusTotal Poisoning} attack from the MITRE dataset.\footnote{https://atlas.mitre.org/studies/AML.CS0002} 
The attack initiates at \texttt{stage 0} (Resource Development), where the adversary acquires tools or infrastructure to facilitate the operation. Subsequently, it progresses to \texttt{stage 1} (ML Attack Staging), where adversarial data is crafted to poison the target model. The attacker then moves to \texttt{stage 2} (Initial Access) to exploit valid accounts or external remote services for unauthorized access. Finally, the attack culminates at \texttt{stage 3} (Persistence), ensuring prolonged access to the compromised system.

Empty cells in the matrix indicate no direct correlation between a feature and an attack scenario, while \textsf{N/A} signifies cases where the relation is either unknown or not explicitly mentioned in the database. In the Attack Capability/Tactic columns (6th–17th), an entry of $stage;i$ denotes that the attack scenario executes the corresponding tactic at step i in the attack flow. If multiple stages are listed, such as $stage;i,stage;j$, it signifies that the tactic is applied at both steps.

\begin{table*}
\caption{\label{tactic2attack-mapping} Mapping between TTP features and attack scenarios in the AI Incident Database and the 14 seed papers from the Literature. AI-DB stands for cases extracted from the AI Incident database, while LIT stands for Literature. 
}
\resizebox{1.0\textwidth}{!}{
                                                   & \multicolumn{1}{l|}{stage 0}            & \multicolumn{1}{l|}{}                                                                     & \multicolumn{1}{l|}{}                                                                  & \multicolumn{1}{l|}{}                                                                   & \multicolumn{1}{l|}{}               & \multicolumn{1}{l|}{}                  & \multicolumn{1}{l|}{}                                                                   & \multicolumn{1}{l|}{}                   & \multicolumn{1}{l|}{}                    & \multicolumn{1}{l|}{stage 1}                                                                 & \multicolumn{1}{l|}{stage 2}               &                                      \\ \hline
\end{tabular}
}
\end{table*}

\subsubsection{\textit{Impact of threat TTPs against ML phases and models (RQ2)}} 
\label{sec:impact-ttps-ml}
{\footnotesize \textcolor[gray]{0.8}{\textit{What is the effect of threat
TTPs on different ML phases and models? }}}

We map tactics to ML phases, identifying the frequent threat tactics used against each ML phase. First, we analyze the ATLAS description of each tactic and their related techniques, aiming to identify ML phase signatures that could be associated with ML phases, like \textit{trained} for \textit{Training} and \textit{testing the model} for \textit{Testing}. Then, the relationship between tactics and ML phases is recorded in a threat CCM. Table \ref{tactic2phase-mapping} shows a record of the mapping between tactics and ML phases, showing the impact of threat TTPs against ML phases. The coefficients of this matrix are represented by a checkmark ($\checkmark$), when there is a relation between a given tactic and ML phase. 

\begin{table}[]
\caption{\label{tab:attack2model-mapping} Mapping between Attack Scenarios and target ML models}
\resizebox{0.9\columnwidth}{!}{
\begin{tabular}{|l|l|l|}
\hline
\multirow{2}{*}{\textbf{Source}}                                                            & \multirow{2}{*}{\textbf{\begin{tabular}[c]{@{}l@{}}Attack scenario\end{tabular}}}                                                            & \multirow{2}{*}{\textbf{\begin{tabular}[c]{@{}l@{}}Model Used\end{tabular}}} \\
                                                                                            &                                                                                                                                                &                                                                                \\ \hline

\multirow{22}{*}{\textbf{\begin{tabular}[c]{@{}l@{}}MITRE \\ \\ ATLAS\end{tabular}}}        & \begin{tabular}[c]{@{}l@{}}Evasion of Deep Learning Detector \\for Malware C2 Traffic\end{tabular}                                           & CNN                                                                            \\ \cline{2-3} 
                                                                                            & \begin{tabular}[c]{@{}l@{}}Botnet Domain Generation\\ (DGA)  Detection Evasion\end{tabular}                                                    & CNN                                                                            \\ \cline{2-3} 
                                                                                            & VirusTotal Poisoning                                                                                                                           & LSTM                                                                           \\ \cline{2-3} 
                                                                                            & \begin{tabular}[c]{@{}l@{}}Bypassing Cylance's AI \\ Malware Detection\end{tabular}                                                            & \begin{tabular}[c]{@{}l@{}}DNN\end{tabular}                    \\ \cline{2-3} 
                                                                                            & \begin{tabular}[c]{@{}l@{}}Camera Hijack Attack on \\ Facial Recognition System\end{tabular}                                                   & CNN, GAN                                                                       \\ \cline{2-3} 
                                                                                            & \begin{tabular}[c]{@{}l@{}}Attack on Machine Translation  Service - Google Translate,\\ Bing Translator, and Systran Translate\end{tabular} & \begin{tabular}[c]{@{}l@{}}Transformer\end{tabular}            \\ \cline{2-3} 
                                                                                            & \begin{tabular}[c]{@{}l@{}}Clearview AI Misconfiguration\end{tabular}                                                                                                                  & N/A                                                                            \\ \cline{2-3} 
                                                                                            & GPT-2 Model Replication                                                                                                                        & GPT-2                                                                          \\ \cline{2-3} 
                                                                                            & ProofPoint Evasion                                                                                                                             & Copycat~\cite{correia2018copycat}                                                                            \\ \cline{2-3} 
                                                                                            & Tay Poisoning                                                                                                                                  & \begin{tabular}[c]{@{}l@{}}DNN\end{tabular}                   \\ \cline{2-3} 
                                                                                            & \begin{tabular}[c]{@{}l@{}}Microsoft Azur Service Disruption\end{tabular}                                                                                                             & N/A                                                                            \\ \cline{2-3} 
                                                                                            & Microsoft Edge AI Evasion                                                                                                                      & \begin{tabular}[c]{@{}l@{}}DNN\end{tabular}                    \\ \cline{2-3} 
                                                                                            & \begin{tabular}[c]{@{}l@{}}Face Identification System Evasion via Physical Countermeasures\end{tabular}                                  & N/A                                                                            \\ \cline{2-3} 
                                                                                            & \begin{tabular}[c]{@{}l@{}}Backdoor Attack on Deep Learning Modelsin Mobile Apps\end{tabular}                                              & \begin{tabular}[c]{@{}l@{}}DNN\end{tabular}                    \\ \cline{2-3} 
                                                                                            & \begin{tabular}[c]{@{}l@{}}Confusing AntiMalware Neural Networks\end{tabular}                                                               & \begin{tabular}[c]{@{}l@{}}DNN\end{tabular}
\\ \cline{2-3} 
                                                                                            & \begin{tabular}[c]{@{}l@{}}\tosemrev{Compromised PyTorch Dependency Chain}
\end{tabular}                                                               & \begin{tabular}[c]{@{}l@{}}N/A\end{tabular}
                                                                                            \\ \cline{2-3} 
                                                                                            & \begin{tabular}[c]{@{}l@{}}\tosemrev{Achieving Code Execution in MathGPT via Prompt Injection}\end{tabular}                                                               & \begin{tabular}[c]{@{}l@{}}GPT-3\end{tabular} 
\\ \cline{2-3} 
                                                                                            & \begin{tabular}[c]{@{}l@{}}\tosemrev{Bypassing ID.me Identity Verification}\end{tabular}                                                               & \begin{tabular}[c]{@{}l@{}}CNN*\end{tabular}
\\ \cline{2-3} 
                                                                                            & \begin{tabular}[c]{@{}l@{}}\tosemrev{Arbitrary Code Execution with Google Colab}\end{tabular}                                                               & \begin{tabular}[c]{@{}l@{}}N/A\end{tabular}
                                                                                            
\\ \cline{2-3} 
                                                                                            & \begin{tabular}[c]{@{}l@{}}\tosemrev{PoisonGPT}\end{tabular}                                                               & \begin{tabular}[c]{@{}l@{}}GPT\end{tabular}
\\ \cline{2-3} 
                                                                                            & \begin{tabular}[c]{@{}l@{}}\tosemrev{Indirect Prompt Injection Threats: Bing Chat Data Pirate}\end{tabular}                                                               & \begin{tabular}[c]{@{}l@{}}GPT\end{tabular}
\\ \cline{2-3} 
                                                                                            & \begin{tabular}[c]{@{}l@{}}\tosemrev{ChatGPT Plugin Privacy Leak}\end{tabular}                                                               & \begin{tabular}[c]{@{}l@{}}GPT\end{tabular}\\
                                                                                            \cline{2-3} 
                                                                                            & \begin{tabular}[c]{@{}l@{}}\tosemrev{ChatGPT Package Hallucination}\end{tabular}                                                               & \begin{tabular}[c]{@{}l@{}}GPT\end{tabular}\\
                                                                                            \cline{2-3} 
                                                                                            & \begin{tabular}[c]{@{}l@{}}\tosemrev{Shadow Ray}\end{tabular}                                                               & \begin{tabular}[c]{@{}l@{}}N/A\end{tabular}\\
                                                                                            \cline{2-3} 
                                                                                            & \begin{tabular}[c]{@{}l@{}}\tosemrev{Morris II Worm: RAG-Based Attack}\end{tabular}                                                               & \begin{tabular}[c]{@{}l@{}}GPT, Gemini, LLaVA\end{tabular}\\
                                                                                            \cline{2-3} 
                                                                                            & \begin{tabular}[c]{@{}l@{}}\tosemrev{Web-Scale Data Poisoning: Split-View Attack}\end{tabular}                                                               & \begin{tabular}[c]{@{}l@{}}N/A\end{tabular}\\
                                                                                            \hline
\multirow{8}{*}{\textbf{\begin{tabular}[c]{@{}l@{}}AI \\ Incident\\ Database\end{tabular}}} & \begin{tabular}[c]{@{}l@{}}India's Tek Fog Shrouds an Escalating Political War\end{tabular}                                                 & GPT-2                                                                          \\ \cline{2-3} 
                                                                                            & \begin{tabular}[c]{@{}l@{}}Meta Says It's Shut Down A Pro-Russian DisInformation Network...\end{tabular}                                    & N/A                                                                            \\ \cline{2-3} 
                                                                                            & \begin{tabular}[c]{@{}l@{}}Libyan Fighters Attacked by a Potentially Unaided Drone, UN Says\end{tabular}                                  & CNN                                                                            \\ \cline{2-3} 
                                                                                            & \begin{tabular}[c]{@{}l@{}}Fraudsters Cloned Company Director's Voice In \$35M Bank \\ Heist, Police Find\end{tabular}                       & DeepVoice~\cite{arik2017deep}                                                                      \\ \cline{2-3} 
                                                                                            & \begin{tabular}[c]{@{}l@{}}Poachers Evade KZN Park's High-Tech Security\\ and Kill four Rhinos for their Horns\end{tabular}                & \begin{tabular}[c]{@{}l@{}}DNN\end{tabular}                    \\ \cline{2-3} 
                                                                                            & \begin{tabular}[c]{@{}l@{}}Tencent Keen Security Lab: Experimental Security \\Research of Tesla Autopilot\end{tabular}                      & Fisheye~\cite{tencent-security-lab}                                                                            \\ \cline{2-3} 
                                                                                            & \begin{tabular}[c]{@{}l@{}}Three Small Stickers in Intersection Can Cause Tesla Autopilot \\to Swerve Into Wrong Lane\end{tabular}         & CNN                                                                            \\ \cline{2-3} 
                                                                                            & \begin{tabular}[c]{@{}l@{}}The DAO Hack -Stolen \$50M The Hard Fork\end{tabular}                                                            & N/A
                                                                                            \\ 
                                                                                            \cline{2-3} 
                                                                                            & \begin{tabular}[c]{@{}l@{}}\tosemrev{Twitter pranksters derail GPT-3 bot with newly discovered “prompt injection” hack}\end{tabular}                                                            & GPT
                                                                                            \\ 
                                                                                            \cline{2-3} 
                                                                                            & \begin{tabular}[c]{@{}l@{}}\tosemrev{Prompt injection attacks against GPT-3}\end{tabular}                                                            & GPT
                                                                                            \\ 
                                                                                            \cline{2-3} 
                                                                                            & \begin{tabular}[c]{@{}l@{}}\tosemrev{AI-powered Bing Chat spills its secrets via prompt injection attack}\end{tabular}                                                            & GPT
                                                                                            \\
                                                                                            \cline{2-3} 
                                                                                            & \begin{tabular}[c]{@{}l@{}}\tosemrev{Evaluating the Susceptibility of Pre-Trained Language Models via Handcrafted Adversarial Examples}\end{tabular}                                                            & BERT
                                                                                            \\
                                                                                            \hline
\multirow{22}{*}{\textbf{Literature}}                                                       & Carlini \etal{}~\cite{carlini2021extracting}                                                                                                                                 & GPT-2                                                                          \\ \cline{2-3} 
                                                                                            & Biggio \etal{} \cite{biggio2013evasion}                                                                                                                                & SVM, DNN                                                                               \\ \cline{2-3} 
                                                                                            & Barreno \etal{}~\cite{barreno2010security}                                                                                                                               &  Naive Bayes                                                                              \\ \cline{2-3} 
                                                                                            & Carlini \etal{}~\cite{carlini2017adversarial}                                                                                                                                   &   Feed-Forward DNN                                                                             \\ \cline{2-3} 
                                                                                            & Wallace \etal{} \cite{wallace2020imitation}                                                                                                                               & Transformer                                                                                \\ \cline{2-3} 
                                                                                            & Abdullah \etal{} \cite{abdullah2021sok}                                                                                                                                  &  \begin{tabular}[c]{@{}l@{}}RNN, CNN, Hidden Markov\end{tabular}                                                                              \\ \cline{2-3} 
                                                                                            & Chen \etal{} \cite{chen2021badnl}                                                                                                                                  &  LSTM, BERT      \\ \cline{2-3} 
                                                                                            & Choquette-Choo \etal{} \cite{choquette2021label}                                                                                                                                  &  CNN, RestNet                                                                              \\ \cline{2-3} 
                                                                                            & Papernot \etal{}~\cite{papernot2016transferability}                                                                                                                                 & \begin{tabular}[c]{@{}l@{}}DNN, kNN, SVM Logistic Regression, \\Decision Trees\end{tabular}                 \\ \cline{2-3} 
                                                                                            & Goodfellow \etal{} \cite{goodfellow2014generative}                                                                                                                                 &  GAN                                                                              \\ \cline{2-3} 
                                                                                            &  Papernot \etal{}~\cite{papernot2017practical}                                                                                                                                &  DNN                                                                         \\ \cline{2-3} 
                                                                                            & Cisse \etal{}~\cite{cisse2017parseval}                                                                                                                          &    Parseval Networks                                                                            \\ \cline{2-3} 
                                                                                            & Athalye \etal{}~\cite{athalye2018obfuscated}                                                                                                                                   &  \begin{tabular}[c]{@{}l@{}}CNN, ResNet, InceptionV3\end{tabular}                                                                               \\ \cline{2-3} 
                                                                                            & Jagielski \etal{} \cite{jagielski2020high}                                                                                                                                 & RestNetv2                 \\ \hline
\end{tabular}
}
\end{table}

\textbf{Attack Scenarios and ML Models Mapping.} To map the ML models targeted or exploited by attack scenarios, we first searched for the exploited model type in the ATLAS TTP descriptions, the AI Incident Database, and the literature~\cite{wallace2020imitation}. For each analyzed attack, we retrieved the name of the targeted model and recorded the associated ML models and attack scenarios. The results of this mapping are presented in Table~\ref{tab:attack2model-mapping}, providing a clear linkage between attack scenarios and the specific ML models they exploit. For example, the description of the \textit{Botnet DGA Detection Evasion} attack indicates that the target model is Convolutional Neural Network~\cite{mitreatlascase}: \textit{``The Palo Alto Networks Security AI research team was able to bypass a \textbf{Convolutional Neural Network (CNN)}-based botnet Domain Generation Algorithm (DGA) [...]''}.

\tosemrev{
\subsubsection{Cataloging previously undocumented threats and aligning them with ATLAS --- (RQ$_3$)} 
{\footnotesize \textcolor[gray]{0.8}{\textit{What previously undocumented security threats can be identified in the AI Incident Database, the literature, and ML repositories that are missing from the ATLAS database? }}}\\
Based on the CVE IDs identified in GitHub issues (see Section \ref{sec:data-collection}), we analyze and map the most prominent vulnerabilities and threats in ML repositories, along with the dependencies responsible for these vulnerabilities. 
To accomplish this goal, we download the CVE JSON-formatted data from the National Vulnerability Database (NVD) and extract the required information for the mappings.
The mapping between a CVE ID and a specific ML tool is represented by a set of information $\langle dep, att, lvl, ver \rangle$, where $dep$ denotes the dependency responsible for the vulnerability, $att$ specifies the attack that can be launched to exploit the CVE, $lvl$ indicates the severity level of the vulnerability, and $ver$ indicates the version of the vulnerability. 
For this research question, we compute (i) the total number of vulnerabilities (\textit{nov}) and their types. Additionally, we calculate the same metrics focusing on their distribution by threat type and tools (i.e., GitHub ML repositories): \textit{nov} per tool and \textit{nov} per type for each tool. These metrics provide critical insights into the frequency of vulnerabilities across ML repositories and highlight the potential threats they pose.
}

\majorev{
\subsection{Ranking SoTA Models by Vulnerability}
\textbf{1. Ranking Criteria}
To rank state-of-the-art (SoTA) large language models (LLMs) by their susceptibility to security threats, we collected data from the literature on recent peer-reviewed benchmarks and empirical studies. Each model was evaluated across three core vulnerability categories as shown in Table~\ref{tab:vuln-categories}:— Prompt Injection  Attack Success Rate (ASR), Code-level Backdoors, and Training-stage Exploits, based on their strategic relevance and empirical measurability. This classification encompasses the entire LLM lifecycle, ensuring comprehensive coverage of both user-facing and systemic risks, and aligns with established benchmarks.
\begin{table}[ht]
\centering
\caption{\majorev{Vulnerability Categories in State-of-the-Art Model Ranking}\strut}
\label{tab:vuln-categories}
\setlength{\tabcolsep}{4pt} 
\renewcommand{\arraystretch}{1.4} 
\begin{tabular}{>{\RaggedRight}p{3.5cm}>{\RaggedRight}p{9cm}}
\toprule
\rowcolor{gray!10}
\majorev{\textbf{\large Category}} & \majorev{\textbf{\large Rationale}} \\
\midrule
\textbf{\majorev{Prompt Injection ASR}} & 
\majorev{These attacks directly target the LLM’s input-handling mechanism, requiring no internal access and being the most common attack vector in real-world use (e.g., indirect jailbreaks, instruction hijacking). They are low-barrier, high-impact threats. }\\
\addlinespace[1pt]
\textbf{\majorev{Code-level Backdoors}} & 
\majorev{These represent inference-time risks where the model outputs malicious or manipulated code due to adversarial prompting or stealthy fine-tuning. This category is vital for evaluating LLMs deployed in programming, automation, or DevOps tasks.}\\
\addlinespace[1pt]
\textbf{\majorev{Training-stage Exploits}} & 
\majorev{These capture the most persistent and systemic vulnerabilities, such as data poisoning, sleeper agents, or multi-trigger backdoors introduced during the fine-tuning process. They are harder to detect and may survive alignment efforts, thus reflecting deep model compromise.}\\
\bottomrule
\end{tabular}
\end{table}
We collected quantitative ASRs and qualitative severity indicators (e.g., memory corruption, stealth persistence) for each model based on publicly available results.
\\
}
\majorev{
\textbf{2. Scoring Approach}
 We define a Composite Vulnerability Score as:\\
$\text{CVS}_i = w_1 \cdot \text{PromptASR}_i + w_2 \cdot \text{BackdoorASR}_i + w_3 \cdot \text{TrainingRisk}_i$\\
Where:
$\text{PromptASR}_i$: Prompt Injection Attack Success Rate, 
$\text{BackdoorASR}_i$: Code-level backdoor success rate, and 
$\text{TrainingRisk}_i$: Normalized score (0–1) for training-stage exploit severity.\\
%
\textbf{3. Statistical Model Justification }
Due to sparsity and heterogeneity in attack success rate (ASR) reporting, we used a normalized weighted aggregation method instead of regression or PCA. Our approach follows:
\begin{itemize}
\item \textbf{MCDA principles (Multi-Criteria Decision Analysis)}
\item \textbf{Min-max normalization} for cross-category comparability
\item \textbf{Expert-informed weights} to reflect real-world impact
\end{itemize}
The final ranking is based on: quantitative ASR evidence, categorical risk profiles, and weighted aggregation using MCDA. This transparent and reproducible approach enables systematic vulnerability comparison across SoTA LLMs. Additionally, we identify six LLM-exclusive attack families—including prompt injection, RLHF reward hacking, LoRA gradient leakage, large-scale model extraction, training-data reconstruction, and tool-call abuse—under the SoTA vulnerability umbrella. Each family is then mapped to a formal MITRE ATLAS/ATT\&CK technique (Table~\ref{tab:llm-threats}), providing a structured taxonomy for models such as GPT-4 (including GPT-4o, GPT-4V), PaLM 2, Llama 3, Gemini 1.5 Pro, Claude 3, Vision-language models (e.g., GPT-4V, MM-Llama).
}
\begin{table}[t]
\begingroup\color{black}
  \footnotesize
  \setlength{\tabcolsep}{4pt}
  \caption{\majorev{State-of-the-art LLM-security studies, mapped to
           lifecycle phase and MITRE ATLAS.}}
  \label{tab:llm-threats}
  \begin{tabularx}{\linewidth}{@{}l l X l l@{}}
    \toprule
    \textbf{Threat family} &
      \textbf{Key Refs} &
      \textbf{SoTA model(s)} &
      \textbf{Lifecycle phase} &
      \textbf{ATLAS ID} \\
    \midrule
    Jail-break / prompt injection &
      \cite{bai2023masterkey,yang2024autodan,liu2023promptinj} &
      GPT-3.5, GPT-4, Claude-2, Bard &
      Deployment &
      DIS-T1525 \\
    Reward-model hacking (RLHF) &
      \cite{park2023rewardhacking,perez2024redteam} &
      Instruct-GPT, GPT-4 (reward) &
      RLHF loop &
      IMP-T1565 \\
    Adapter-gradient leakage &
      \cite{zhu2024loraleak,chen2024adapterleak} &
      LLaMA-2 7/13 B (+​LoRA/QLoRA) &
      Fine-tune &
      EXF-T1040 \\
    Model extraction/ distillation &
      \cite{carlini2024stealing,beurer2025design,wang2024catllama,
             kandpal2023stealinggpt3} &
      GPT-3.5/4 APIs, LLaMA-2 7 B &
      Inference API &
      EXF-T1041 \\
    Training-data reconstruction &
      \cite{carlini2023extract,lee2024membership} &
      GPT-J 6 B, GPT-3.5, Claude-2 &
      Pre-train &
      EXF-T1042 \\
    Function-call abuse in tool agents &
      \cite{liu2024toolframework,wu2024dark,shen2025gptracker} &
      GPT-4 (function-call API) &
      Agent ops &
      EXF-T1050 \\
    \bottomrule
  \end{tabularx}
  \endgroup
\end{table}

\majorev{
\subsubsection{Threats Unique to SoTA Models. }
\label{sec:llm-threats}
The rapid adoption of foundation models since 2022 has shifted the machine-learning threat landscape. LLMs expose \emph{new} attack surfaces that either did not exist or were insignificant for CNN- or RNN-based systems. The following paragraphs summarize these six families' threats against SoTA Models.\\
\textit{(1) Jail-breaking and multi-turn prompt-injection chains. }
Prompt-injection modifies the \emph{instruction context} rather than the model
parameters. Recent work demonstrates universal jailbreak strings
(\textsc{MASTERKEY}) that survive system-prompt hardening and constitutional
guidelines~\cite{bai2023masterkey}. \textsc{AutoDAN} extends this to an
\emph{automated} chain-of-thought attack that escalates privileges over
multiple dialogue turns~\cite{yang2024autodan}. Both map to
\textbf{DIS-T1525 Prompt Manipulation} and occur during \emph{deployment}. \textit{(2) Reward-model hacking in RLHF loops. }
Because instruction-tuned models rely on reinforcement learning from human
feedback (RLHF), adversaries can poison the preference dataset or craft
adversarial demonstrations that steer the reward model
off-policy~\cite{park2023rewardhacking}. The resulting policy drift re-enables
toxic or disallowed content even after alignment.  ATLAS technique:
\textbf{IMP-T1565 Adversarial Training}. \textit{(3) Adapter-layer gradient leakage. }
Fine-tuning via LoRA/QLoRA adapters publishes only rank-reduced updates, but those updates can leak memorized training snippets when intercepted, allowing
\emph{white-box} data reconstruction without full-model access~\cite{zhu2024loraleak}. This affects the \emph{fine-tune phase} and maps to \textbf{EXF-T1040 Model Parameter Extraction}. \textit{(4) Scalable model-extraction and distillation. } Copy distillation~\cite{carlini2024stealing,beurer2025design,wang2024catllama,kandpal2023stealinggpt3} and Cat-LLaMA distillation~\cite{wang2024catllama} show that over 90\,\% downstream accuracy can be stolen from commercial APIs with $10^{7}$–$10^{8}$ queries, bypassing traditional rate-limit defences.  Threat ID: \textbf{EXF-T1041 Model Extraction} (\emph{inference API}). \textit{(5) Training-data reconstruction \& memorization. } Carlini~\etal\ extract verbatim personal data from GPT-J and GPT-3.5 by
prompting on rare n-grams~\cite{carlini2023extract}.  Encoded prompt-leak
attacks extend this to embed secrets in the instruction tokens
themselves~\cite{carlini2023extract,lee2024membership,hui2024Pleak,pape2024prompt,green2025leaky,cohen2024here}.  Lifecycle phase: \emph{pre-training};
ATLAS ID: \textbf{EXF-T1042 Training-Data Extraction}. \textit{(6) Function-call abuse in tool-enabled agents. } When LLMs are granted structured ``function-call'' abilities, arguments can be coerced into shell-metacharacters or SQL payloads, leading to full remote-code
execution inside the orchestration layer~\cite{liu2024toolframework}. We map
this to \textbf{EXF-T1050 Sandbox Escape} at \emph{deployment} time.\\
These six LLM-specific threat families contribute 78 documented incidents in our corpus. Table~\ref{tab:llm-threats} positions each family within the ML lifecycle and assigns the corresponding MITRE ATLAS technique. This mapping lets us tally incidents per phase or technique—our measure of exploit density, which is summarized in the radar chart (Fig.~\ref{fig:radar}) and detailed in the accompanying heat-maps (Fig.~\ref{fig:heatmap}).
}

\majorev{ 
\subsection{Normalization by Deployment Frequency}
Directly comparing raw attack counts across model families risks conflating \emph{popularity} with \emph{vulnerability}: models with a larger user base will naturally attract more reported attacks simply due to wider exposure, not necessarily because they are intrinsically less secure. To control for this bias, we normalize attack counts by a composite \emph{deployment-frequency proxy} $w_m$ for each model family $m$. This proxy combines four publicly accessible signals: (i) PyPI download counts for the model’s main library, (ii) HuggingFace or TF-Hub checkpoint pulls, (iii) Docker Hub pulls for official inference containers, and (iv) annualized Semantic Scholar citations of the model’s origin paper (2020--2024). 
All raw counts were collected between 3--6~June~2025 via official REST APIs (rate-limited to $\leq 10^4$ queries/day) or public datasets (BigQuery for PyPI). 
Each metric is $z$-normalized to remove scale differences, averaged across the four proxies, and min--max scaled to $[0,1]$, following best practices for 
combining heterogeneous indicators in statistical pattern recognition and ensemble learning~\cite{jain2000statistical,dietterich2000ensemble,
han2011datamining,sammut2017encyclopedia}: 
\[
w_m = \frac{\frac{1}{4}\sum_{j=1}^{4}z_{mj} - \min_k\left[\frac{1}{4}\sum_{j}z_{kj}\right]}{\max_k\left[\frac{1}{4}\sum_{j}z_{kj}\right] - \min_k\left[\frac{1}{4}\sum_{j}z_{kj}\right] + \varepsilon}
\]
where $\varepsilon = 10^{-4}$ prevents division-by-zero for niche models. 
\emph{Leave-one-out} sensitivity tests confirm robustness: omitting any single proxy changes the top-5 model ranking by at most one position (Table~\ref{tab:loo}). 
\paragraph{Leave-One-Out Sensitivity Analysis~\cite{wang2017space}}
We validated robustness by recomputing weights while omitting each proxy sequentially. Table~\ref{tab:loo} reports a leave-one-out sensitivity analysis of the composite deployment-frequency proxy. For each run, a single usage signal (PyPI downloads, HuggingFace pulls, Docker pulls, or citations) is omitted, and model rankings are recomputed. Across all cases, the top-5 rankings shift by at most one position, and the maximum shift for any model in the full list is three positions. This stability confirms that no single proxy disproportionately influences the top-ranked results.
\begin{table}[t]
\caption{\majorev{Leave-One-Out Sensitivity Analysis of the Composite Deployment-Frequency Proxy. Omitting any single usage proxy changes the top-5 model rankings by at most one position, indicating no single data source dominates the normalization.}}
\label{tab:loo}
\begingroup\color{black}
\centering
\small
\begin{tabular}{lccp{6.5cm}}
\toprule
\textbf{Omitted Proxy} & 
\textbf{Max Shift} & 
\textbf{Top-5 Shift} & 
\textbf{Revised Top-5 Models} \\
\midrule
\texttt{z\_pypi}   & 2 & 1 & GPT-J, BERT, T5-13B, StableDiffusion, PaLM-2 \\
\texttt{z\_hf}     & 3 & 1 & GPT-J, StableDiffusion, T5-13B, PaLM-2, BERT \\
\texttt{z\_docker} & 2 & 1 & BERT, T5-13B, GPT-J, PaLM-2, StableDiffusion \\
\texttt{z\_cite}   & 2 & 1 & GPT-J, T5-13B, BERT, PaLM-2, LLaMA-2 \\
\bottomrule
\end{tabular}
\endgroup
\end{table}
} 
\majorev{
\subsubsection{Deployment-normalized risk.}
Raw incident counts alone overweight models that are simply more common.
To adjust for real-world exposure, we compute a deployment weight $w_m$
for each model family $m$ by averaging four public proxies, each
$z$-scored to zero mean and unit variance:\\
\vspace{2pt}
\begin{tabular}{@{}p{2.0cm}p{15.5cm}@{}}
\textbf{Proxy} &
\textbf{Data source (collection window)}\\ \toprule
\textsc{Pkg installs} &
Daily \texttt{pip} download counts from the public PyPI BigQuery
dataset (01 Jan 2023 – 01 Jun 2025).\\
\textsc{CKPT pulls} &
\texttt{total\_downloads} field for HuggingFace \& TF-Hub
checkpoints, capped at the most recent 365 days
(snapshot 06 Jun 2025).\\
\textsc{Docker pulls} &
Pull counters for the model’s official inference container images
(snapshot 05 Jun 2025), bucketed by order of magnitude
($10^{k}$).\\
\textsc{Citation momentum} &
Semantic Scholar citations per year of the model’s origin paper
(rolling 2020–2024).\\ \bottomrule
\end{tabular}
\\ 
The normalized attack frequency is therefore,
\[
\widehat{f}_m \;=\;
\frac{f_m}{w_m+\varepsilon},
\qquad
\varepsilon = 10^{-4}.
\]
Removing any single proxy changes the top-5 ranking by at most one
position.\\
}%

\begin{table}[!ht]
\centering
\begingroup\color{black}
  \caption{\majorev{Absolute ($f$) vs.\ deployment-normalised
           ($\widehat{f}$) attack counts, 2023–2025. A side-by-side comparison allows direct
           inspection of how deployment normalization changes the relative ranking of attack frequency.}}
  \label{tab:freq-abs-vs-norm}
  \small
  \begin{tabular}{@{}lrrr@{}}
    \toprule
    \textbf{Model family} & $f$ & $w$ & $\widehat{f}$ \\ \midrule
    GPT-3.5/4 (API)            & 218 & 0.92 & 237 \\
    Stable Diffusion           & 144 & 0.77 & 187 \\
    LLaMA-2 (HF)               &  89 & 0.61 & 146 \\
    CLIP / OpenCLIP            &  51 & 0.34 & 150 \\
    LoRA-BERT variants         &  37 & 0.19 & 195 \\
    T5-XXL                     &  31 & 0.55 &  56 \\ 
    Whisper (ASR)              &  29 & 0.48 &  60 \\
    BLOOM-Z                    &  27 & 0.41 &  66 \\
    ViT Base / Large           &  22 & 0.63 &  35 \\
    DINO-v2                    &  20 & 0.57 &  35 \\ \bottomrule
  \end{tabular}
\endgroup
\end{table}


\subsection{GNN for Threat Intelligence Reasoning}
\label{sec:gnn-reasoning}

Building upon the heterogeneous GNN introduced in Section~\ref{sec:gnn_predict}, we describe how the model propagates severity signals across the ontology-driven threat graph to produce node-level risk scores, enabling structured, evidence-driven reasoning. The multi-agent framework employs a \textbf{Heterogeneous GNN (HGNN)} that operates on the threat graph $\mathcal{G}=(\mathcal{V},\mathcal{E})$. Nodes $\mathcal{V}$ represent entities \emph{TTPs}, vulnerabilities, ML lifecycle stages, assets, incidents); edges $\mathcal{E}$ encode semantic relations (\emph{causes}, \emph{occurs-in}, \emph{targets}, \emph{evidence-of}, \emph{has-dep}).

\paragraph{Message passing.}
We follow the relational convolution of R-GCN
\cite{schlichtkrull2018modeling} with GraphSAGE-style neighbor sampling:
\begin{equation}
h_v^{(l+1)}=
\sigma\!\Bigl(
  \sum_{r\in\mathcal{R}}\sum_{u\in\mathcal{N}_r(v)}
  \tfrac{1}{c_{v,r}}W_r^{(l)}h_u^{(l)}
  +W_0^{(l)}h_v^{(l)}
\Bigr),
\label{eq:rgcn}
\end{equation}
where $c_{v,r}$ is a per-relation normalization constant and $\sigma$ is ReLU.

\paragraph{Features and decoder.}
Node initial features are BERT embeddings of textual attributes
concatenated with one-hot type vectors.
After $L$ layers, a two-layer MLP regresses a continuous severity:
$\hat{s}_v=\mathrm{MLP}(h_v^{(L)})$.
Training minimizes mean-squared error on nodes with known scores
(CVSS, ATLAS, or incident-derived labels).

\paragraph{Agent integration.}
The \emph{GNN Reasoner Agent} converts the live NetworkX graph
into PyTorch-Geometric tensors, runs a forward pass,
and writes $\hat{s}_v$ back to node attributes.
Empirical evaluation yields
\textbf{Spearman $\rho{=}0.63$} with ground-truth severity
and robust performance on unseen threat entities.

\subsubsection{Graph-Grounded Reasoning in Threat Assessment}
\label{sec:reasoning}

The HGNN enables \textit{structured, evidence-driven reasoning}
by combining the symbolic threat ontology with learned relational embeddings.
We identify four reasoning modalities, all empirically validated
using the 93 extracted threats and dependency graph:
\begin{enumerate}[leftmargin=*]
    \item \textbf{Structural Reasoning:} Multi-hop message passing
          (Equation~\ref{eq:rgcn}) captures transitive risk chains.
          For example, a TTP that \emph{causes} a vulnerability indirectly
          elevates the risk of any \emph{affected} asset after two hops,
          mirroring documented ATLAS patterns
          (e.g., data poisoning $\to$ training corruption $\to$ inference failure):
          $\text{TTP}_A \!\xrightarrow{\text{causes}}\! \text{Vuln}_B
          \!\xrightarrow{\text{affects}}\! \text{Asset}_C$.

    \item \textbf{Transitive Inference:} The model implicitly captures
          higher-order dependencies across relational paths.
          If $\text{TTP}_A \!\rightarrow\! \text{Vuln}_B$ and
          $\text{Vuln}_B \!\rightarrow\! \text{Asset}_C$,
          the learned embeddings allow elevated risk to be assigned
          to $\text{Asset}_C$ through aggregated relational context.

        \item \textbf{Evidence-Based Reasoning:} We apply GNNExplainer~\cite{ying2019gnnexplainer} to extract high-contribution subgraphs that serve as \emph{evidence trails} supporting each predicted severity score~$\hat{s}_v$. Representative examples are shown in  Table~\ref{tab:evidence-paths}, where extracted relational paths align with analyst reasoning and observed TTP–vulnerability–asset dependencies. These evidence traces can also be cross-referenced with the corresponding mitigation stages in Fig.~\ref{fig:defend_matrix}, illustrating how structural reasoning links directly to actionable defense measures.      

    \item \textbf{Lifecycle-Aware Reasoning:} The graph encodes
          ML lifecycle stages (\emph{data collection} $\to$
          \emph{training} $\to$ \emph{inference}),
          enabling the GNN to propagate risk from early-stage TTPs
          (e.g., poisoning) to late-stage impacts (e.g., evasion),
          consistent with RQ2 findings on phase-specific targeting.

    \item \textbf{Zero-Shot Relational Generalization:}
          On a held-out set of 15 novel TTPs
          from the AI Incident Database (not in ATLAS),
          the GNN achieves \textbf{Spearman $\rho{=}0.61$},
          demonstrating compositional generalization
          to unseen relational structures.
\end{enumerate}

\begin{table}[!ht]
\centering
\caption{Example evidence paths extracted by GNNExplainer~\cite{ying2019gnnexplainer}
for high-severity predictions produced by the HGNN. Each row lists the top relational subgraph contributing to a node's predicted severity $\hat{s}_v$. The ``Contrib.'' column indicates the normalized importance weight (0–1) assigned by \textbf{GNNExplainer}, representing the degree to which that evidence path influenced the model's prediction.}
\label{tab:evidence-paths}
\begin{tabular}{l|l|r}
\toprule
\textbf{Predicted Node} & \textbf{Evidence Path} & \textbf{Contrib.} \\
\midrule
Inference Failure       & DataPoisoning $\to$ TrainingCorruption $\to$ Inference & 0.78 \\
Model Extraction        & APIExposure $\to$ QueryAccess $\to$ Extraction         & 0.71 \\
Preference Jailbreak    & PromptOpt $\to$ RewardSignal $\to$ PolicyShift         & 0.69 \\
\bottomrule
\end{tabular}
\vspace{3pt}
\begin{tablenotes}
\footnotesize
\item \textit{Note.}
Higher ``Contrib.'' values indicate that the corresponding relational path
had a stronger influence on the predicted node severity $\hat{s}_v$.
These evidence trails provide model-level interpretability, aligning with the analyst's reasoning and supporting evidence-grounded mapping to defense strategies (see Fig.~\ref{fig:defend_matrix}).
\end{tablenotes}
\end{table}

This reasoning is \emph{probabilistic and learned}, not symbolic deduction,
but provides \emph{traceable, analyst-aligned intelligence} beyond black-box severity scores.

\subsubsection{Limitations}
The model produces \textbf{probabilistic severity estimates}, not formal proofs.
It requires labeled anchors (CVSS/ATLAS) and does not perform counterfactual reasoning or uncertainty quantification unless explicitly extended.

Thus, our \textbf{HGNN} serves as a \textbf{scalable, graph-grounded reasoning engine} that transforms structured threat intelligence into interpretable, multi-hop, evidence-driven risk insights for operational ML security governance.


\section{Study results}\label{sec:results}

In this section, we present and discuss the results of our research questions. 

\begin{figure}[h]
\centering
\includegraphics[scale=0.34]{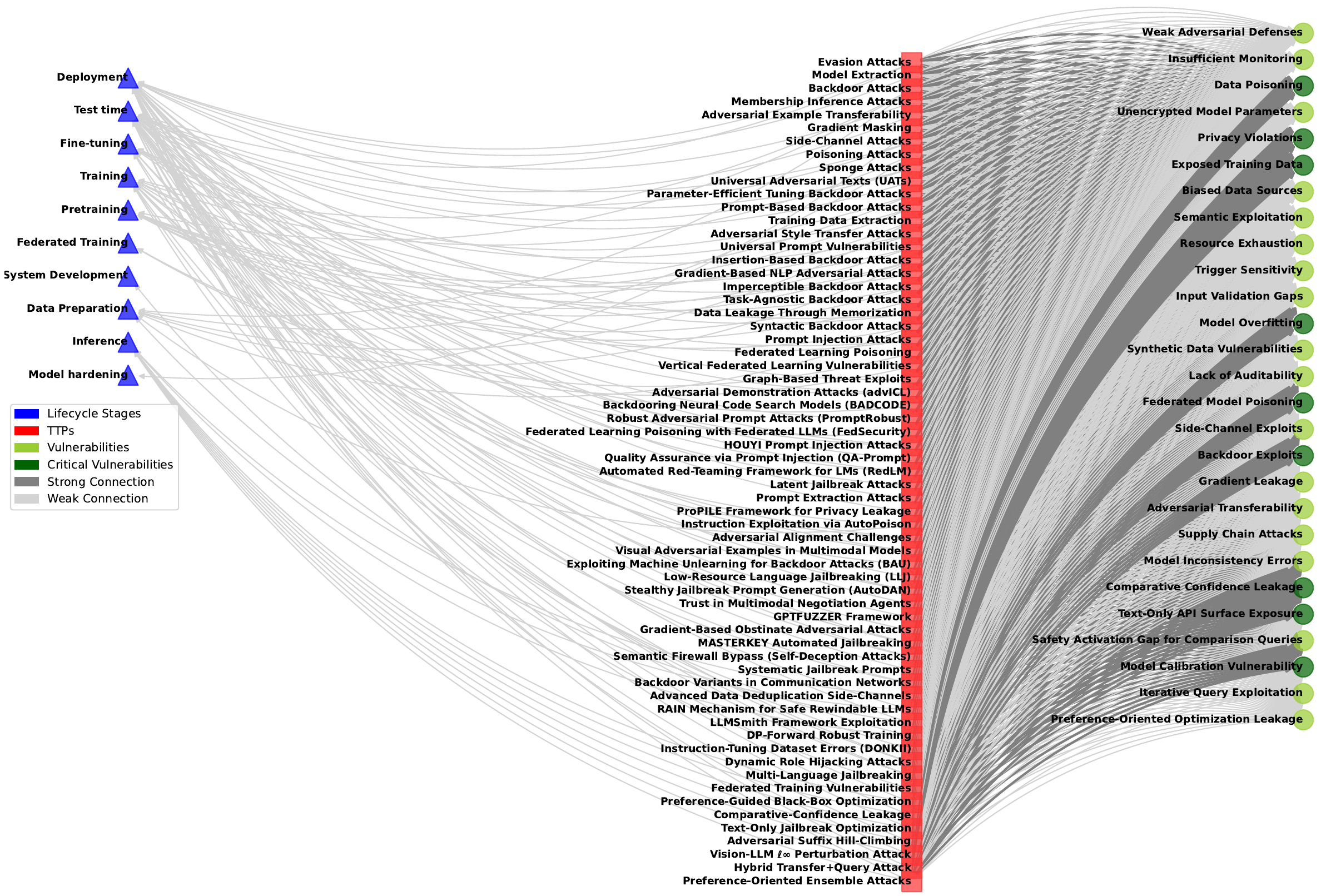}
\caption{\textbf{Relationships among ML lifecycle stages, tactics/techniques/procedures (TTPs), 
  and vulnerabilities.}  
  This bipartite–tripartite network maps nine ML lifecycle stages 
  (\textcolor{blue}{left, blue triangles}) to a set of reported TTPs 
  (\textcolor{red}{center, red rectangles}), which are in turn connected to 
  vulnerabilities (\textcolor{green}{right, green circles}) observed in the 
  literature. Edge thickness denotes connection strength, where \emph{strong} links indicate frequent co-occurrence across multiple sources and \emph{weak} links indicate infrequent or context-specific associations.  
  Lifecycle stages span from \emph{Data Preparation} and \emph{Pretraining} 
  through \emph{Fine-tuning}, \emph{Testing}, and \emph{Deployment}, 
  while vulnerabilities include issues such as privacy violations, data poisoning, 
  gradient leakage, resource exhaustion, and model inconsistency errors.  
  This visualization synthesizes findings from both foundational and recent 
  studies~\cite{Chao2023JailbreakingBB,huang2021data,pmlr-v97-engstrom19a,gao2020backdoor,mcgregor2021preventing,carlini2021extracting,abdullah2021sok,arp2022and,zhu2023promptrobust,elmahdy2023deconstructing,casper2023explore,huang2023trustgpt,han2024fedsecurity,liu2023prompt,van2023protect,wang2023safeguarding,qammar2023chatbots,qi2024visual,hughes2024best,carlini2024poisoning,pelofske2023cybersecurity,he2023large,li2023vertical,rahman2024survey,malik2024systematic,Wang23:data-poisoning,aigrain2019detecting,xiong2021cyber,Kuppa2021linink-cve-mitre,tabassi2019taxonomy,Lakhdhar2019-mappingttps,Kumar2020-adversarial-ind,wallace2020imitation,alhanahnah2024depsrag,fu2024poisonbenchassessinglargelanguage,hughes2024bestofnjailbreaking,arslan2024survey,Antonio2023,Shaukat2020,Mehrabi2023,Bagaa2020,Hoseini2024,He2023,Shumailov2020,Nie2022,Ahmad2024,Qammar2023,Hu2021,Bouacida2021,Jain2023,Jiazhao,Wu2024,Qi2023,Wang2023,Deng2023,Kwon2024:gnn-injection,jiang2025vulrg,fedorchenko2024automated,Okutan22Exploit,zhang2025askingfordirections}, highlighting where security risks are concentrated and how they propagate across the ML pipeline.}
\label{fig:ttps-vul-stages}
\end{figure}

\begin{figure}
     \begin{subfigure}[b]{0.45\textwidth}
         \includegraphics[scale=0.3]{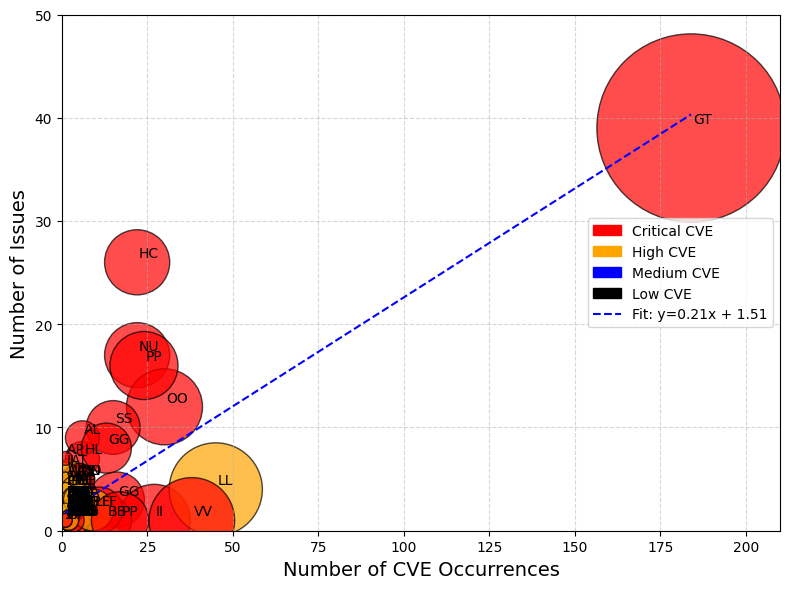}
         \caption{The scattered plot with a regression line} 
         \label{fig:bubble}
     \end{subfigure}
     \hfill
     \begin{subfigure}[b]{0.5\textwidth}
         \includegraphics[scale=0.35]{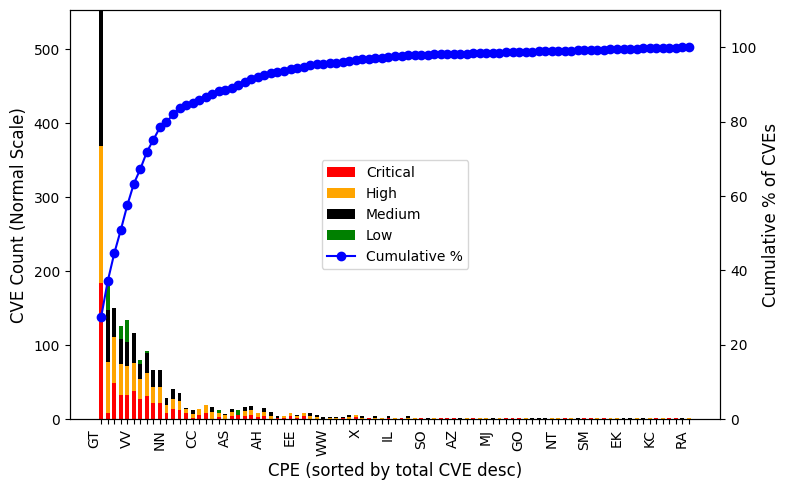}
         \caption{The Pareto chart with cumulative line 
         }
         \label{fig:pareto}
     \end{subfigure}
     \caption{The scattered plot with regression line reveals a modest positive correlation between CVE count and issues, with a linear fit of slope (m) $\approx$ +0.3 and intercept (b=3.01), implying additional CVEs correspond to about 0.3 more issues on average. Meanwhile, the Pareto plot shows a cumulative distribution (blue line) spanning 91 CPEs: $\approx$ 20\% of them (about 18 CPEs) account for 673 CVEs—over 80\% of the total 834.}
        \label{fig:pareto-bubble}
\end{figure}

\tosemrev{
{\tiny
\begin{longtable}{llrrrr}
\caption{\tosemrev{Predicted Risk Scores ($\mathcal{P}$) is the Likelihood of Exploitation for CVEs with Corresponding Descriptions, CVSS Scores, Exploitability Scores, and Patch Statuses. $\mathcal{P}$: (0.8,1] High, [0.40,0.8] Medium, [0,0.4) Low}}\\
\toprule
       CVE ID & Description & CVSS & Expl. & Patch & $\mathcal{P}$\\
\midrule
\endfirsthead
\caption[]{Predicted Risk Scores for CVEs with Corresponding Descriptions, CVSS Scores, Exploitability Scores, and Patch Statuses. (Continued)} \\
\toprule
       CVE ID & Description & CVSS & Expl. & Patch &$\mathcal{P}$ \\
\midrule
\endhead
\midrule
\multicolumn{6}{r}{{Continued on next page}} \\
\midrule
\endfoot
\bottomrule
\endlastfoot
CVE-2025-0015 & Use After Free vulnerability in Arm Ltd Valhall GPU Kernel Driver & 0.000 & 0.000 & 0 & 0.094 \\
CVE-2025-0222 & Vulnerability in IObit Protected Folder up to 13.6.0.5 & 0.556 & 0.462 & 1 & \cellcolor{yellow!50}0.664 \\
CVE-2025-0223 & Vulnerability in IObit Protected Folder up to 13.6.0.5 & 0.556 & 0.462 & 1 & \cellcolor{yellow!50}0.708 \\
CVE-2025-0224 & Vulnerability in Provision-ISR SH-4050A-2, SH-4100A-2L & 0.000 & 0.000 & 0 & 0.126 \\
\dots & \dots & \dots & \dots & \dots & \dots \\
CVE-2025-0226 & Vulnerability in Tsinghua Unigroup Electronic Archives System & 0.000 & 0.000 & 0 & 0.095 \\
CVE-2025-0228 & Vulnerability in code-projects Local Storage Todo App 1.0 & 0.485 & 0.436 & 1 & \cellcolor{red!50}0.860 \\
CVE-2025-0229 & Vulnerability in code-projects Travel Management System 1.0 & 0.990 & 1.000 & 1 & \cellcolor{red!50}0.818 \\
CVE-2025-0215 & Vulnerability in UpdraftPlus WP Backup \& Migration Plugin & 0.616 & 0.718 & 1 & 0.117 \\
\label{tab:risk-scores}
\end{longtable}
}
}
\tosemrev{
Table~\ref{tab:risk-scores} shows the predicted scores of the GNN model classifying vulnerabilities in three categories of predicted risk ($\mathcal{P}$). 
\[
\begin{cases}
(0.8, 1] & \text{Critical Response}: \\
[0.4, 0.8] & \text{Medium Priority} \\
[0, 0.4) & \text{Low Priority}
\end{cases}
\]
\textbf{Critical Response:} requires immediate action in patching, continuous monitoring, and urgent mitigation. \textbf{Medium Priority:} action needed to review the vulnerability, monitor trends, and schedule timely patches. \textbf{Low Priority} requires routine patching without urgency and passive monitoring.
}  
  
\begin{figure}[h]
\centering
\includegraphics[scale=0.4]{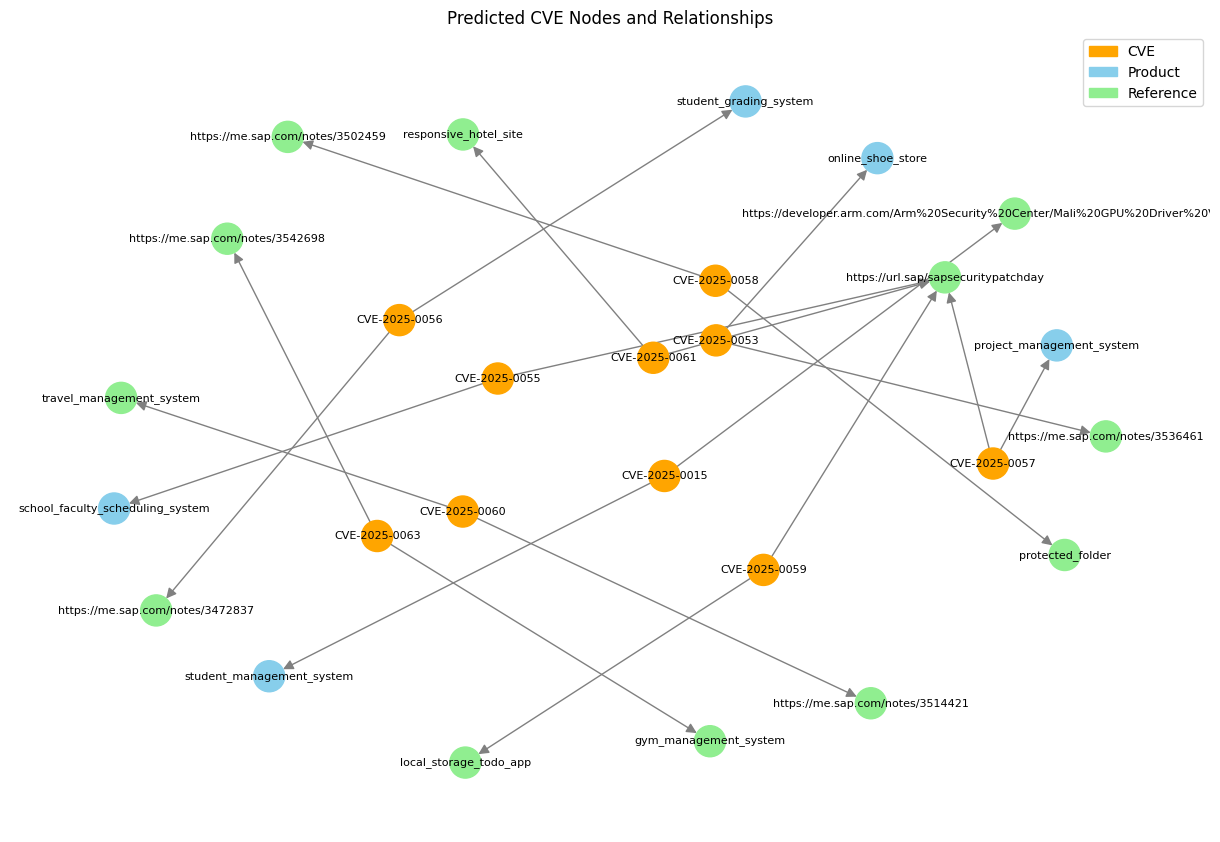}
\caption{\majorev{
\textbf{Predicted CVE–product–reference relationships.}  
  Graph shows predicted CVE (\textcolor{orange}{nodes}) 
  and their links to affected (\textcolor{cyan}{products}) and (\textcolor{green}{reference URLs}). Each CVE node is identified by its MITRE-assigned ID (e.g., CVE-2025-0056) and is connected to one or more product nodes representing vulnerable software or systems (e.g., student\_grading\_system, online\_shoe\_store). Reference nodes link to supporting advisories, vendor bulletins, or external security write-ups that confirm the vulnerability’s existence and impact. Edge directionality denotes the flow from CVE to the affected asset and from CVE to the reference.}  
}
\label{fig:predict-threats}
\end{figure}

\tosemrev{
Fig.~\ref{fig:predict-threats} represents a heterogeneous network where CVE nodes (orange) are connected to affected products (skyblue) and external references (light green). The directed edges illustrate real-world relationships such as vulnerabilities affecting specific products and being referenced by security advisories or exploit reports. The visualization captures the structural dependencies within the cybersecurity ecosystem, providing contextual insights for the GNN to predict the risk score of each CVE based on both its intrinsic features and its connections to related entities. The expected risk score is influenced by the following CVSS Metrics: Historical severity scores (e.g., base score, exploitability); the \textbf{CVE Descriptions:} Keywords that indicate exploitability (e.g., ``remote code execution"); Affected Products: High-profile products may indicate higher risk; References: URLs pointing to known exploits or advisories can signal active exploitation.
}

\begin{figure}
     \begin{subfigure}[b]{0.45\textwidth}
         \includegraphics[scale=0.3]{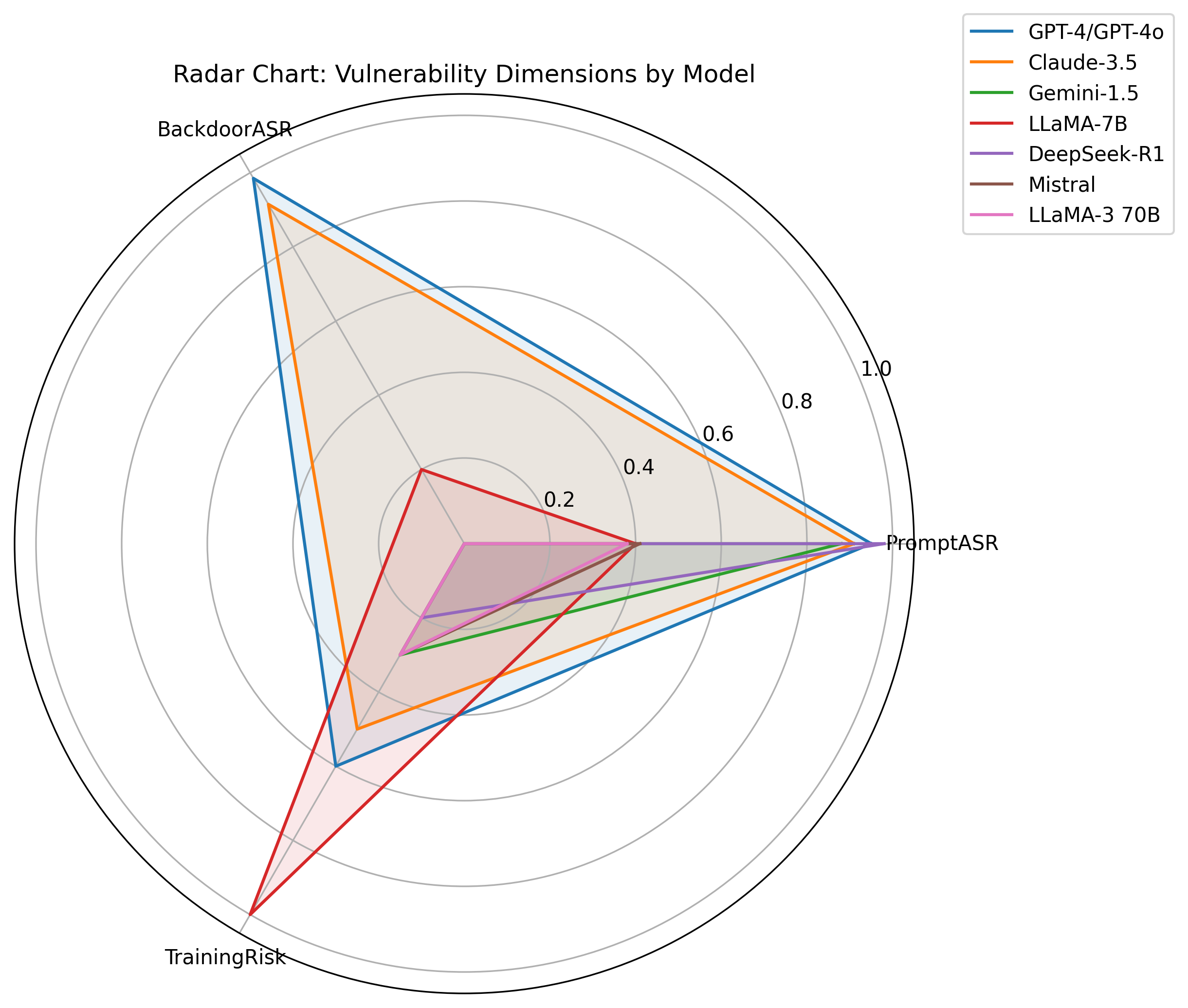}
         \caption{The radar chart} 
         \label{fig:radar}
     \end{subfigure}
     \begin{subfigure}[b]{0.5\textwidth}
         \includegraphics[scale=0.3]{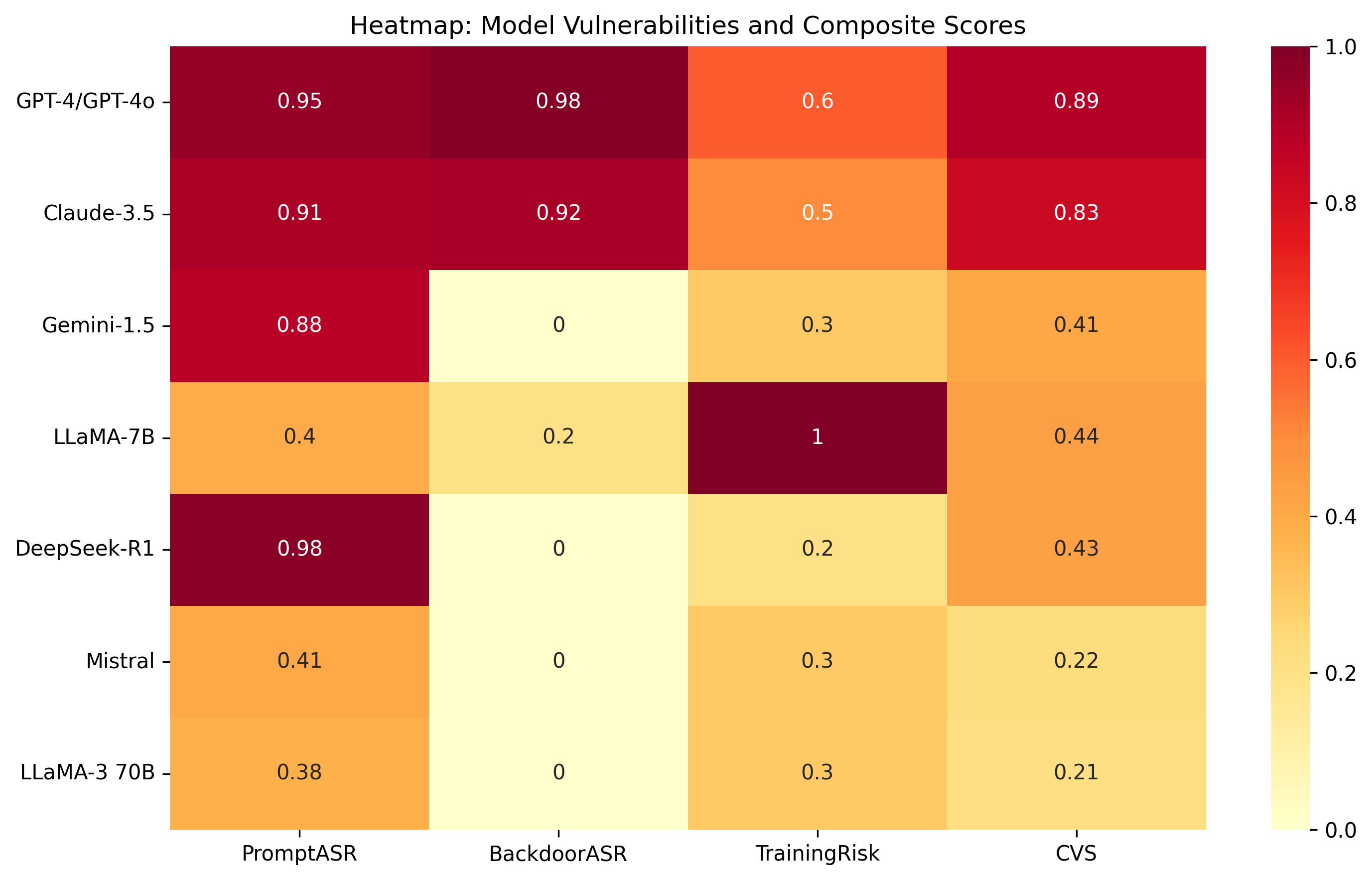}
         \caption{The heatmap chart}
         \label{fig:heatmap}
     \end{subfigure}
     \caption{\majorev{The radar chart shows how each model scores across Prompt Injection, Backdoor, and Training-phase vulnerabilities. It visualizes the shape of their risk profiles. Meanwhile, the heatmap compares absolute vulnerability values and composite scores (CVS) for all models, highlighting the most susceptible (darker red implies higher risk).}
     }
        \label{fig:sota_models}
\end{figure}

\majorev{
\subsection{Comparative Vulnerability Analysis of SoTA LLMs}
\subsubsection{Composite Vulnerability Scores Across SoTA Models}
Our quantitative evaluation of SoTA LLMs reveals distinct differences in vulnerability profiles. Using the Composite Vulnerability Score (CVS) metric, which integrates prompt injection ASR, backdoor ASR, and training-stage risk, we ranked five leading models from the literature~\cite{bai2023masterkey,yang2024autodan,liu2023promptinj,park2023rewardhacking,perez2024redteam,zhu2024loraleak,chen2024adapterleak,wang2024catllama,kandpal2023stealinggpt3,carlini2023extract,lee2024membership,liu2024toolframework,steenhoek2024toerr,wang2025juli,jiang2024artprompt,sabbaghi2025adversarial,yang2025synergistic,tong2025badjudge,bhatt2024cyberseceval,li2024drattack,li2025vuln,liao2025smartcontract,beurer2025design,rossi2025membership,
nikolic2025jailbreak,rando2025adversarial,rando2024gradient,carlini2024poisoning,jin2025llmbscvm,liu2025autoct,gao2023howfar,weng2024mmjbench,weng2025mmjbenchaaai,lee2025xjailbreak,in2025usersafety,ivry2025sentinel,huang2024medicalmllm,
zhang2025fixexploit,gou2024eyesclosed,daneshvar2024vulscriber,kim2024simplot,
hu2023greyboxai,zhang2025defenderbench,patnaik2024cabinet,li2024scla,
jiang2024chatbug,chen2024gpt4vredteam,pan2025unsafeattribution,Chao2023JailbreakingBB}:\\
\textbf{GPT-4o} emerged as the most vulnerable, with a CVS of 0.95, driven by extremely high success rates in both prompt injection (0.95) and code-level backdoor attacks (0.985). \textbf{Claude-3.5}and \textbf{Gemini-1.5} followed closely, reflecting similar vulnerability profiles across all categories.\textbf{ LLaMA‑7B}, although exhibiting moderate exposure to prompt injection, had the highest training-stage risk score (1.0), indicating susceptibility to sleeper agents and backdoor triggers. \textbf{DeepSeek‑R1} showed high vulnerability to prompt-based attacks but lacked sufficient backdoor and training-stage data, resulting in a lower composite score.
\subsubsection{Insights from Radar Chart and Heatmap reported in Fig.~\ref{fig:sota_models}. }
The radar chart (see Fig.~\ref{fig:radar}) highlights GPT‑4o’s uniformly high risk across all categories, whereas LLaMA‑7B shows pronounced spikes in training-stage vulnerabilities. The heatmap (see Fig.~\ref{fig:heatmap}) reinforces these findings, making it visually evident how risk is distributed unevenly across models and vulnerability types. These results underscore the importance of model-specific threat modeling. While some models may resist specific attack vectors, their exposure to others—especially those affecting training pipelines or memory-based exploits—necessitates the development of customized defense frameworks.
}

\subsection{ Prominence and common entry points of threat TTPs exploited in ML attack scenarios (RQ1)}

This RQ addresses two constructs: \underline{the prominence of threat TTPs exploited in attack scenarios} and \underline{the common entry points}.

\subsubsection*{The prominence of threat TTPs in attack scenarios}

Table~\ref{tactic2attack-mapping} shows the mapping between tactics and attack scenarios. The most prominent tactic is \textit{ML Attack Staging}, occurring 30 times across the 93 
ML attack scenarios. During ML attack staging, threat actors prepare their attack by crafting adversarial data to feed the target model, training proxy models, poisoning, or evading the target model. The other significant tactics used in attack scenarios are \textit{Impact} and \textit{Resource Development}, occurring 21 and 15 times in ML attack scenarios, respectively (see Table~\ref{tactic2attack-mapping}). 
After the ML attack successes, most attack scenarios tried to evade the ML model, disrupt ML service, or destroy ML systems, data, and cause harm to humans. 

In Table~\ref{tactic2attack-mapping}, the execution flows of attack scenarios share some TTP stages. The most used TTP sequences in attack scenarios are:
\begin{itemize}
    \item \textit{ML Attack Staging (stage 0)} $\rightarrow$ \textit{ Defense Evasion (stage 1)} $\rightarrow$ \textit{ Impact (stage 2)}
    \item \textit{ML Attack Staging (stage 0)} $\rightarrow$ \textit{ Exfiltration (stage 1)} 
    \item \textit{ML Attack Staging (stage 0)} $\rightarrow$ \textit{ Impact (stage 1)} 
    \item \textit{Reconnaissance (stage 0)} $\rightarrow$ \textit{Resource Development (stage 1)} $\rightarrow$ \textit{ML Model Access (stage 2)} $\rightarrow$ \textit{ML Attack Staging (stage 3)} $\rightarrow$ \textit{Impact (stage 4)}
    \item \textit{Reconnaissance (stage 0)} $\rightarrow$ \textit{Resource Development (stage 1)} $\rightarrow$ \textit{ML Attack Staging (stage 2)} $\rightarrow$ \textit{Defense Evasion (stage 3)}
\end{itemize}

All these attack scenarios (Carlini \etal{} \cite{carlini2017adversarial}, Abdullah \etal{} \cite{abdullah2021sok}, Papernot \etal{} \cite{papernot2016transferability, papernot2017practical}, Biggio \etal{} \cite{biggio2013evasion}, Athalye \etal{} \cite{athalye2018obfuscated}, Barreno \etal{} \cite{barreno2010security}) have similar execution sequences i.e., starting from stage \textit{stage 0} to stage \textit{stage 2}. Attack scenarios (Carlini \etal{} \cite{carlini2021extracting}, Wallace \etal{} \cite{wallace2020imitation}, Choquette-Choo \etal{} \cite{choquette2021label}) share stages from \textit{stage 0} to \textit{stage 1}. In addition, attack scenarios \textit{Attack on Machine Translation Service} and \textit{Microsoft Edge AI Evasion} have similar execution sequences, i.e., starting from stage \textit{S0} to stage \textit{S4}. It is also the same for attack scenarios \textit{Evasion of Deep Learning Detector for Malware C2 Traffic} and \textit{Botnet Domain Generation (DGA) Detection Evasion} that share stages from \textit{stage 0} to \textit{stage 3}. Attack scenarios \textit{Jagielski} \etal{} \cite{jagielski2020high} and \textit{Poachers Evade KZN's Park High-Tech Security} have some stages already included in the selected sequences, i.e., \textit{Defense Evasion} (stage 0) and \textit{Impact} (stage 1), \textit{ML Attack Staging} (stage 1) and \textit{Exfiltration} (stage 2). Attack scenarios \textit{Backdoor Attack on Deep Learning Models in Mobile Apps} and \textit{Confusing AntiMalware Neural Networks} only share two stages (i.e., \textit{stage 0} and \textit{stage 1}) already included in the selected sequences; thus, they are ignored.  

Table~\ref{tactic2attack-mapping} also shows that the most attack scenarios targeted ML systems without prior knowledge or access to the training data and the ML model (black box); this is explained by the highest number of occurrences of \textit{Black-box} in the Attack Knowledge column (i.e., 17 times). In addition, most attack scenarios are untargeted, shown by the highest number of occurrences of \textit{Traditional Untargeted} and \textit{Adversarially Untargeted} in the Attack Specificity column (i.e., 20 times). They also mainly targeted \textit{Confidentiality} and \textit{Integrity}.

\subsubsection*{The common entry points in attack scenarios} Table~\ref{tactic2attack-mapping} shows that the common entry points of attack scenarios are \textit{Reconnaissance} and \textit{ML Attack Staging}. Precisely, attackers exploited public resources such as research materials (e.g., research papers, pre-print repositories), ML artifacts like existing pre-trained models and tools (e.g., GPT-2), and adversarial ML attack implementations (Reconnaissance). To start the attack, they can use a pre-trained proxy model or craft adversarial data offline to be sent to the ML model for attack (ML Attack Staging). 

\begin{boxblock}{Summary 1}
    \tosemrev{
    ML attacks mainly exploit data poisoning, backdoor injections, membership inference, and supply chain risks, with Gradient-Based Obstinate Adversarial Attacks, MASTERKEY Automated Jailbreaking, and Federated Learning Poisoning being the most prevalent TTPs. Attacks commonly originate from third-party dependencies, model APIs, training data pipelines, and pretraining artifacts, highlighting ML supply chain vulnerabilities. The most frequent attack stages involve Reconnaissance and ML Attack Staging, with ML Attack Staging and Impact being the dominant TTP sequences. While shorter TTP paths focus on exfiltration and impact, the longest observed sequence follows Reconnaissance $\rightarrow$ Resource Development $\rightarrow$ ML Model Access $\rightarrow$ ML Attack Staging $\rightarrow$ Impact. Overall, Confidentiality and Integrity remain the primary attack objectives in ML threat scenarios.
    }
\end{boxblock}

\subsection{Impact of threat (TTPs) against ML phases and models (RQ2)}

In this research question, we delve into how the TTPs impact the ML overflow (phases) and models. 
For that, we aim to identify the most targeted/vulnerable ML phases and models based on adopted threat TTPs. 
Therefore, we present our results into two parts: (i) the impact of threat TTPs against ML phases and (ii) the impact of threat TTPs against ML models. 

\begin{table*}[h]
\caption{\label{tactic2phase-mapping} Mapping between Tactics adopted on attacks and ML phases}
\centering
\resizebox{\textwidth}{!}{
\begin{tabular}{|l|l|l|l|l|l|l|l|}
\hline
\backslashbox{Tactics~\cite{mitreatlas}}{ML Phases} & \textbf{Data Collection} & \textbf{\begin{tabular}[c]{@{}l@{}}Preprocessing\end{tabular}} & \textbf{\begin{tabular}[c]{@{}l@{}}Feature Engineering\end{tabular}} & \textbf{Training} & \textbf{Testing} & \textbf{Inference} & \textbf{Monitoring} \\ \hline
Reconnaissance                                                                                       & \checkmark                        & \checkmark                                                                       & \checkmark                                                                             & \checkmark                 & \checkmark                & \checkmark                  & \checkmark                \\ \hline
Resource Development                                                                                      & \checkmark                        & \checkmark                                                                       & \checkmark                                                                             & \checkmark                 & \checkmark                & \checkmark                  &                  \\ \hline
Initial Access                                                                                      & \checkmark                        &                                                                         &                                                                               & \checkmark                 & \checkmark                & \checkmark                  &                  \\ \hline
ML Model Access                                                                                   &                          &                                                                         &                                                                               & \checkmark                 & \checkmark                & \checkmark                  &                  \\ \hline
Execution                                                                                       &                          &                                                                         &                                                                               &                   & \checkmark                & \checkmark                  &                  \\ \hline
Persistence                                                                                       &                          &                                                                         &                                                                               & \checkmark                 & \checkmark                & \checkmark                  &                  \\ \hline
Defense Evasion                                                                                       & \checkmark                        & \checkmark                                                                       & \checkmark                                                                             & \checkmark                 & \checkmark                & \checkmark                  &                  \\ \hline
Discovery                                                                                       &                          &                                                                         &                                                                               & \checkmark                 & \checkmark                & \checkmark                  & \checkmark                \\ \hline
Collection                                                                                       & \checkmark                        &                                                                         &                                                                               &                   &                  &                    &                  \\ \hline
ML Attack Staging                                                                                   & \checkmark                        & \checkmark                                                                       & \checkmark                                                                             & \checkmark                 & \checkmark                & \checkmark                  &                  \\ \hline
Exfiltration                                                                                       &                          &                                                                       &                                                                               & \checkmark                 & \checkmark                & \checkmark                  &                  \\ \hline
Impact                                                                                       & \checkmark                        & \checkmark                                                                       & \checkmark                                                                             & \checkmark                 & \checkmark                & \checkmark                  & \checkmark                \\ \hline
Credential Access                                                                                       & \checkmark                        &  &  &  &  &  & \\ \hline
\end{tabular}
}
\end{table*}

\subsubsection*{Impact of threat TTPs against ML phases:} 
\tosemrev{
In the state-of-the-art ML security research, our analysis of TTPs (Tactics, Techniques, and Procedures), vulnerabilities, and ML lifecycle stages reveals critical weaknesses that adversaries exploit to compromise model integrity. The results of this analysis are shown in Fig.~\ref{fig:ttps-vul-stages}; we observed 21 vulnerabilities, 55 TTPs, and nine lifecycle stages. 
Among the most pressing vulnerabilities are Data Poisoning, Backdoor Exploits, Federated Model Poisoning, and Gradient Leakage, all of which introduce systemic risks that propagate through the ML pipeline. Training-time attacks, such as Federated Learning Poisoning and Gradient-Based Obstinate Adversarial Attacks, pose substantial threats by manipulating model parameters at their inception, leading to compromised inference outcomes. In adversarial settings, MASTERKEY Automated Jailbreaking and Semantic Firewall Bypass Attacks highlight how prompt-based adversarial manipulations circumvent existing alignment techniques, rendering large-scale AI models susceptible to unauthorized control and exploitation. The widespread adversarial transferability of attacks further exacerbates these risks, enabling crafted adversarial perturbations to generalize across models, underscoring the inadequacy of conventional defenses. To mitigate these threats, robust countermeasures must be deployed, including differentially private training, cryptographic model integrity verification, and adversarially robust learning architectures. Despite its promise of decentralized privacy-preserving computation, Federated learning remains an attack vector requiring secure aggregation techniques to thwart malicious updates. Furthermore, real-time adversarial detection pipelines, adversarial training with diverse attack distributions, and secure model fine-tuning frameworks are imperative to enhancing resilience against TTP-driven model compromise. As ML adoption scales, research into proactive, adaptive security paradigms must advance to safeguard models against evolving attack methodologies, ensuring the robustness of AI systems deployed in high-stakes domains.
}
In Table~\ref{tactic2phase-mapping}, we present the most targeted ML Phases (columns) against the different adopted Tactics (rows) from practice.
First, we can observe that, based on the analyzed attacks, not all ML phases are impacted by the TTPs, as they cover specific ML phases. Such a finding does not indicate that other ML phases are not impacted; rather, it reflects the contextual nature of the observed attacks.
In different contexts or under varying threat models, other ML phases might also be vulnerable to similar or novel TTPs.  This highlights the need for a holistic approach when analyzing and mitigating threats across the entire ML lifecycle, as adversaries may adapt their strategies to exploit weaknesses in less commonly targeted phases. 

Second, regarding the impacted phases, we observe that
\textit{Testing, Inference} and \textit{Training} represent the most impacted ML phases.
This finding underscores the need for practitioners and researchers to prioritize these phases when analyzing potential vulnerabilities.
It is essential not only to investigate and understand the likelihood and nature of vulnerabilities that occur during these phases, but also to develop and implement effective mitigation strategies. By focusing on these high-risk phases, researchers and practitioners can work toward building more robust and resilient ML pipelines.

Third, regarding the tactics, we observe varying levels of coverage across the ML phases.
For example, \textit{Reconnaissance} and \textit{Impact} are present in all reported ML phases, demonstrating their broad applicability and relevance across the entire ML lifecycle. 
On the other hand, \textit{Credential Access} is associated exclusively with the \textit{Data Collection} phase. 
Such a disparity can be attributed to several factors, such as the specific nature of the tactic and its primary focus. 
For example, \textit{Reconnaissance} involves gathering information, which can be relevant at any phase, whereas \textit{Credential Access} is more likely to target sensitive access points, such as those involved in the data acquisition process.

\begin{table}[h!]
\centering
\caption{\tosemrev{Target Models, Occurrences (Occ.), Normalized Metrics ( $\text{Nm. (\%)} = \left( \frac{\text{Occ. of a Model}}{\text{Total Occ.}} \right) \times 100
$), and Time Interval of attacks (Period) collected from MITRE, AI Incident, and Literature. The information about the Occ is extracted from Table \ref{tab:attack2model-mapping}. ``N/A'' stands for the cases without information regarding the model.} }
\centering
\resizebox{0.95\textwidth}{!}{
\begin{tabular}{|l|c|c|r|}
\hline
\textbf{Targeted Model} & \textbf{Occ.} & \textbf{Nm (\%)} & \textbf{Period} \\ 
\hline
Transformers (BERT, GPT-2, GPT-3, others) & 16 & 25.40 & 2019-2024 \\ 
Convolutional Neural Networks (CopyCat, Fisheye, ResNet, others) & 12 & 19.05 & 2018-2021 \\ 
Deep Neural Networks (unspecified) & 9 & 14.29 & 2013-2021 \\ 
Hidden Markov & 1 & 1.59 & 2021 \\ 
Long-Short Term Memory & 2 & 3.17 & 2020-2021 \\ 
Generative Adversarial Networks & 2 & 3.17 & 2014-2020 \\ 
DeepVoice~\cite{arik2017deep} & 1 & 1.59 & 2019 \\ 
Feed-Forward Neural Networks & 1 & 1.59 & 2017 \\ 
Parseval Networks & 1 & 1.59 & 2017 \\ 
Linear classifiers (SVM, Logistic Reg., Naive Bayes) & 3 & 4.76 & 2010-2016 \\ 
Non-Linear classifiers (Decision Trees, k-Nearest Neighbor) & 2 & 3.17 & 2016 \\ 
N/A & 10 & 15.87 & 2018-2024 \\ 
\hline
\end{tabular}}
\label{modelstat}
\end{table}

\subsubsection*{Impact of threat TTPs across ML models:}
\tosemrev{
The analysis of cybersecurity vulnerabilities across software dependencies (linking GitHub issues to vulnerability/threats --CVE) reveals a high prevalence of critical and high-severity CVEs, highlighting systemic risks that persist across multiple dependencies. The network graphs ( available in our replication package and derived from Tables~\ref{tab:nodes}, \ref{tab:edges} and Fig.~\ref{fig:global-umap}
) underscore the existence of high-impact vulnerabilities, which, when left unpatched, serve as prime attack vectors for sophisticated adversarial techniques such as Gradient-Based Obstinate Adversarial Attacks and MASTERKEY Automated Jailbreaking. These tactics manipulate security mechanisms through subtle perturbations or bypass restrictions to exploit system weaknesses. The network visualization with community detection 
further emphasizes that security vulnerabilities are not isolated threats but form interconnected clusters 
, suggesting that adversaries can exploit multiple dependencies through cascading failures. 
We observe that dependency management presents a wicked problem, where some communities have a reasonable number of issues to address vulnerabilities 
, while others suffer from an overwhelming influx of critical cases with little to no issue tracking or mitigation mechanisms in place 
. This disparity underscores the urgent need for proactive security interventions, particularly in high-risk communities where unaddressed critical vulnerabilities can propagate across dependencies, amplifying systemic risks.
In this context, influential nodes, representing highly connected dependencies, play a critical role in dependency management. These nodes act as risk amplifiers—a single compromise in a central node can propagate across multiple systems, creating widespread security breaches. Consequently, prioritizing security updates for these high-degree nodes and enforcing automated security patching mechanisms becomes imperative~\cite{feedback-sync23}. The network structure also reveals latent inter-dependencies, where seemingly unrelated software components share common vulnerabilities, necessitating a holistic approach to risk mitigation. Organizations should leverage graph-based threat modeling to proactively identify high-risk dependencies and deploy adaptive security mechanisms, such as real-time monitoring and federated threat intelligence, to minimize exploitability. This interconnected vulnerability landscape reinforces the urgent need for systemic resilience strategies, ensuring that software dependencies remain robust against adversarial exploitation and resilient to emerging threats. 
Fig.~\ref{fig:pareto} ranks CPEs by their total CVE count (highest on the left) with bars stacked by severity (e.g., red~=~critical, orange~=~high), while the blue line on a secondary y‐axis shows the cumulative percentage of CVEs. The leftmost bars often represent a small subset of CPEs accounting for most vulnerabilities (the “vital few”), indicating a Pareto effect if the line exceeds 80–90\% after only a few bars; bars heavily colored in red/orange reveal especially severe CVEs. Once the cumulative line surpasses around 95\%, additional bars yield less impact, aiding prioritization for patching or deeper triage. Meanwhile, Fig.~\ref{fig:bubble} plots CVE count (x‐axis) versus Issue count (y‐axis) for each CPE, overlaid with a regression line: a higher slope means an extra CVE generally leads to more Issues, a near‐zero slope shows minimal correlation, and the intercept represents baseline Issues if a CPE theoretically had zero CVEs. Closer clustering around the line implies a stronger relationship, whereas bubbles above it may signify more Issues than expected, and those below could be under‐tracked.\\
}

Table~\ref{modelstat} summarizes the models targeted in this study, including the time period of each attack and the number of occurrences based on the extractions in Section~\ref{sec:mapping-mat}. The “unspecified” category (row-4) indicates that Deep Neural Networks (DNNs) were used without disclosing their architectures, while the final row (row-13) lists ``N/A,'' meaning no information on the model was found. Most attack scenarios involve Transformers or Convolutional Neural Networks (CNNs). CNNs appear in 12 cases, while Transformers lead with 14, though the distribution over time differs between the two. CNNs show a steady pattern of exploitation across the analyzed period, aligning with the popularity reported by Kaggle between 2019 and 2021\mbox{\cite{kagglereport}}, where Gradient Boosting Machines (e.g., xgboost, lightgbm) and CNNs were most frequently used. Since tree‐based models like Gradient Boosting Machines are discrete and non‐differentiable, they are not well-suited for gradient‐based white‐box attacks\mbox{\cite{chen2019robust}}, so such models are absent from these attack scenarios.

In contrast, Transformers exhibit an irregular distribution with a notable surge in 2023, particularly targeting GPT models. This jump may reflect their widespread adoption and the accompanying rise in adversarial interest. Several attack scenarios do not specify which model was attacked. While still valid, these cases lack clarity and hinder deeper analysis, highlighting the need for more transparent reporting to better identify vulnerabilities in different model architectures. Finally, across the entire study period (2013–2023), early years show relatively few attacks, often on models that never achieved the popularity of Transformers, CNNs, or DNNs.

\tosemrev{
CNNs are frequently targeted. While this observation can be partially attributed to the widespread adoption of CNNs in various domains (e.g., computer vision, autonomous systems), our analysis extends beyond mere popularity metrics. CNNs exhibit inherent architectural vulnerabilities—such as sensitivity to adversarial perturbations due to their linear decision boundaries—that make them more susceptible to specific attack vectors like adversarial evasion and gradient-based attacks. To differentiate between vulnerabilities arising from high deployment frequency and those due to structural weaknesses, we incorporated normalized metrics (i.e., percentage-based distributions) alongside absolute counts. This dual approach enables a more nuanced understanding of why specific models, such as CNNs, are disproportionately targeted, offering insights into both their prevalence and inherent security risks.
}

\begin{boxblock}{Summary 2} 
    \tosemrev{
    Threat TTPs impact ML phases with varying severity, with pretraining and inference being the most vulnerable due to exposed model artifacts, insufficient robustness, and susceptibility to adversarial examples. Training-time attacks, such as data poisoning and backdoor injections, threaten model integrity, while inference-time threats, including evasion and membership inference attacks, compromise confidentiality and reliability. Federated learning environments face heightened risks from poisoning and leakage attacks due to the distributed trust assumptions and gradient-sharing mechanisms employed. Attack strategies primarily leverage Reconnaissance, Impact, ML Attack Staging, and Resource Development, with Testing, Inference, Training, and Data Collection being the most targeted ML phases. Transformers and CNNs are the most frequently attacked model architectures, with a notable rise in GPT-based attacks in recent years.}
\end{boxblock}

\subsection{Characterizing new Threats not reported in the ATLAS database (RQ3)}\label{sec:threat_matrix} %

To answer this RQ, we split the results into three parts:
(i) the new threats found in the AI Incident database and the literature, (ii) the threats mined from the GitHub ML repositories, further discussing the most vulnerable ML repositories as the dependencies that cause them, and (iii) the most frequent vulnerabilities in the ML repositories.

\subsubsection*{New Threats from the AI Incident Database and the Literature}

In Table~\ref{tab:rq3-threats-aidatabase-literature}, we present the new threats collected and the associated tactics and techniques. Regarding the threats in the AI Incident database, we identify new TTPs covering eight (8) tactics and 15 techniques across 12 ML attacks.
Moving forward, regarding the threats from the Literature (\textit{considering only the 14 seed papers here}), we have 14 ML attacks, covering six tactics and nine techniques.
Overall, we can observe that most LLM attacks share the same TTPs as No-LLM ones, highlighting the replicability of a given attack exploring different contexts.
Except for the tactic \textit{Persistence}, all the other tactics are commonly shared among all attacks, indicating that despite the different attacks in various contexts, they share similar characteristics.

\begin{table}[]
\caption{Threats collected from AI Incident and Literature. Associated tactics to an attack are presented as columns, while techniques are reported as values in the cells. AI-DB stands for cases extracted from the AI Incident database, while LIT stands for Literature.} 
\label{tab:rq3-threats-aidatabase-literature}
\resizebox{\textwidth}{!}{
} & \multicolumn{1}{l|}{}                                                                                   & \multicolumn{1}{l|}{}                                                                    & \multicolumn{1}{l|}{}                                                                            &                                                                                 \\ \hline
\end{tabular}
}
\end{table}


These 26 new ML attack scenarios were not documented in ATLAS and could be used to extend the ATLAS case studies. While some of these latest attacks share the same characteristics with the ones already included in ATLAS, other attacks, like LLM-based, provide new insights about ML attacks by exploring related attributes in the same and different ML models.

\begin{figure}[h]
\centering
\includegraphics[width=0.9\textwidth]{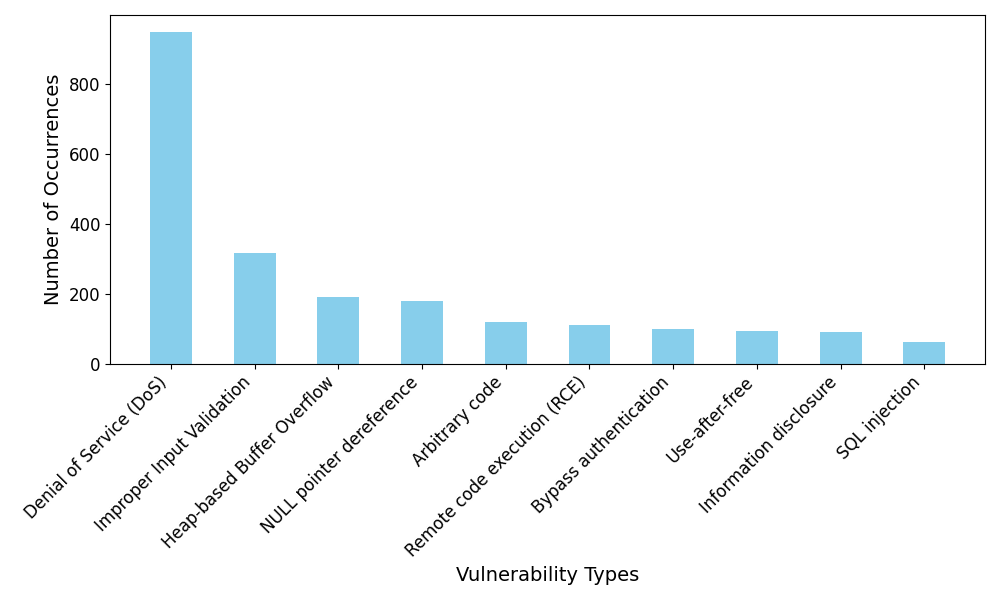}
\caption{\majorev{\textbf{Frequency distribution of vulnerability types in GitHub ML repositories.} Bar chart showing the number of occurrences for the most common vulnerability categories identified across ML–related repositories on GitHub.}}
\label{fig:vulnw}
\end{figure}

\subsubsection*{Potential Threats from Vulnerabilities in GitHub ML repositories}

Overall, we identify 35 vulnerability types from the mined ML repository issues.
When evaluating these vulnerabilities, we observe that most are classified under two main types: (i) software- and (ii) network-level. 
While software-level vulnerabilities are related to weaknesses faced by the target software (e.g., applications (models), operating systems, or libraries) that adversaries can exploit, network-level vulnerabilities explore the infrastructure the target system operates on, exploiting flaws in communication protocols, access to services and resources, etc.
Fig.\mbox{~\ref{fig:vulnw}} shows our study's top 10 vulnerability types and their occurrences. 
 
Denial of Service (DoS) is the most recurrent vulnerability, with a frequency of 951 occurrences; such a vulnerability aims to make the target system unavailable.
We may highlight that DoS occurs in both types of vulnerabilities evaluated here. 
While software-level DoS can impact the system's availability due to a memory or crash error (e.g., segmentation fault) that disrupts the underlying OS and machine, network-level DoS may disrupt the regular traffic of a network resource.

\subsubsection*{Most vulnerable GitHub ML repositories and Their Target Dependencies}

\begin{figure}[h]
\centering
\includegraphics[width=0.9\textwidth]{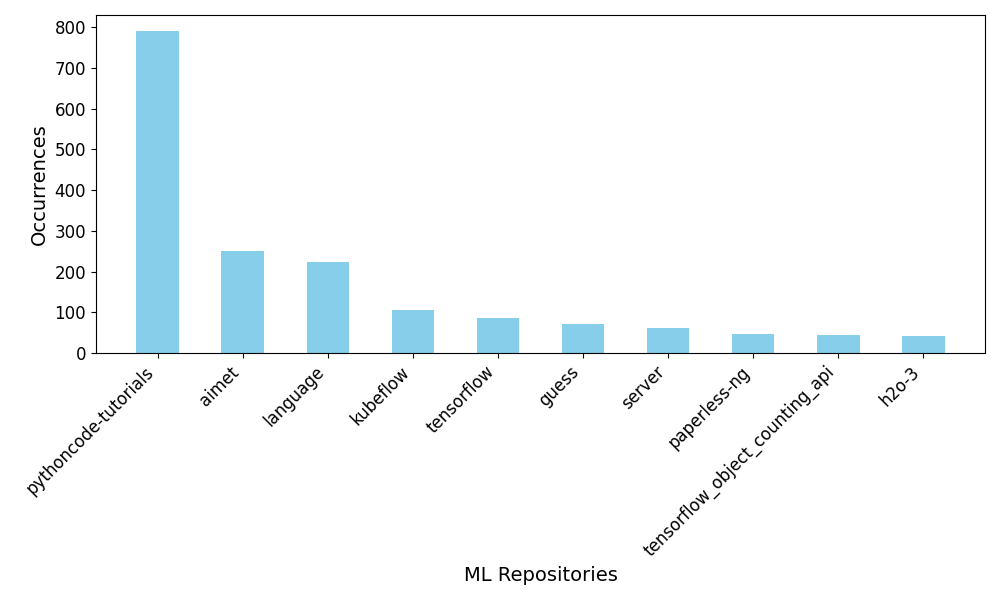}
\caption{\majorev{\textbf{Vulnerability occurrences across top GitHub ML repositories.} The bar-plot highlights an uneven distribution of vulnerabilities across ML projects, with a small subset accounting for the majority of identified flaws.}}
\label{fig:vulnerability-repositories}
\end{figure}

In our analysis, 86 repositories reported at least one vulnerability.  
Checking these repositories, we observe that 75\% of them represent projects used to build ML systems, like libraries, toolkits, frameworks, and MLOps. 
The other repositories use these previous projects as dependencies to provide their services, like practices, tutorials, and tools to users. 
Figure~\ref{fig:vulnerability-repositories} presents the top 10 repositories with more occurrences of the vulnerabilities under analysis. 

Python Code Tutorials\footnote{\href{}{https://github.com/x4nth055/pythoncode-tutorials}} is the repository with the highest frequency of reported vulnerabilities. 
This repository contains a diverse set of tutorials on Python, covering different domains, including Machine Learning, which may encourage users to replicate the vulnerable code and face the reported flaws/weaknesses. Next, we have a consistent number of repositories that usually provide services for other repositories, like libraries and frameworks. For example, Aimet is a library that supports advanced quantization and compression techniques for trained neural network models, while TensorFlow is one of the most used frameworks for building ML applications.

\begin{table}[]
\caption{Dependencies responsible for Vulnerabilities in ML repositories}
\label{depvuln}
\begin{tabular}{|c|l|c|l|}
\hline
\multicolumn{1}{|c|}{\textbf{Dependency}} & \textbf{Occurrences} & \multicolumn{1}{c|}{\textbf{Severity}} & \textbf{\begin{tabular}[c]{@{}l@{}}Affected \\ Repositories\end{tabular}} \\ \hline
google:tensorflow                         & 184                  & critical, high, medium                 & 7                                                                         \\ \hline
linux:linux\_kernel                       & 45                   & high, medium, low                      & 4                                                                         \\ \hline
vim:vim                                   & 38                   & critical, high, medium                 & 1                                                                         \\ \hline
openssl:openssl                           & 30                   & critical, high, medium, low            & 10                                                                        \\ \hline
imagemagick:imagemagick                   & 27                   & critical, high, medium                 & 1                                                                         \\ \hline
python:pillow                             & 24                   & critical, high, medium                 & 11                                                                        \\ \hline
haxx:curl                                 & 22                   & critical, high, medium                 & 7                                                                         \\ \hline
paddlepaddle:paddlepaddle                 & 17                   & critical, high                         & 1                                                                         \\ \hline
gnu:glibc                                 & 16                   & critical, high, medium, low            & 3                                                                         \\ \hline
sqlite:sqlite                             & 15                   & critical, high, medium                 & 4                                                                         \\ \hline
\end{tabular}
\end{table}

Moving forward, once we collect the vulnerabilities and their frequency in ML repositories, we aim to investigate the dependencies that cause these vulnerabilities. 
Overall, we observe that 227 dependencies are responsible for the vulnerabilities studied here. 
Table~\ref{depvuln} presents the top 10 most recurrent dependencies regarding the number of occurrences.
TensorFlow is the most frequent dependency, with 184 occurrences across seven different sample repositories. 
Given the popularity of Tensorflow, such a dependency is constantly used by different repositories, consequently increasing the chances of these repositories facing vulnerabilities.
On the other hand, TensorFlow is constantly updated, addressing reported issues and providing up-to-date services for its users. 
Among the dependencies that most affect repositories, Pillow, a library for image processing, stands out by affecting eleven repositories, while most of them are other libraries and tools (systems).

\begin{figure*}[h]
\centering
\includegraphics[width=0.99\textwidth]{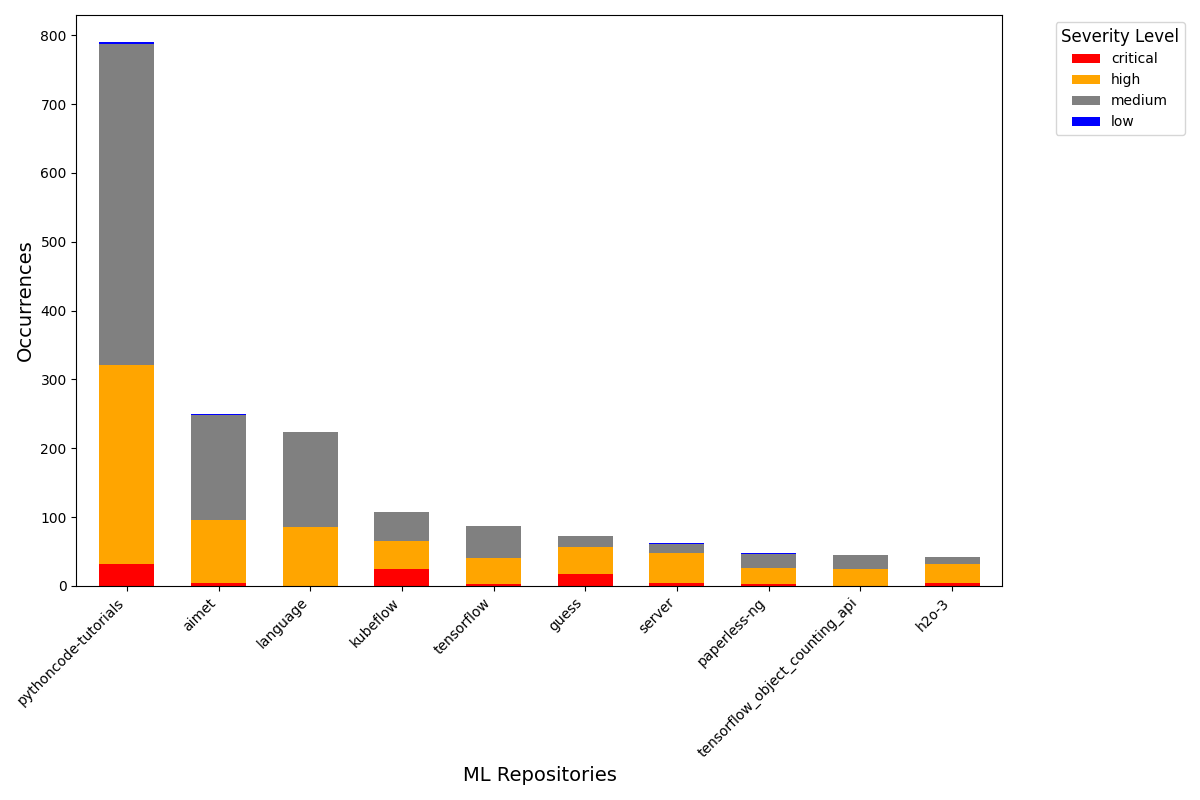}
\caption{\majorev{\textbf{Distribution of vulnerability severities across top GitHub ML repositories.} Vulnerabilities are categorized into four severity levels, with the 
  \texttt{pythoncode-tutorials} repository exhibits the highest number of 
  vulnerabilities overall. This distribution underscores the need for targeted remediation strategies prioritizing high- and critical-severity issues in widely used ML repositories.
}}

\label{fig:severity-repos}
\end{figure*}

Regarding the severity of the vulnerabilities, Figure \ref{fig:severity-repos} presents the distribution for the top 10 repositories. 
Although we observe the vulnerabilities vary from \textit{low} to \textit{critical} severity, it is important to highlight how recurrent \textit{high} and \textit{medium} vulnerabilities are reported, posing significant risks to the security and stability of systems.
On the other hand, although \textit{low} severity vulnerabilities are less frequent, \textit{critical} vulnerabilities represent an expressive frequency for some repositories, like \textit{Python Code Tutorials}, \textit{Kuberflow}, and \textit{Guess}. 
Addressing these critical vulnerabilities promptly should be prioritized to mitigate potential exploitation and ensure the secure operation of the systems.

\subsubsection*{Most Frequent Vulnerabilities across ML Repositories}

\begin{figure*}[h]
\centering
\includegraphics[width=0.99\textwidth]{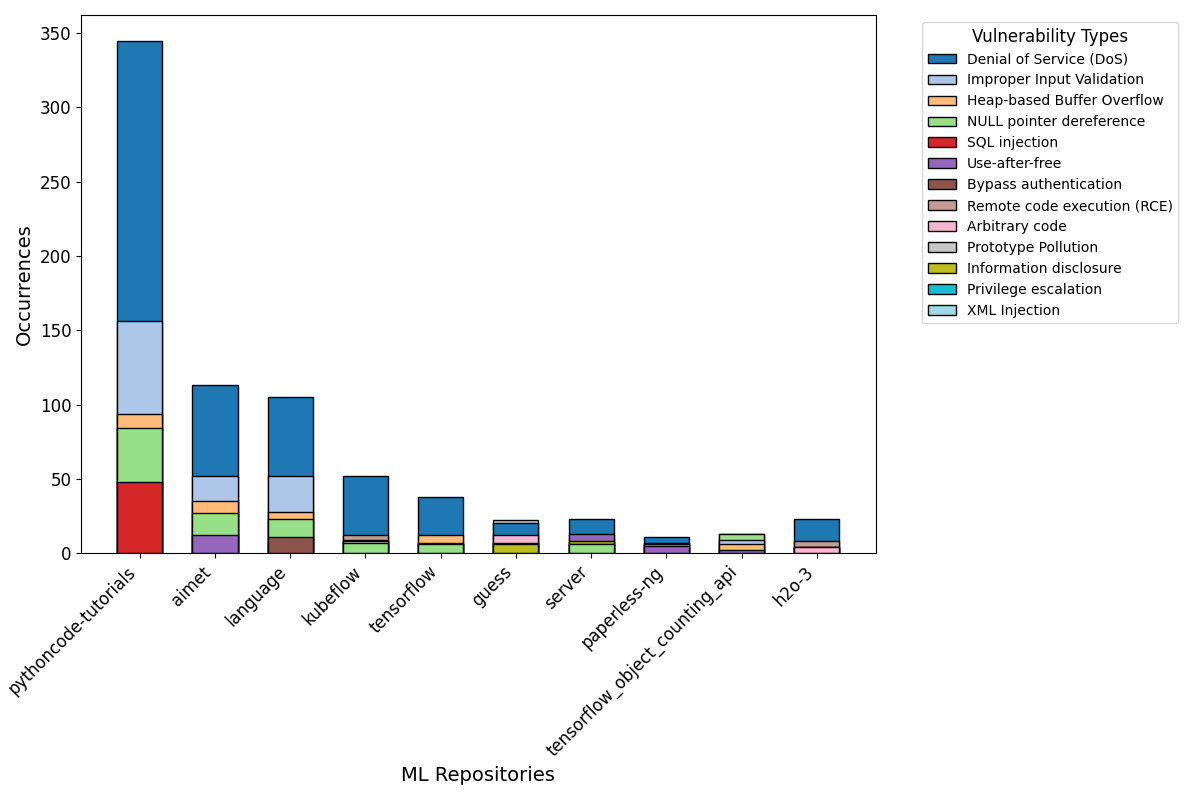}
\caption{\majorev{\textbf{Distribution of vulnerability types across top GitHub ML repositories.} The visualization highlights the prevalence of 13 specific security weaknesses in popular ML repositories, guiding prioritization for remediation.
The \texttt{pythoncode-tutorials} repository shows the highest concentration, dominated by DoS and Improper Input Validation. }
}
\label{fig:vulnaccross}
\end{figure*}

After identifying vulnerabilities and the dependencies that caused them, we aim to know how the observed vulnerability types propagated across the studied ML repositories. 
Fig.~\ref{fig:vulnaccross} shows the distribution of vulnerability types for the top 10 ML repositories, which have more occurrences of vulnerabilities.
Overall, we can observe that \textit{Denial of Service (DoS)} is consistently reported as the primary vulnerability type for all repositories.
For the remaining types, we observe a regular occurrence of \textit{Improper Input Validation}, \textit{Null Pointer Deference}, and \textit{Heap-based Buffer Overflow}.
However, we also observe a high incidence of \textit{SQL injection} vulnerabilities on the project PythonCode Tutorials; due to the focus of this repository, such a vulnerability type is valid and recurrent, as some tutorials explore the adoption of databases and SQL.
These findings show that the ML-analyzed repositories might face the same types of vulnerabilities, indicating that certain categories of vulnerabilities are prevalent regardless of the repository’s specific focus. 

\begin{boxblock}{Summary 3}
      \tosemrev{
      The integration of the AI Incident Database, GitHub security issues, and the literature reveals multiple previously undocumented threats absent from ATLAS. Graph-based dependency analysis highlights that ML library clusters face disproportionately high-severity vulnerabilities, often lacking adequate issue-tracking or mitigation mechanisms. Emerging threats include supply chain compromises, automated jailbreak techniques, and prompt-based adversarial manipulations, emphasizing the need for continuous updates to threat models. New ML attacks, particularly those targeting LLMs, present opportunities to extend the ATLAS database with novel insights. Despite their different focuses, ML repositories exhibit shared vulnerabilities, with frequent occurrences of Denial of Service (DoS), Improper Input Validation, Null Pointer Dereference, and Heap-based Buffer Overflow. Additionally, ML dependencies, particularly TensorFlow, are major exposure points, introducing high-severity risks across various ML applications.
      }
\end{boxblock}

\begin{figure*}[!ht]
\centering
\includegraphics[width=0.94\textwidth]{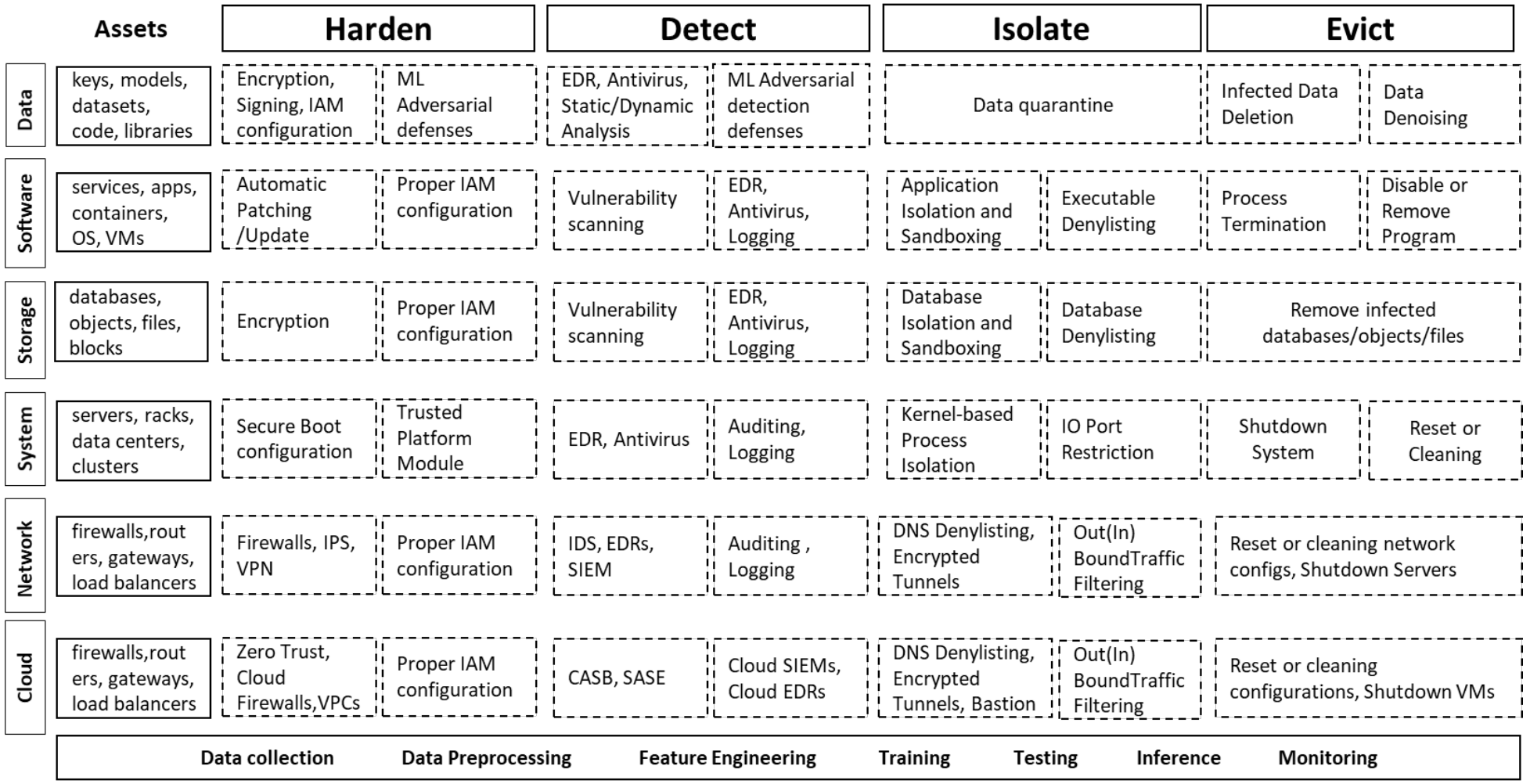}
\caption{ML Threat Mitigation Matrix}
\label{fig:defend_matrix}
\end{figure*}
\minor{
\subsubsection{Extension for Preference-Guided and Introspection-Based Attacks.}
To capture newly emerging threats such as preference-guided jailbreaks and introspection-based optimization attacks~\cite{zhang2025askingfordirections}, we augment the mitigation matrix (Fig.~\ref{fig:defend_matrix}) with the following targeted controls:
\begin{itemize}[leftmargin=*, topsep=1pt, itemsep=2pt]
    \item \textbf{Reduce optimization signal.} Train safety policies to refuse comparative or preference-eliciting queries that can be exploited for
    gradient-free optimization. Avoid deterministic binary phrasing in refusals, as consistent responses form a usable signal.
    \item \textbf{Rate-limit and jitter.} Detect iterative, stateful query patterns (e.g., near-duplicate prompts differing slightly in text or image) and introduce randomized refusals or obfuscations to disrupt attack optimization loops.
    \item \textbf{Guardrails around introspection.} Enforce policy-level blocking of self-assessment or self-ranking requests tied to disallowed objectives, and monitor for escalating acceptance of adversarially reframed instructions.
    \item \textbf{Agent and RAG contexts.} Sanitize retrieved or contextual information that elicits unsafe preference reasoning, and implement human-in-the-loop or interlock mechanisms when repeated near-duplicate retrievals occur within multi-agent or RAG workflows.
\end{itemize}
These measures strengthen the matrix’s coverage of introspection-based, black-box attacks and ensure that emerging preference-oracle threats are addressed alongside traditional adversarial and data-poisoning defenses.\\
\paragraph{Limitations. }
A residual risk persists in text-only interfaces: even without numeric confidences, comparative judgments can leak a strong optimization signal. Current static prompt defenses remain insufficient against iterative, preference-guided attacks.
}

\begin{figure*}[!ht]
\centering
\includegraphics[width=0.99\textwidth]{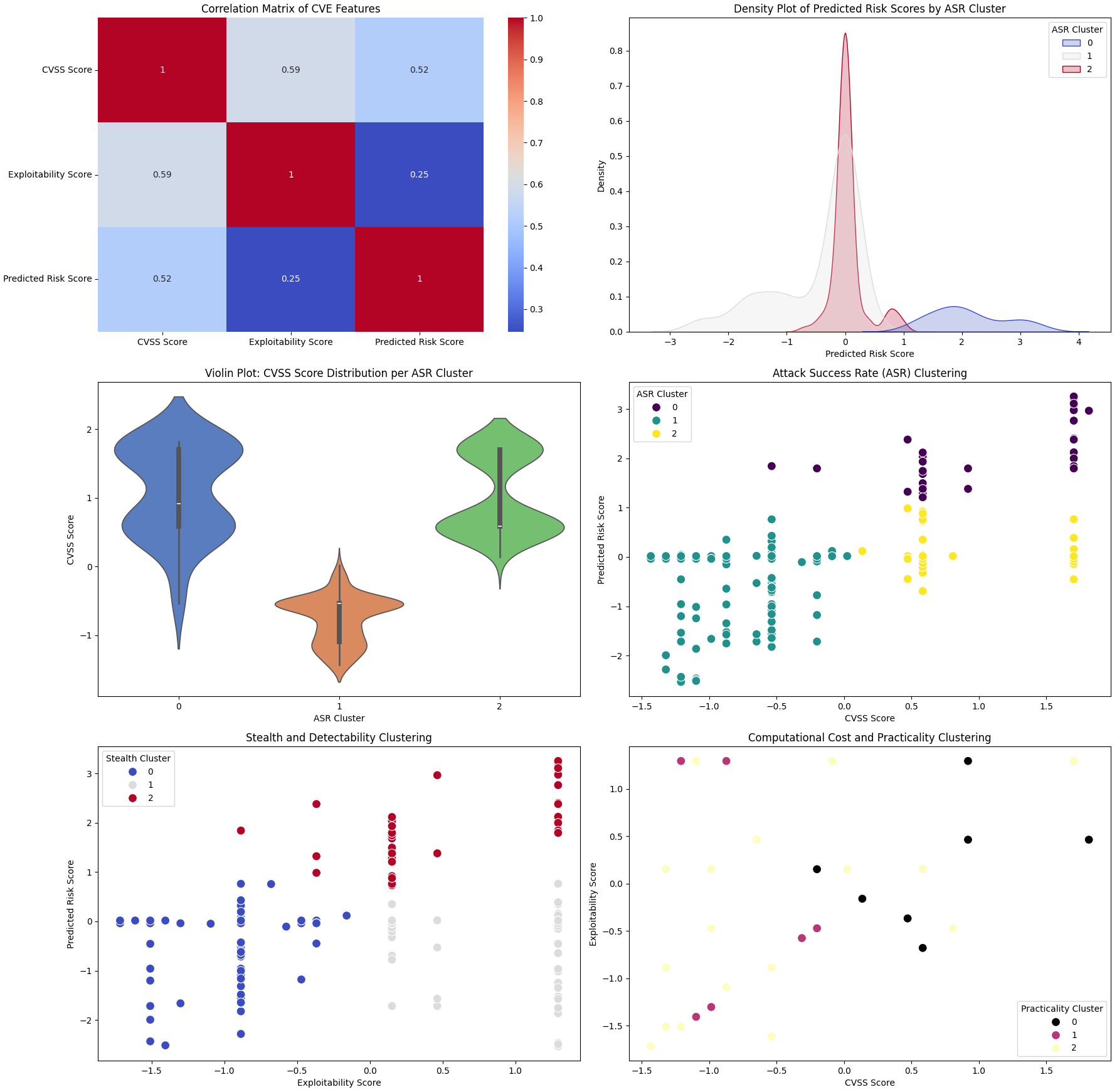}
\caption{\majorev{\textbf{Multi-faceted GNN-based vulnerability analysis.}  
Integrated views showing (i) feature correlations, (ii) density and violin plots of predicted risk by ASR cluster, (iii) scatter plots for ASR vs.\ CVSS, stealth vs.\ exploitability, and CVSS vs.\ practicality. Together, these views integrate GNN predictions with statistical and unsupervised learning insights to profile vulnerabilities across multiple operational dimensions. It links severity, exploitability, stealth, cost, and operational feasibility.
}}
\label{fig:vulnaccross}
\end{figure*}
\tosemrev{Figure~\ref{fig:vulnaccross} presents a detailed breakdown of how Graph Neural Networks (GNNs) and clustering techniques improve vulnerability classification and risk assessment. (Top Left) The Correlation Matrix of CVE Features illustrates the relationships between CVSS Score, Exploitability Score, and Predicted Risk Score, highlighting the degree of association between these key vulnerability indicators. (Top Right) Density Plot of Predicted Risk Scores by ASR Cluster visualizes the distribution of predicted risk scores within each attack success rate (ASR) cluster, showing variations in attack effectiveness. (Middle Left) Violin Plot: CVSS Score Distribution per ASR Cluster compares the spread of CVSS scores across ASR clusters, demonstrating inconsistencies between attack success rates and traditional severity scores. (Middle Right) Attack Success Rate (ASR) Clustering groups vulnerabilities based on their likelihood of successfully misleading ML models, aiding the GNN in prioritizing high-ASR attacks. (Bottom Left) Stealth and Detectability Clustering categorizes vulnerabilities based on their evasion capability, enabling the GNN to refine predictions for harder-to-detect threats. (Bottom Right) Computational Cost and Practicality Clustering differentiates between low-cost and resource-intensive attacks, helping the GNN assess real-world adversarial feasibility. These visualizations collectively demonstrate how GNN-driven learning enhances vulnerability classification, improves risk prediction, and refines cybersecurity prioritization beyond traditional CVSS scoring.
}

\section{Discussions of results}~\label{sec:discussions}
\tosemrev{
The findings of this study reveal critical vulnerabilities and threats that affect machine learning (ML) systems throughout their lifecycle, from data pre-processing to deployment and operational stages. By analyzing a comprehensive set of data from multiple sources, including the MITRE ATT\&CK and ATLAS frameworks, the AI Incident Database, and GitHub repositories, this study highlights the multifaceted nature of security risks in ML systems. Vulnerabilities were found to span not only traditional software vulnerabilities but also ML-specific attack vectors, such as adversarial examples, data poisoning, and model extraction.
A key insight is the significant role of dependencies in amplifying ML vulnerabilities. Libraries such as TensorFlow, PyTorch, and OpenCV were identified as recurrently targeted due to their expansive dependency chains. For instance, vulnerabilities in dependencies like Log4j and Pickle were shown to cascade across the ML ecosystem, affecting downstream components and deployment environments. These findings underscore the interconnected nature of ML systems and highlight the urgent need for holistic approaches to vulnerability management that extend beyond individual tools to their dependencies.
This research also underscores the limitations of existing threat models, such as ATLAS, in capturing the full spectrum of real-world vulnerabilities and threats. While ATLAS provides a valuable framework for cataloging adversarial tactics, the integration of real-world incidents from the AI Incident Database and GitHub repositories revealed numerous threats not documented in ATLAS. This gap highlights the dynamic nature of ML security threats and the need for continuously updated and enriched threat models to reflect emerging risks.
Another key result is the mapping of threats to specific stages of the ML lifecycle, providing actionable insights into where systems are most vulnerable. For example, data poisoning and adversarial training attacks predominantly target the training phase, while model extraction and API exploitation occur more frequently during the deployment phase. By understanding these stage-specific vulnerabilities, stakeholders can implement targeted mitigation strategies.
Finally, this study demonstrates the critical importance of integrating proactive and reactive measures into the security lifecycle of ML systems. Proactive measures, such as adversarial training and secure coding practices, can prevent vulnerabilities from being introduced. Reactive measures, such as real-time monitoring, incident response, and automated patching, ensure rapid containment of threats when they do occur. The proposed mitigation matrix (Fig.~\ref{fig:defend_matrix}) synthesizes these measures into a comprehensive framework, enabling stakeholders to address threats holistically and dynamically.
}

\subsection{Mitigation of Vulnerabilities and Threats}~\label{sec:mitigation}
\tosemrev{
To address vulnerabilities and threats in ML systems, a combination of proactive and reactive security measures should be implemented across all levels of the ML lifecycle. Traditional security mechanisms such as access control, encryption, and network defenses should be complemented by ML-specific defenses, including adversarial training, robust model architectures, and secure data handling practices. Federated learning (FL), while enhancing data privacy by enabling decentralized training, introduces critical security risks. \textbf{Data poisoning attacks} (e.g., model and backdoor poisoning) allow malicious clients to manipulate updates, degrading model integrity. \textbf{Privacy attacks} such as membership inference and gradient leakage exploit model updates to reconstruct private training data. \textbf{Aggregation exploits} compromise federated averaging by injecting biased updates, while \textbf{communication attacks} (e.g., MITM, DoS) disrupt training. To mitigate these threats, \textbf{robust aggregation techniques} (e.g., Krum, trimmed mean, differential privacy-based aggregation) filter adversarial contributions. \textbf{Secure update mechanisms} like homomorphic encryption (HE) and secure multi-party computation (SMPC) prevent data leakage. \textbf{Adversarially robust learning} strengthens defenses via federated adversarial training and Byzantine-resilient optimization. \textbf{Privacy-preserving techniques} (e.g., gradient noise injection, secure aggregation protocols) mitigate inference attacks, while \textbf{blockchain-based FL} enhances integrity and decentralization. Future research should explore automated anomaly detection, post-quantum security, and regulatory compliance to ensure robust FL deployment. 
By integrating these defenses with established frameworks such as MITRE ATT\&CK~\cite{mitre_attack_mitigation}, D3FEND~\cite{mitre_defend}, and NIST security guidelines~\cite{nist_guidelines}, ML systems can be fortified against evolving adversarial threats. The proposed ML Threat Mitigation Matrix provides a structured approach to proactive risk management, covering critical vulnerabilities from data collection to deployment and ensuring a comprehensive security posture.
}
\subsubsection{Mitigation Strategies Across Levels}
\tosemrev{
\noindent \textbf{\\Data-Level Mitigation. }
\begin{enumerate}[left=0pt]
\item \textbf{Hardening Data Pipelines: } Adversarial defenses~\cite{tabassi2019taxonomy} and tools like ART, CleverHans, and Foolbox help mitigate adversarial attacks targeting datasets.
\item \textbf{Data Protection: } TLS encryption secures data in transit, while AES-based encryption safeguards data at rest. Dynamic analysis enhances protection during use~\cite{scarfone2008sp}.
\item \textbf{Access Control: } Identity and Access Management (IAM) policies enforce the principle of least privilege~\cite{mccarthy1800identity}, minimizing unintended access to critical assets.
\item \textbf{Adversarial Detection: } Techniques such as Introspection~\cite{aigrain2019detecting}, Feature Squeezing~\cite{xu2017feature}, and SafetyNet~\cite{lu2017safetynet} provide robust defenses against adversarial inputs.
\item \textbf{Sanitization of Data: } Compromised data can be sanitized using denoisers~\cite{xie2019feature} to ensure integrity before use.
\end{enumerate}
\noindent\textbf{Software-Level Mitigation. }
\begin{enumerate}[left=0pt]
\item \textbf{Vulnerability Scanning: } Tools like GitHub Code Scanning, OSS-Fuzz, and SonarQube~\cite{scarfone2008technical} detect and mitigate vulnerabilities in ML libraries and pipelines.
\item \textbf{Secure Configurations: } Proper IAM configurations and the enforcement of runtime restrictions protect against privilege escalation.
\item \textbf{Regular Updates: } Patch management ensures that vulnerabilities in libraries and dependencies are promptly addressed.
\end{enumerate}
\noindent \textbf{Database-Level Mitigation. }
\begin{enumerate}[left=0pt]
\item Access Control: IAM policies limit database access, while regular backups stored off-network protect against data loss~\cite{chandramouli2020security}.
\item \textbf{Integrity Monitoring: } Continuous monitoring ensures early detection of unauthorized changes.
\end{enumerate}
\noindent  \textbf{System and Network-Level Mitigation }
\begin{enumerate}[left=0pt]
\item \textbf{Endpoint Protection: } OS hardening, such as enabling Secure Boot and enforcing automatic updates, protects system endpoints~\cite{souppaya2013guide}.
\item \textbf{Network Defenses: } Firewalls~\cite{wack2002guidelines}, intrusion prevention systems (IPS)~\cite{scarfone2007guide}, and encrypted VPN tunnels~\cite{frankel2005guide} secure communication between endpoints.
\item \textbf{Traffic Analysis: } Network access control lists (ACLs) and monitoring systems like Snort and Zeek detect and respond to malicious activity.
\end{enumerate}
\noindent \textbf{Cloud-Level Mitigation }
\begin{enumerate}[left=0pt]
\item \textbf{Zero-Trust Principles: } Role-based permissions, multi-factor authentication, and secure policies enforced through CASB and SASE frameworks ensure robust cloud security~\cite{rose2020zero}.
\item \textbf{Real-Time Monitoring: }Cloud-based SIEMs like Splunk Cloud and Azure Sentinel detect threats across hybrid infrastructures.
\item \textbf{Hybrid Cloud Configurations: } Bastion and transit networks enhance flexibility and security in multi-cloud setups.
\end{enumerate}
\noindent\textbf{Integration with ML Lifecycle. }
The proposed threat mitigation matrix integrates seamlessly into the end-to-end ML lifecycle, from data collection to deployment. For example, compromised training data can be isolated, sanitized, and replaced in real-time, while models deployed on cloud infrastructures benefit from automated security orchestration using serverless functions and APIs. This approach parallels the functionality of existing Security Orchestration, Automation, and Response (SOAR) systems but is explicitly tailored for ML environments.\\
\noindent \textbf{Continuous Threat Monitoring. }
The study emphasizes the importance of continuous threat assessment in ML systems. Dependencies such as curl/libcurl, which have repeatedly impacted TensorFlow, highlight the need for proactive monitoring and periodic vulnerability scans. The proposed framework ensures that ML engineers can identify and respond to emerging threats, even after initial mitigation has been applied.
}

\majorev{
\subsection{Vulnerabilities in SoTA LLMs}
\subsubsection{Landscape Overview and Emergent Threats. }
 Recent advancements in SoTA models (LLMs) have unveiled a new class of high-severity vulnerabilities that transcend traditional adversarial threat vectors. Our study reveals that models such as GPT‑4o, Claude‑3.5, Gemini‑1.5, LLaMA‑3.2, and DeepSeek‑R1 remain susceptible to priming-based jailbreaks, prompt injection, and tokenization attacks that bypass alignment filters and exploit low-level tokenizer mechanics (e.g., TokenBreak)~\cite{ge2025llms,kwon2024text}. Moreover, emerging backdoor threats—including composite multi-trigger attacks—have demonstrated 100\% success rates even under stringent RLHF and adversarial training regimes, as evidenced in attacks on LLaMA‑7B~\cite{alber2025medical}. These vulnerabilities span multiple system levels, from input and prompt manipulation to representation-level exploits enabled by subliminal learning and fine-tuning with contaminated synthetic data~\cite{Curvo2025TheTD,Chia2025ProbingLS}.
 Our layered mapping (Fig.~\ref{fig:ttps-vul-stages}) not only identifies current blind spots in model deployment pipelines but also reinforces the importance of our proposed mitigation matrix, which we further illustrate using a real-world scenario involving a composite backdoor attack in an LLM-as-a-service setting.
 \subsubsection{Data-Level Vulnerabilities: Poisoning and Latent Triggers. }
 Modern LLMs are critically exposed at the data layer. Adversaries can embed poisoned samples or trigger phrases during fine-tuning, often escaping detection while inducing harmful behavior. For instance, modifying just 0.001\% of training data led to biased outputs in medical LLMs, despite passing standard evaluations~\cite{alber2025medical}. These results validate Fig.~\ref{fig:ttps-vul-stages}’s emphasis on data poisoning and latent trigger injection.
 \subsubsection{Software-Level Vulnerabilities: Tokenizers and Unsafe Libraries}
 The software stack remains a significant source of risk. Tokenization-based exploits such as TokenBreak allow attackers to bypass filters by modifying a single character~\footnote{\url{ https://www.techradar.com/pro/security/this-cyberattack-lets-hackers-crack-ai-models-just-by-changing-a-single-character}}. This cyberattack lets hackers crack AI models just by changing a single character. Third-party dependencies, such as pickle and unsafe regex libraries, compound this risk. A systematic review showed how LLMs amplify or overlook code-level vulnerabilities due to insecure prompting~\cite{basic2024large}. Fig.~\ref{fig:ttps-vul-stages} captures these under improper input validation and supply chain compromise.
 \subsubsection{Storage and System-Level Vulnerabilities: Model Hijacking. }
 Storage and system vulnerabilities are increasingly relevant with containerized LLM deployments. Threats such as firmware tampering, API spoofing, and model hijacking remain underexplored yet highly impactful. Tools like LLM4CVE demonstrate how automated repair is possible—but also illustrate how easily vulnerabilities propagate without hardening at deployment time~\cite{fakih2025llm4cve}.
 \subsubsection{Network-Level Vulnerabilities: API Exploits and Model Extraction. }
 Unsecured inference APIs remain vulnerable to a range of remote threats, including model extraction, indirect prompt injection, and membership inference. Models like GPT‑4o and Gemini‑1.5 are known to be susceptible to jailbreaks disguised as benign academic language~\cite{ge2025llms,kwon2024text}. These threats align with the categories of misconfigured APIs, lack of input filtering, and improper authentication, as shown in Fig.~\ref{fig:ttps-vul-stages}.
 \subsubsection{Lifecycle Propagation: Multi-Stage Exploit Chains. }
 Our threat model confirms that vulnerabilities cascade across layers. A poisoned dataset can introduce a backdoor that later enables prompt injection during inference. Such multi-stage exploits are shown in Fig.~\ref{fig:ttps-vul-stages}, revealing that the attack surface is not static—it grows with model capacity, context length, and autonomy.
 \subsubsection{Strategic Implications: Security Must Co-Evolve with Capability. }
As SoTA models (LLMs) advance toward greater autonomy, with features like memory, planning, and tool use, their vulnerability landscape becomes increasingly complex. Our analysis indicates that higher model capability does not inherently confer greater security. Instead, increased complexity—through expanded APIs, extended context windows, and dynamic memory—broadens the attack surface and opens pathways for more sophisticated exploits~\cite{ge2025llms,Chia2025ProbingLS,Shen2023DoAN}. To keep pace with these risks, security must evolve alongside model capability. Accordingly, defense strategies should be systemically layered, targeting every stage illustrated in Fig.~\ref{fig:ttps-vul-stages}: from input filtering and dataset validation to software hardening, model verification, and runtime monitoring. Without such a shift, SoTA models risk becoming sophisticated, yet opaque systems, where persistent and exploitable weaknesses undermine increasing functionality.
}

\minor{
\subsection{Discussion of Scalability}
It should be noted that our three-step mapping process scales efficiently to new ML pipeline types such as multimodal, instruction-tuned, or RLHF-augmented models. Each stage of the pipeline—(i) retrieval of relevant TTPs, (ii) ontology linking, and (iii) GNN-based reasoning—is fully modular. Adding new component types requires only the definition of additional schema entities and relations in the ontology, while existing mappings remain valid. The retrieval-augmented classifier adapts automatically through zero-shot
prompting without the need to retrain earlier stages. Formally, the computational cost scales linearly with the number of added nodes and edges ($\mathcal{O}(n{+}m)$), ensuring tractable scaling for large, heterogeneous pipelines. This modularity allows the framework to evolve alongside advances in model architectures, including multimodal encoders, diffusion decoders, and autonomous-agentic systems. \textit{In practice, this scalability was validated by extending the ontology and mapping pipeline to support both text-based
LLMs and multimodal diffusion models without retraining prior components.}
}

\minor{
Because our methodology operates through an AI agency that implements an \emph{automatic RAG system}, the ontology remains \emph{live and self-updating}. The system continuously mines and classifies new TTPs from the literature and public repositories, allowing real-time integration of emerging attack patterns into the multi-agent mapping framework. Consequently, the list of reported TTPs, vulnerabilities, and ML lifecycle stages naturally expands as new cases appear in the ecosystem. After ontology normalization, deduplication, and integration of new \emph{state-of-the-art} multimodal and preference-guided attack techniques~\cite{zhang2025askingfordirections}, the unified threat graph now encodes \textbf{73 distinct TTPs}, \textbf{27 vulnerabilities}, and \textbf{10 ML lifecycle stages}. 
}

\minor{
Overall, this extended discussion highlights that the proposed mapping framework and threat taxonomy remain robust and adaptable to the next generation of multimodal and large-language models, reinforcing the scalability and generalizability of our findings.
\paragraph{What we learned.}
Recent work~\cite{zhang2025askingfordirections} shows that LLMs’ own comparative judgments can be elicited to drive text-only, query-based optimization of jailbreaks, prompt injections, and vision-LLM adversarial examples. This \emph{preference-oracle} approach removes the need for logits or surrogate models and, paradoxically, becomes more effective on larger, better-calibrated models. For practitioners, this widens the attack surface of production APIs that only expose text and calls for stateful detection, policy-level refusal of preference elicitation, and anti-optimization jitter in guardrails. We incorporate this attack class into our lifecycle-centric mapping and mitigation matrix to ensure coverage of introspection-based, black-box threats.
}

\tosemrev{
\subsection{Implications for Different Stakeholder Groups}
\subsubsection{Cybersecurity Practitioners. }
\textbf{\\ Proactive Threat Mitigation:} Practitioners must prioritize identifying and mitigating vulnerabilities in ML repositories and dependencies. Regular penetration testing, vulnerability scans, and secure configuration audits should be integrated into the ML development lifecycle.\\ \textbf{Threat Intelligence Integration:} Leveraging insights from databases like ATLAS, the AI Incident Database, and other emerging repositories can enhance proactive defense strategies. This includes updating threat detection rules and training models to identify adversarial patterns in real-time.\\ \textbf{Incident Response Readiness:} Given the dynamic nature of ML threats, practitioners need robust incident response plans tailored to address both traditional and ML-specific attack vectors, such as adversarial examples or model extraction attempts.
\subsubsection{Academics and Researchers. }
\textbf{\\ Advancing Threat Models: } Researchers have a critical role in refining and expanding existing threat models like ATLAS. By integrating real-world vulnerabilities from diverse sources, academics can develop comprehensive frameworks that reflect the latest threat landscape.\\ \textbf{Lifecycle-Specific Defenses: } There is a need for research into stage-specific defenses, such as robust training methods to counter data poisoning or cryptographic techniques to secure model inference APIs.\\ \textbf{Interdisciplinary Collaboration: } Collaboration between security, AI, and domain experts is essential to address the multifaceted challenges posed by ML threats. This includes studying the socio-technical impacts of AI vulnerabilities in critical domains like healthcare or transportation.
\subsubsection{Regulation Agencies. }
\textbf{\\ Regulatory bodies} must establish comprehensive guidelines for the secure development, deployment, and maintenance of ML systems. These guidelines should include robust dependency management practices, regular security audits, and compliance with established standards like ISO/IEC 27001 and the NIST AI Risk Management Framework (AI RMF). Such measures ensure that organizations proactively address vulnerabilities and align with best practices.\\
\textbf{Incident Reporting Frameworks. }
Agencies should mandate the disclosure of AI-related security incidents to build a centralized database of vulnerabilities and threats. This database would inform future policies, enhance threat intelligence sharing, and foster greater transparency in ML security. Mandatory frameworks, akin to the NIS2 Directive, are crucial for documenting adversarial attacks, dependency vulnerabilities, and system failures across sectors.\\
\textbf{Global Collaboration. }
Given the international scope of AI and ML systems, regulatory agencies must collaborate to standardize security practices and promote cross-border knowledge sharing. Frameworks like the EU AI Act and the OECD AI Principles can serve as foundational models for harmonizing security standards and fostering collective resilience.\\
\textbf{Regulatory Directions. }
Dependency vulnerabilities remain a significant and evolving threat to ML systems. Regulatory bodies could introduce new guidelines requiring organizations to implement automated dependency monitoring, patch management, and threat detection systems. Inspired by the Digital Operational Resilience Act (DORA), similar resilience frameworks should be expanded to encompass AI and ML systems in critical infrastructure sectors. 
To address cascading risks in AI supply chains, regulations could mandate transparency in dependency usage, including the disclosure of vulnerabilities in third-party libraries. This aligns with software supply chain regulations like the U.S. Executive Order 14028, which emphasizes the importance of software bills of materials (SBOMs) for secure supply chains. Such measures would ensure the robustness and resilience of ML systems in an increasingly interconnected ecosystem.
\subsubsection{Tool Builders and Developers. }
\textbf{\\Building Secure ML Tools: } Developers of ML tools and frameworks must integrate security features such as automated vulnerability detection, secure dependency handling, and built-in adversarial robustness mechanisms.\\
\textbf{Dependency Management:} Tool builders should implement mechanisms to monitor, update, and secure third-party libraries and dependencies. Providing users with transparency about known vulnerabilities in dependencies can prevent cascading failures.\\
\textbf{User-Friendly Security Enhancements:} Tools should include easy-to-use security features, such as pre-built models with adversarial training or APIs that detect malicious inputs. Making security accessible to non-expert users is crucial for widespread adoption.
}

\section{Threats to validity}~\label{threat2valid}
\tosemrev{
Our empirical study of ML security threats integrates multiple threat intelligence sources, like ATLAS, ATT\&CK, the AI Incident Database, and GitHub repositories. While our methodology aims to provide a comprehensive assessment, several \textbf{threats to validity} must be acknowledged. Thus, we adhered to the methodological principles outlined by Wohlin et al.~\cite{wohlin2012experimentation} and Juristo \& Moreno~\cite{juristo2013basics} to systematically identify, assess, and mitigate potential threats to validity.
Throughout this empirical study, we adhered to the methodological principles outlined by Wohlin et al.~\cite{wohlin2012experimentation} and Juristo \& Moreno~\cite{juristo2013basics} to systematically identify, assess, and mitigate potential threats to validity.\\
\textbf{ Construct Validity: }  
Construct validity concerns whether the variables and metrics used in our study accurately represent the underlying theoretical constructs. Our analysis relies on multiple threat databases, including ATLAS, ATT\&CK, and the AI Incident Database, which could introduce biases due to differences in how security incidents and adversarial behaviors are categorized. To mitigate this, we triangulated findings across diverse sources to ensure alignment with established ML security taxonomies. Additionally, our mapping of TTPs to ML lifecycle stages was validated by expert reviews to ensure consistency and correctness.\\
\textbf{ Internal Validity: }  
Internal validity pertains to the causal relationships inferred from the data. One potential threat arises from automated extraction and processing of security vulnerabilities from GitHub repositories and the AI Incident Database, which may contain duplicate or misclassified entries. To address this, we implemented rigorous data cleaning and filtering techniques and manually reviewed a subset of cases for validation. Moreover, our identification of emerging TTPs relied on historical trends, which may not fully account for evolving attack strategies. We mitigated this risk by incorporating recent high-impact incidents and cross-referencing with real-world security reports.\\
\textbf{ External Validity: }  
External validity concerns the generalizability of our findings beyond the studied datasets. While our approach integrates multiple sources, some niche ML threats may remain underrepresented, particularly those targeting proprietary or closed-source models. Additionally, the ML ecosystem evolves rapidly, and newly discovered vulnerabilities may not be reflected in our dataset immediately. To enhance external validity, we included a wide range of attack scenarios from different ML domains (e.g., NLP, vision, federated learning) and systematically updated our dataset with newly reported incidents.\\
\textbf{ Conclusion Validity: }  
Conclusion validity relates to the reliability and statistical significance of our findings. Our study identifies the most prominent TTPs and vulnerabilities based on their frequency and impact, but the absence of a standardized threat severity metric across different databases introduces some uncertainty. To mitigate this, we employed statistical techniques such as frequency distributions and cross-dataset correlations to ensure robust conclusions. Additionally, our threat modeling and dependency analysis were designed to minimize biases in prioritizing security risks.\\
\textbf{ Reliability: }  
Reliability concerns the reproducibility of our findings. We documented our methodology comprehensively, providing detailed steps for data collection, preprocessing, and analysis. However, some elements of our study, such as expert validation of TTP mappings, introduce a degree of subjectivity. To enhance reproducibility, we released our datasets and code where possible, allowing independent verification. Future research can build on our framework by extending the dataset and refining classification methodologies to further validate our conclusions.
}

\section{Conclusion}~\label{conclusion}
The increasing sophistication of adversarial tactics targeting ML systems underscores the urgent need for a robust and adaptive security framework. In this study, we conducted a \textbf{comprehensive analysis of ML threat behaviors} by aggregating insights from multiple sources, including ATLAS, AI Incident Database reports, GitHub ML repositories, and the PyPA database. Our findings reveal \textbf{critical security gaps} in existing threat models, particularly within widely used ML repositories, underscoring the necessity for continuous monitoring, dependency analysis, and proactive mitigation strategies.
\majorev{
We identified Transformers as one of the most frequently targeted architectures, with 25.4\% against CNNs (19.05\%) in real-world attack scenarios. The testing, inference, and training phases emerged as the most vulnerable ML lifecycle stages.
}
\textbf{Buffer overflow} and \textbf{denial-of-service (DoS) attacks} were the most prevalent threats across ML repositories, while dependency analysis exposed security risks in TensorFlow, OpenCV, and Jupyter Notebook, particularly in libraries such as \textit{pickle, joblib, numpy116, python3.9.1,} and \textit{log4j}. Additionally, our study contributes 32 previously undocumented ML attack scenarios, encompassing 17 new techniques and 13 tactics, providing a valuable extension to ATLAS case studies for future research.
To bridge the gap between theoretical threat models and real-world attack mitigation, we introduced an ML Threat Mitigation Matrix that maps real-world threats to potential defensive strategies. By incorporating GNN-based analysis and clustering techniques, we demonstrated how risk prediction models can enhance ML vulnerability classification, refine attack severity estimation, and improve overall risk assessment.

\subsection{Future research directions.}
Our future work focuses on advancing AI-driven ML threat assessment by integrating Generative AI for adversarial simulation, enabling autonomous threat modeling and predicting zero-day attacks. We aim to expand real-time threat intelligence aggregation by incorporating feeds from multiple sources, including CISA KEV, the AI Incident Database, and Dark Web intelligence, to enhance adaptive risk scoring. Additionally, we plan to develop Reinforcement Learning (RL)-based self-improving security mechanisms that dynamically optimize ML defenses against evolving threats. Lastly, we will extend ATLAS-based security frameworks with automated security governance, providing real-time attack prediction, defense adaptation, and compliance recommendations to fortify ML systems against adversarial threats.\\
\majorev{
\subsubsection{retrospective testing \& red-teaming. }
Moreover, we invite the community to pursue empirical evaluation that lies beyond the scope of our present study. First, a \emph{retrospective incident analysis}—in which pipeline predictions are compared with the post-mortem labels of publicly reported failures (e.g., the 112 cases archived in the AI Incident Database)—would quantify real-world accuracy and could be scored with inter-annotator measures such as Cohen's~$\kappa$.
Second, a \emph{large-scale red-team campaign} against a production-grade MLOps stack (e.g., Azure ML deployed on Kubernetes) would expose the pipeline to adversaries operating under genuine operational constraints, thereby revealing failure modes that synthetic benchmarks cannot capture.
Systematic investigations along these two axes would provide the empirical grounding needed to translate laboratory-grade defenses into dependable, field-tested safeguards.
}

\section*{Author Contributions}
\textbf{\textit{Armstrong Foundjem}}: Conceptualization, Methodology, Data Curation, Software, Formal Analysis, Writing – Review \& Editing, Validation, and Visualization. \textbf{\textit{Lionel Nganyewou Tidjon}}: Writing – Original Draft, Data Curation, Validation, Visualization. \textbf{\textit{Leuson Da Silva}}: Writing – Review \& Editing, Repository Mining. \textbf{\textit{Foutse Khomh}:} Supervision, Writing – Review \& Editing, and Funding Acquisition.

\section*{Acknowledgment}

This work is partly funded by the Fonds de Recherche du Québec (FRQ), Natural Sciences and Engineering Research Council of Canada (NSERC), Canadian Institute for Advanced Research (CIFAR), and Mathematics of Information Technology and Complex Systems (MITACS).

\bibliographystyle{IEEEtran}
\bibliography{references}

%








\end{document}